\documentclass[prl,twocolumn,superscriptaddress,showpacs,notitlepage,floatfix,bibnotes,nofootinbib]{revtex4-2}
\usepackage{float}
\usepackage{graphicx}% Include figure files
\usepackage{dcolumn}% Align table columns on decimal point
\usepackage{bm}% bold math
\usepackage[figurename=Figure]{caption}
\usepackage{verbatim} % For comments and other
\usepackage{amsmath}  % For math
\usepackage{amssymb}  % For more math
\usepackage{listings} % For source code
\usepackage{subfig}   % For subfigures
\usepackage{graphics}
\usepackage{latexsym,epsfig}

\usepackage[colorlinks=true,breaklinks=true,allcolors=blue]{hyperref}
\usepackage{lipsum}
\usepackage{dsfont}

\usepackage[scr=boondox]{mathalfa}

\def\Oo{\ensuremath{{\cal O}}} %Order sign
\def\Hh{\ensuremath{{\cal H}}} %the Hamiltonian

\usepackage[scr=boondox]{mathalfa}

\usepackage[dvipsnames]{xcolor}
\usepackage[normalem]{ulem}
\usepackage{xcolor}

\usepackage{tikz,bm} % Useful for drawing plots
\usepackage{circuitikz}
\usepackage{makecell}

\definecolor{mycolor}{rgb}{1,0.2,0.3}
\definecolor{brightgreen}{rgb}{0.4, 1.0, 0.0}
\definecolor{britishracinggreen}{rgb}{0.0, 0.26, 0.15}
\definecolor{cadmiumgreen}{rgb}{0.0, 0.42, 0.24}
\definecolor{ceruleanblue}{rgb}{0.16, 0.32, 0.75}
\definecolor{darkelectricblue}{rgb}{0.33, 0.41, 0.47}
\definecolor{darkpowderblue}{rgb}{0.0, 0.2, 0.6}
\definecolor{darktangerine}{rgb}{1.0, 0.66, 0.07}
\definecolor{emerald}{rgb}{0.31, 0.78, 0.47}
\definecolor{palatinatepurple}{rgb}{0.41, 0.16, 0.38}
\definecolor{pastelviolet}{rgb}{0.8, 0.6, 0.79}

\begin{document}
	
	\title{Temporal Entanglement Transitions in the Periodically Driven Ising Chain}% Force line breaks with \\
	%\thanks{A footnote to the article title}%
	\author{Karun Gadge}
	\email{karun.gadge@uni-goettingen.de}
	\affiliation{Institute for Theoretical Physics, Georg-August-Universit\"{a}t G\"{o}ttingen, 37077 G\"ottingen, Germany}
	\author{Abhinav Prem}
	\email{aprem@bard.edu}
	\affiliation{Physics Program, Bard College, 30 Campus Road, Annandale-on-Hudson, NY 12504, USA}
	\author{Rishabh Jha}
	\email{rishabh.jha@uni-goettingen.de}
	\affiliation{Institute for Theoretical Physics, Georg-August-Universit\"{a}t G\"{o}ttingen, 37077 G\"ottingen, Germany}

	\begin{abstract}
Periodically driven quantum systems can host non-equilibrium phenomena without static analogs, including in their entanglement dynamics. Here, we discover \textit{temporal entanglement transitions} (TET) in a Floquet spin chain, which correspond to a quantum phase transition in the spectrum of the entanglement Hamiltonian and are signaled by dynamical spontaneous symmetry breaking. We identify the symmetry principles underlying these transitions: they appear when the driven Hamiltonian preserves global symmetry (here, $\mathbb{Z}_2$), the initial state respects this symmetry, and the reduced density matrix carries weight in both subsystem-parity sectors, with TET occurring precisely when the sector weights become equal (given the previous two conditions are also satisfied). Intriguingly, we find these transitions across a broad range of driving frequencies (from adiabatic to high-frequency regime) and independently of drive details, where they manifest as periodic, sharp entanglement spectrum reorganizations marked by the Schmidt-gap closure, a vanishing entanglement echo, and symmetry-quantum-number flips, while remaining invisible to conventional local observables. At high frequencies, the entanglement Hamiltonian acquires an intrinsic timescale decoupled from the drive period, rendering the transitions genuine steady-state features. Finite-size scaling reveals universal critical behavior with correlation-length exponent $\nu=1$, matching equilibrium Ising universality despite its emergence from purely dynamical mechanisms decoupled from static criticality. Our work establishes TET as novel features in Floquet quantum matter.
	\end{abstract}
	
	%\date{\today}
	
	\maketitle
	
	%%%%%%%%%%%%%%%%%%%%%%%%%%%%%%%%%
	%%%%%%%%%%%%%%%%%%%%%%%%%%%%%%%%%

	\textit{Introduction.---} Periodically driven quantum many-body systems are a powerful platform for exploring quantum phases inaccessible in equilibrium, opening the door to the coherent control and engineering of quantum matter via time-periodic fields~\cite{goldman2014periodically,oka2019floquet,khemani2019review,else2020review,harper2020review}. The study of Floquet driven systems has unveiled remarkable phenomena ranging from Floquet topological insulators~\cite{rudner2020floquet, cayssol2013floquet} and time crystals~\cite{Berdanier2018Aug, ippoliti2021many, riera2019time} to dynamical localization~\cite{tiwari2024dynamical, nag2014dynamical} and prethermalization~\cite{PhysRevLett.109.257201, Abanin2015Dec, Machado2019Dec}. 
	While extensive research has focused on conventional observables like magnetization and transport properties, exploring the entanglement structure of driven quantum systems remains an active area of research, given its fundamental role in characterizing quantum phases and phase transitions~\cite{zhao2022scaling, cho2017quantum}. Of particular interest is the entanglement spectrum (ES)~\cite{lihaldane}, which provides direct access to the \textit{entanglement Hamiltonian} and encodes information beyond that contained in the entanglement entropy alone. The ES serves as a powerful diagnostic for capturing the universal features of quantum phases in equilibrium, especially in the context of gapped ground states but also for critical points~\cite{regnault1,papic,chandran2011,dubail2012,cano2015,qi2012,swingle2012,prodan2010,pollmann2010,turner2010,fidkowski2010,alba2012,choo2018,ho1,ho2,berg2017,stringnetespec,premespec,redon2024realizing, schneider2022entanglement,dalmonte2022entanglement, demidio2024universal}. There is growing recognition that the ES also reveals distinctive signatures of quantum chaos, thermalization, and criticality out of equilibrium~\cite{baykusheva2023witnessing, lewis2019unifying, hahn2024eigenstate,wen2025critical}.

	Recent work has begun exploring the interplay between driving and entanglement, including for Floquet-driven conformal field theories~\cite{wen2018floquet, fan2021floquet, berdanier2017floquet, lin2025local} and in driven-dissipative systems~\cite{stannigel2012driven, zippilli2013entanglement, chen2024periodically}. Particularly relevant are studies of Page curves~\cite{kehrein2024page, jha2025page, Li2025Jul}, where entanglement transitions emerge from competition between entanglement generation and relaxation in the presence of conserved charges. While equilibrium quantum phase transitions are intimately connected to entanglement scaling and the structure of the many-body ground state~\cite{wei2018linking, bhattacharyya2015signature, cho2017quantum}, the analogous relationship in driven systems remains unexplored. Since driven systems can access dynamical phases with no equilibrium counterpart, this suggests new entanglement phenomena waiting to be uncovered. 
	However, whether non-analytic entanglement transitions can occur in generic periodically-driven systems remains an open question.
	
	In this Letter, we report the discovery of \textit{temporal entanglement transitions} (TET) via dynamical spontaneous symmetry breaking in the entanglement Hamiltonian (EH) of a periodically driven spin chain.
	These transitions require $\mathbb{Z}_2$-symmetric driven Hamiltonian and initial state, with critical times occurring at sector-weight degeneracies (made precise below), and manifest through three synchronized signatures: Schmidt gap closure~\cite{dechiara2012entanglement, wald2020closure, bayat2014order}, vanishing entanglement echo~\cite{poyhonen2021entanglement}, and parity flips~\cite{jaeger2003entanglement, liu2023symmetry, alba2021entanglement}, while remaining invisible to conventional observables (magnetization, Loschmidt echo).
	Crucially, these transitions occur across all driving frequencies and exhibit universal critical behavior with correlation length exponent $\nu=1$, matching the equilibrium Ising universality class~\cite{Koziol2021Jun, Sachdev_2011}, despite emerging purely dynamically, decoupled from any underlying equilibrium criticality. 
	
	Our findings establish TET as a distinct class of non-equilibrium phenomenon that is uniquely detectable through entanglement measures while remaining hidden from conventional observables, revealing that the EH itself can undergo quantum phase transitions. At high frequencies, we find that the EH gains an intrinsic timescale decoupled from the drive, rendering its Floquet-periodic transitions as genuine steady-state features.
	The universal critical behavior persists whenever the above $\mathbb{Z}_2$-symmetry condition and sector-weight equality are satisfied (including across different equilibrium phases that preserve $\mathbb{Z}_2$), confirming a fundamentally non-equilibrium character decoupled from the underlying equilibrium phase diagram (see Supplemental Material (SM)~\cite{Supplemental} for details).

	%%%%%%%%%%%%%%%%%%%%%%%%%%%%%%%%%
	%%%%%%%%%%%%%%%%%%%%%%%%%%%%%%%%%
	
	\begin{figure}[htbp!]
		\centering
		\includegraphics[width=0.9\columnwidth]{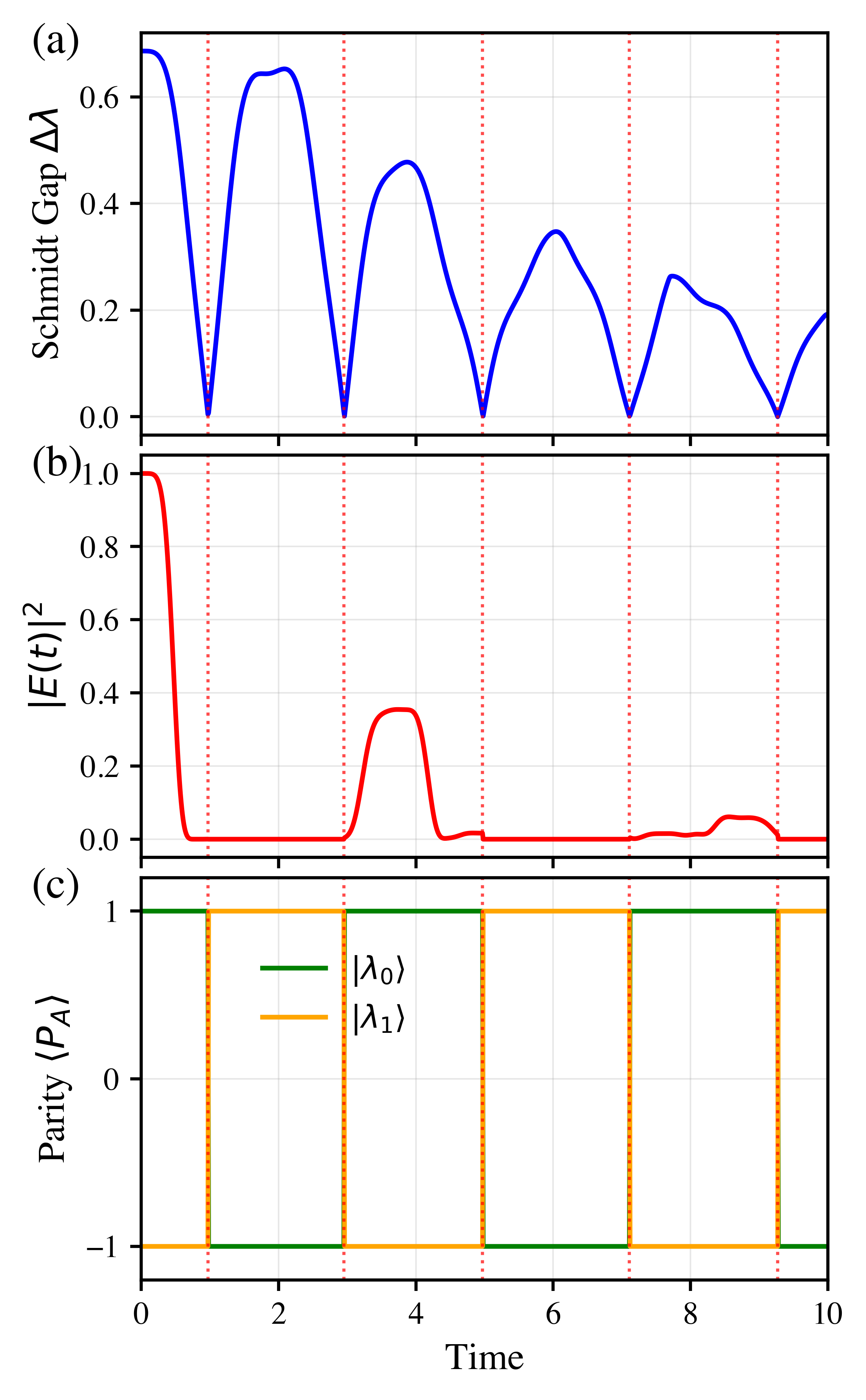}
		\caption{
			Temporal entanglement transitions (TET) in the driven TFIM. \textbf{(a)} Schmidt gap $\Delta\lambda = \lambda_0 - \lambda_1$ closing periodically at critical times. \textbf{(b)} Entanglement echo $|E(t)|^2$ vanishing at odd critical times $t_c^{(k=\text{odd})}$, as the dominant Schmidt vector $|\lambda_0\rangle$ orthogonalizes into a distinct symmetry sector. \textbf{(c)} Subsystem parity expectations showing dynamical spontaneous $\mathbb{Z}_2$ symmetry breaking in the entanglement ground state (green) and complementary behavior in the first-excited state (orange). The alternating parity pattern physically underlies the echo's behavior. Red vertical lines mark critical times of symmetry breaking, and are overlaid across all panels to provide a synchronized, consistent picture. Parameters: $L=24$, $L_A=9$, $J=1.0$, $h_0=2.0$, $\omega=5.0$, $dt=0.01$.
		}
		\label{fig:temporal_transitions}
	\end{figure}

	\textit{Model and Results.---}
	We consider the transverse-field Ising model (TFIM) with open boundary conditions, consisting of $L$ spin-$1/2$ particles with time-dependent Hamiltonian
	\begin{equation}
		\Hh(t) = -J\sum_{i=1}^{L-1} \sigma_i^z \sigma_{i+1}^z - h(t)\sum_{i=1}^{L} \sigma_i^x,
	\end{equation}
	where $J$ is the Ising coupling, $h(t) = (h_0/2)\cos(\omega t)$ is the oscillating transverse field, and $\sigma_i^{\alpha}$ are Pauli matrices. We initialize the system in the ground state of the static Hamiltonian $\Hh_{\text{static}} = \Hh(t)|_{t=0}$ and evolve under $\Hh(t)$ using time-dependent variational principle (TDVP)~\cite{tdvp-1, tdvp-2, tdvp-3} for matrix product states (MPS)~\cite{schollwock2011density}, as implemented in the \texttt{ITensors} library~\cite{Fishman2022Aug} (see SM~\cite{Supplemental}). Note that our results are independent of the specific Floquet drive.
	
	Our primary focus is the entanglement dynamics of subsystem $A$, comprising the first $L_A$ spins. From the reduced density matrix $\rho_A = \text{Tr}_{\bar{A}}|\psi(t)\rangle\langle\psi(t)|$ ($\bar{A}$ is the remaining $L-L_A$ sites), we extract Schmidt values ${\lambda_i(t)}$ (which satisfies $\sum_i \lambda_i = 1$ at each instant of time $t$) and define the \textit{entanglement Hamiltonian} (EH) $\Hh_\text{ent}(t)$ via $\rho_A(t) \equiv e^{-\Hh_\text{ent}(t)}$~\cite{lihaldane,dalmonte2022entanglement}. Since they commute, the eigenstates are shared by $\Hh_{\text{ent}}(t)$ and $\rho_A(t)$ at each instant of time. We probe the dynamics of $\Hh_\text{ent}(t)$ by examining its instantaneous eigenspectrum, $\Hh_\text{ent}(t) |\lambda_n(t)\rangle = \epsilon_n(t) |\lambda_n(t)\rangle$, and the expectation values of subsystem operators within these states. Here, the index $n=0,1,2,\ldots$ labels the eigenstates with eigenvalues $\epsilon_n$ in ascending order, where $\epsilon_0$ denotes the ``ground-state" eigenvalue. The transformation $\lambda_n=e^{-\epsilon_n}$ gives the eigenvalues of the reduced density matrix $\rho_A$ (the Schmidt values), with $\lambda_0$ being the largest. Note that the reduced density matrix $\rho_A$ allows us to evaluate all R\'enyi entropies $S_n\left(\rho_A\right)=\frac{1}{1-n} \ln \left(\sum_{i=1}^r \lambda_i^n\right)$, where the min-entropy ($n \to \infty$) is given by $S_{\text{min}} = -\ln \lambda_0 = \epsilon_0$ (ground state energy of the EH). Moreover, a fictitious temperature $1/n$ can be associated to the EH (see Ref.~\cite{jha2025page}).
	
TET are a universal consequence at all driving frequencies provided (i) the driven Hamiltonian and (ii) the initial state preserve the global $\mathbb{Z}_2$ symmetry, and (iii) $\rho_A(t)$ has support in both subsystem-parity sectors. We quantify (iii) by the sector weights $w_\pm(t)=\operatorname{Tr}[P_\pm\rho_A(t)]$, with $P_\pm=(\mathds{1}_A\pm P_A)/2$ and $P_A=\prod_{i\in A}\sigma_i^x$. Conditions (i)–(ii) imply $[\rho_A(t),P_A]=0$, which we verify using $C(t)=\|[\rho_A(t),P_A]\|$. Within this symmetry-preserving regime, the critical times $t_c^{(k)}$ are selected by the operational criterion $w_+(t_c^{(k)})=w_-(t_c^{(k)})$ (within the evolution window), at which the Schmidt gap closes and the leading Schmidt vectors exchange subsystem-parity discontinuously. By contrast, an isolated suppression of the Schmidt gap or a smooth parity drift—even if accompanied by a sector-weight crossing—indicates a rounded crossover rather than a TET (see Sec.~III of the SM \cite{Supplemental}).

	A crossing in the two largest Schmidt values, $\lambda_0(t)$ and $\lambda_1(t)$, signifies a fundamental reorganization of the ES, corresponding to a non-analyticity in the ground state energy $\epsilon_0(t)$ of $\Hh_\text{ent}(t)$ (with fictitious temperature $T_{\text{fict}} = \lim_{n \to \infty} \frac{1}{n} = 0$), signaling a \textit{sharp} temporal quantum phase transition in the EH.
	Surprisingly, we observe this behavior from the adiabatic to the high-frequency regime. This behavior emerges when the global $\mathbb{Z}_2$ symmetry is preserved by the driven Hamiltonian and respected by the initial state, with critical times occurring at sector-weight crossings $w_+(t_c^{(k)}) = w_-(t_c^{(k)})$.
	{Conversely, $\mathbb{Z}_2$-breaking initial states (including symmetry-broken ferromagnetic ground states as well as product states such as domain wall and random product states) do not exhibit TET, independent of the drive protocol (see Sec.~III in SM \cite{Supplemental}).

		To diagnose dynamical transitions within the entanglement spectrum, we introduce and track specific observables designed to detect symmetry breaking. The central quantity is the \textit{entanglement echo} $E(t) = \langle\lambda_0(0)|\lambda_0(t)\rangle$, which measures the fidelity of the instantaneous entanglement ground state to its initial configuration \cite{poyhonen2021entanglement}. A vanishing echo signals an orthogonalization of $|\lambda_0(t)\rangle$, suggesting a crossing into a distinct symmetry sector. To directly test this, we compute the expectation values of symmetry operators within the entanglement eigenstates. Specifically, we monitor the subsystem parity $\langle \lambda_n(t) | P_A | \lambda_n(t) \rangle$. A spontaneous change in the parity of the dominant state $|\lambda_0(t)\rangle$, coinciding with a vanishing entanglement echo and a vanishing Schmidt gap $\lambda_0 - \lambda_1$, constitutes the hallmark of a TET. For each subsequent interval, the dynamics alternate between two dwell times: $T_{-}$, during which the dominant state carries $\langle P_A \rangle=-1$, and $T_{+}$, during which it carries $\langle P_A \rangle = +1$. At intermediate drives ($\omega \gtrsim$ 7), these intervals become individually regular yet remain unequal, $T_{-} \neq T_{+}$, evidencing partial ``Floquet inheritance'' by the EH. For higher frequencies ($\omega \gtrsim$ 10), the alternation persists but the intervals synchronize, $T_{-} \approx T_{+} \equiv T_c$, and $T_c$ saturates to an $\omega$-independent constant value (Fig.~\ref{fig:critical_time_vs_omega}). This synchronization and saturation indicate complete Floquet inheritance: the EH dynamics are governed by a high-frequency effective (Floquet-Magnus) description, in which the entanglement transitions form a uniformly spaced temporal lattice set by an emergent, drive-induced time-scale rather than the bare period. See Table~II, Sec.~III and Sec.~IV.B in SM \cite{Supplemental} for further details.
		
		Unless otherwise noted, we use $L=24$ (open boundaries), $L_A = 4$--$12$, $J=1.0$, $h_0=2.0$ (corresponding to equilibrium criticality), and $\omega=5.0$. Time steps $dt \leq 0.1/\omega$ ensure numerical stability; convergence is verified in SM \cite{Supplemental}.

		%%%%%%%%%%%%%%%%%%%%%%%%%%%%%%%%%
		%%%%%%%%%%%%%%%%%%%%%%%%%%%%%%%%%

		\textit{Temporal Entanglement Transitions.---} 
		The combined signatures of a TET are unambiguously observed in the driven TFIM (Fig.~\ref{fig:temporal_transitions}, $L_A=9$). 
		The \textit{Schmidt gap} $\Delta\lambda = \lambda_0 - \lambda_1$ (Fig.~\ref{fig:temporal_transitions}a) closes to numerical precision at critical times $t_c^{(k)}$ (with first critical time $t_c^{(k=1)} \equiv t^*$), signaling energy-level crossing in $\Hh_\text{ent}(t)$.
		Concurrently, the entanglement echo $E(t)=\langle\lambda_0(0) \mid \lambda_0(t)\rangle$ (Fig.~\ref{fig:temporal_transitions}b) vanishes at odd $t_c^{(k=\text{odd})}$, confirming that $|\lambda_0(t_c)\rangle$ orthogonalizes into a distinct symmetry sector. 
		The subsystem parity $\langle P_A \rangle$ (Fig.~\ref{fig:temporal_transitions}c) flips discontinuously at each $t_c^{(k)}$ between $+1$ and $-1$, evidencing dynamical spontaneous $\mathbb{Z}_2$ breaking with respect to the largest Schmidt vector $|\lambda_0(t)\rangle$, while the first excited state $|\lambda_1(t)\rangle$ carries opposite parity.

		We observe these signatures across all non-zero driving frequencies, indicating the generic nature of this phenomenon in the periodically driven Ising chain.
		While the transitions remain sharp, the spacing of critical times ${t_c^{(k)}}$ becomes nearly periodic only at higher driving frequencies, indicating a crossover in the \emph{timing} (Floquet inheritance) rather than in the transition itself (see SM~\cite{Supplemental}).

		The synchronization of these three diagnostics: the closing of the Schmidt gap, the vanishing of the entanglement echo, and the discontinuous flip in parity, provides definitive evidence of a TET. These are not independent events but are intrinsically linked manifestations of the same underlying phenomenon: a periodically-driven, dynamical quantum phase transition of the EH $\Hh_\text{ent}(t)$.
		
		%\onecolumngrid

		%\twocolumngrid

		%%%%%%%%%%%%%%%%%%%%%%%%%%%%%%%%%
		%%%%%%%%%%%%%%%%%%%%%%%%%%%%%%%%%

		\textit{Finite-Size Scaling.---} The TET are marked by non-analytic kinks in the ground state energy $\epsilon_0(t) = -\ln \lambda_0(t)$ of the EH. To establish the critical nature and universality of TET for the critical initial state (where all three conditions are robustly satisfied in the short-time dynamics), we perform comprehensive finite-size scaling analysis across subsystem sizes $L_A = 4-12$. 
		
		Fig.~\ref{fig:scaling} demonstrates the scaling properties of the first critical time $t_c^{(1)}=t^*$ and the minimum Schmidt gap density $S_{\text{min}}/L_A$ at criticality. The critical time exhibits power-law scaling $t^*/L_A \propto L_A^{-1/\nu}$ with critical exponent $\nu \simeq 1.00$ (Fig.~\ref{fig:scaling}a) that corresponds to the divergence of correlation length. Simultaneously, the critical entropy density follows $S_{\text{min}}/L_A \propto L_A^{-a}$ with $a \simeq 1$ (Fig.~\ref{fig:scaling}b). Surprisingly, the correlation length exponent $\nu = 1$ establishes this as a continuous quantum phase transition for all ranges of driving frequencies, belonging to the same universality class as the equilibrium 2D classical Ising/1D quantum TFIM. 
		Note that this exponent match $\nu = 1$ appears coincidental; TET seem to emerge independently of the equilibrium phase diagram, as confirmed in SM \cite{Supplemental}.

		The universality of these transitions for the critical initial state is demonstrated through data collapse using the scaling ansatz (recall $\epsilon_0=S_{\text{min}}$):
		\begin{equation}
			\frac{\epsilon_0}{L_A} = \frac{1}{L_A^{a}} \mathcal{F}\left[ \left( \frac{t}{L_A} - \frac{t^*}{L_A} \right) L_A^{1/\nu } \right],
			\label{eq:scaling ansatz}
		\end{equation}
		where $\mathcal{F}$ is a universal scaling function, and $\nu$ is the critical exponent corresponding to diverging correlation length. 
		For our critical initial state, $t^*$ occurs within the simulated time window for all driving frequencies studied, enabling a uniform finite-size scaling analysis from the adiabatic to the high-frequency regime.
		Fig.~\ref{fig:scaling}c shows excellent collapse of data from all subsystem sizes onto a universal curve, validating the scaling hypothesis. The agreement is further verified in Fig.~\ref{fig:scaling}d, which shows the raw data near the critical point.
		
		Significantly, the same exponent describes all subsequent kinks at $t_c^{(k\geq 1)}$ for higher driving frequencies ($\omega \gtrsim 5.0$ for our parameter choices), where the EH inherits Floquet-like periodicity. This universality across multiple transitions suggests a common underlying fixed point. For the critical initial state, the exponent $\nu \simeq 1.0$ defines a new universality class for non-equilibrium entanglement dynamics in driven systems, which appears to be completely decoupled from the underlying equilibrium criticality (more on this below).

		%\onecolumngrid
		\begin{figure}[htbp!]
			\centering
			\includegraphics[width=1.1\columnwidth]{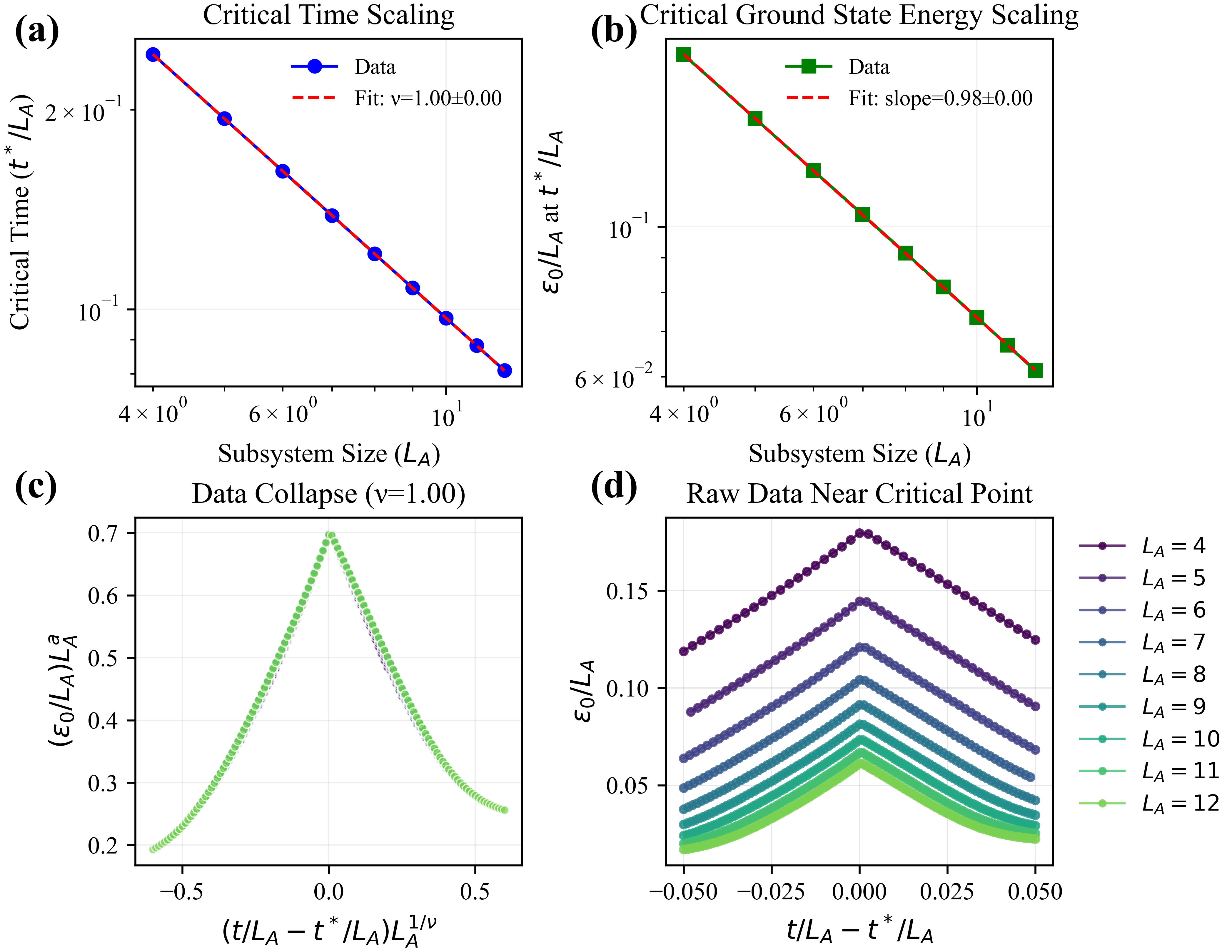}
			\caption{Finite-size scaling analysis of TET starting with critical initial state $(h_0/2=1.0)$. 
				\textbf{(a)} Critical time scaling $t^*/L_A \propto L_A^{-1/\nu}$ with $\nu = 1.00$. 
				\textbf{(b)} Critical entropy density scaling $\epsilon_0/L_A \propto L_A^{-a}$ with $a = 1.00$. 
				\textbf{(c)} Universal data collapse using scaling ansatz in Eq.~\eqref{eq:scaling ansatz}. 
				\textbf{(d)} Raw data before scaling collapse. 
				The exponent $\nu = 1$ establishes the same universality class for our driven non-equilibrium entanglement spectrum as equilibrium 2D classical Ising/1D TFIM. Parameters: $L = 24$, $J = 1.0$, $h_0 = 2.0$, $\omega = 5.0$, $dt=0.01$.}
			\label{fig:scaling}
		\end{figure}
		%\twocolumngrid 

		\begin{figure}[htbp!]
			\centering
			\includegraphics[width=0.95\columnwidth]{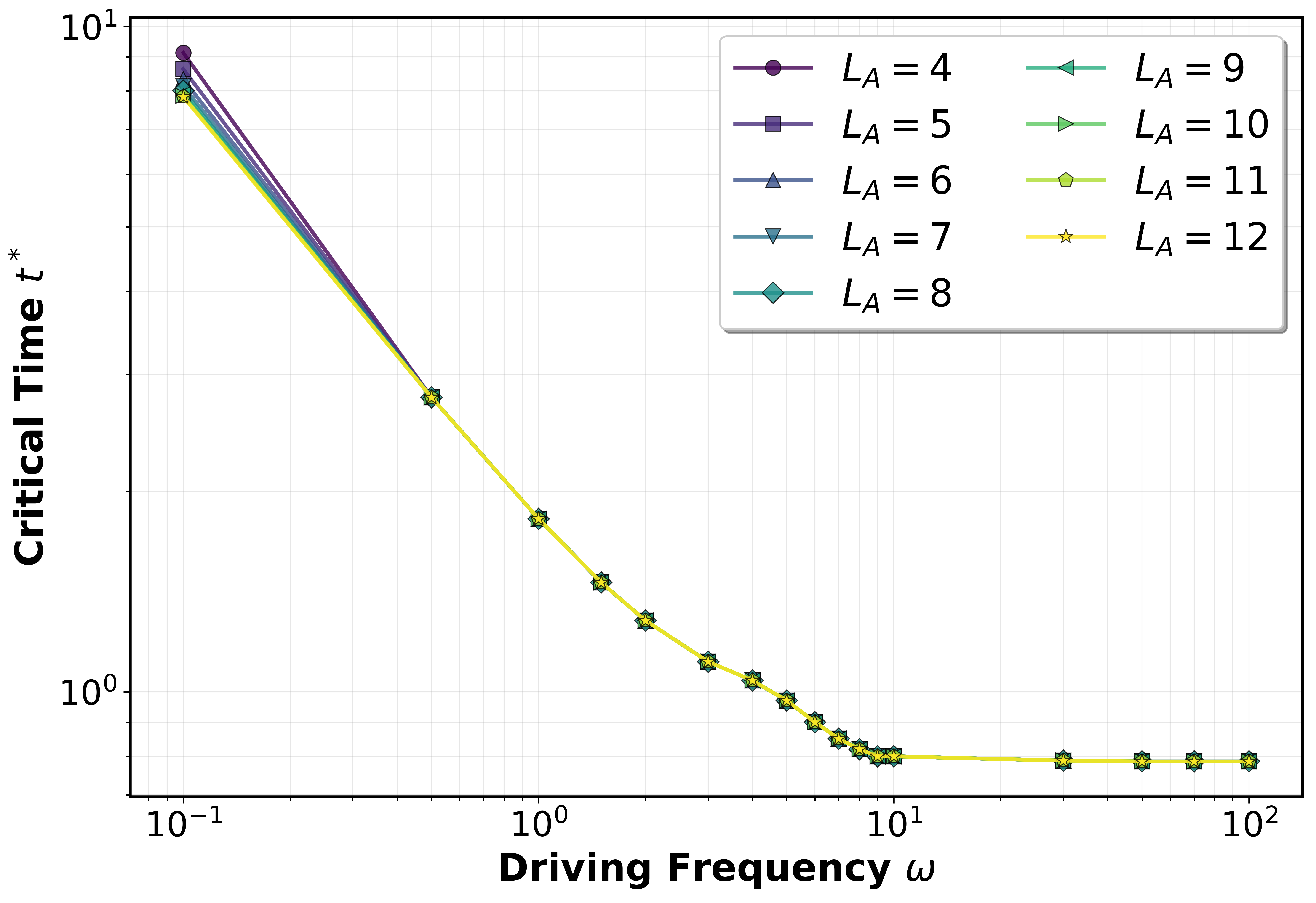}
			\caption{Frequency dependence of the first critical time $t^*$ across subsystem sizes $L_A = 4-12$ for critical initial state $(h_0/2=1.0)$. At low frequencies, $t^* \propto \omega^{-1}$ (adiabatic regime), while at high frequencies ($\omega \gtrsim 10$), $t^*$ saturates to frequency-independent values determined solely by subsystem size, indicating a crossover to Floquet steady-state behavior where the EH develops an intrinsic timescale. The universal saturation indicates that TET become intrinsic properties of the effective time-independent dynamics rather than driven phenomena, validating the Floquet-Magnus description and establishing characteristic timescales independent of the external driving protocol. Parameters: $L=24$, $J=1.0$, $h_0=2.0$. Time steps are chosen as $dt=0.01$ for $\omega=[0.1, 10.0]$, $dt=0.002$ for $\omega=\{30, 50\}$ and $dt=0.001$ for $\omega=\{70,100\}$. A total of 153 data points in this plot show excellent collapse (see Fig.~9 of SM \cite{Supplemental}).} 
			\label{fig:critical_time_vs_omega}
		\end{figure}

		%%%%%%%%%%%%%%%%%%%%%%%%%%%%%%%%%
		%%%%%%%%%%%%%%%%%%%%%%%%%%%%%%%%%

		\textit{Effective Steady State at High Frequency.---}
		At high driving frequencies, TET acquire an intrinsic timescale explained by Floquet-Magnus theory \cite{Bukov2015Mar, Sen2021Aug, KUWAHARA201696}. We decompose the time-dependent Hamiltonian as $\mathcal{H}(t) = \mathcal{A} + \mathcal{B}\cos(\omega t)$, where $\mathcal{A} = -J\sum_{i} \sigma_i^z \sigma_{i+1}^z$ and $\mathcal{B} = -(h_0/2)\sum_i \sigma_i^x$. The effective time-independent Hamiltonian emerges as a systematic expansion $\mathcal{H}_{\text{eff}} = \mathcal{H}_0 + \mathcal{H}_1 + \mathcal{H}_2 + \cdots$ [$\Hh_{(n)}$ is of order $\Oo(\omega^{-n})$], where the leading corrections are
		\begin{equation}
			\begin{aligned}
				\Hh_{\mathrm{eff}}=&-J\left(1+\frac{h_0^2}{2 \omega^2}\right) \sum_{i=1}^{L-1} \sigma_i^z \sigma_{i+1}^z+\frac{h_0^2 J}{2 \omega^2} \sum_{i=1}^{L-1} \sigma_i^y \sigma_{i+1}^y\\
				&-\frac{2 h_0 J^2}{\omega^2}\left(\sigma_1^x+\sigma_L^x\right)-\frac{4 h_0 J^2}{\omega^2} \sum_{i=2}^{L-1} \sigma_i^x \\
				&-\frac{4 h_0 J^2}{\omega^2} \sum_{i=2}^{L-1} \sigma_{i-1}^z \sigma_i^x \sigma_{i+1}^z + \Oo(\omega^{-3}).
			\end{aligned}
			\label{eq:effective hamiltonian}
		\end{equation}
		Beyond renormalizing the Ising coupling, the expansion generates new terms: YY interactions, enhanced transverse fields at boundaries, and crucially, three-body interactions $\sim \sigma_{i-1}^z \sigma_i^x \sigma_{i+1}^z$ that couple neighboring bonds through local spin flips.

		As demonstrated in SM~\cite{Supplemental}, exact time evolution under $\mathcal{H}(t)$ shows good agreement with that under $\mathcal{H}_{\text{eff}}$ at higher frequencies, with entanglement entropy, Schmidt gap dynamics, and state fidelity exhibiting near-perfect overlap. TET persist in the effective evolution with identical critical exponent $\nu$, confirming that these phenomena represent genuine features of the driven steady state where the EH develops an intrinsic timescale.

		\begin{table}[htbp!]
			\centering
			\caption{\label{tab:comparison}Conditions for TET.}
			\begin{ruledtabular}
				\begin{tabular}{llll}
					System & Drive $\mathbb{Z}_2$ & Initial state $\mathbb{Z}_2$ & TET \\ \hline
					Equilibrium TFIM & --- & --- & No \\
					Driven TFIM & Yes & No (symmetry-broken) & No \\
					Driven TFIM & Yes & Yes & Yes$^{\mathrm{a}}$ \\
					\makecell[tl]{Conserved-charge\\ models} & --- & Any & Yes$^{\mathrm{b}}$ \\
				\end{tabular}
			\end{ruledtabular}
			\vspace{-1ex}
			\begin{flushleft}
				$^{\mathrm{a}}$ Provided $\rho_A(t)$ carries weight in both subsystem-parity sectors, with TET occurring when $w_+(t_c^{(k)}) = w_-(t_c^{(k)})$ while $[\rho_A,P_A]$ remains negligible, see SM Sec.~III. Many standard product-state preparations are $\mathbb{Z}_2$-breaking and thus excluded, while $\mathbb{Z}_2$-symmetric (often entangled) initial states are allowed. \\
				
				$^{\mathrm{b}}$ Only when $\rho_A$ is dynamically forced between disconnected sectors \cite{kehrein2024page, jha2025page}.
			\end{flushleft}
		\end{table}

		%%%%%%%%%%%%%%%%%%%%%%%%%%%%%%%%%
		%%%%%%%%%%%%%%%%%%%%%%%%%%%%%%%%%
		
		\textit{Discussion and Conclusion.---} We have demonstrated that the Floquet driven Ising chain hosts temporal entanglement transitions (TET) characterized by dynamical spontaneous symmetry breaking within the EH. These transitions exhibit universal critical behavior with exponent $\nu = 1$ for critical initial state ($h_0/2 = 1.0$), identical to equilibrium quantum criticality in the transverse-field Ising universality class, yet emerge from non-equilibrium dynamics that are decoupled from static criticality, as shown by their persistence across different equilibrium phases ($h_0 = 1.6$ to $2.4$,  see SM~\cite{Supplemental}).
		
		This work establishes that: (1) the ES develops non-analyticities for all $\omega$, with periodic reorganizations at higher frequencies; (2) transitions are diagnosed by synchronized Schmidt gap closure, entanglement echo vanishing, and parity flipping, evidencing dynamical spontaneous symmetry breaking; (3) finite-size scaling for the critical initial state reveals universal $\nu=1$ critical behavior across frequencies; (4) at high frequencies, Floquet-Magnus theory provides an effective description accurately capturing the transition; (5) TET occur precisely when the sector weights become equal, $w_+(t_c^{(k)}) = w_-(t_c^{(k)})$, provided the driven Hamiltonian and initial state preserve global $\mathbb{Z}_2$ symmetry (Table~\ref{tab:comparison}; also see SM, Sec.~III \cite{Supplemental}). Experimental access to these transitions is feasible: the key observables (Schmidt gap, parity expectations, entanglement echo) require only low-lying eigenstates of the EH, measurable via randomized measurements~\cite{Brydges2019Apr, Elben2020Nov} or Ramsey spectroscopy~\cite{Pichler2016Nov}.
		
		Fig.~\ref{fig:critical_time_vs_omega} shows critical times $t^*$ saturating to $\omega$-independent values at high frequencies, indicating intrinsic timescales of the driven steady state. The collapse of $t^*$ curves across subsystem sizes demonstrates universal Floquet inheritance, where the EH acquires periodicity and full data collapse is satisfied by the same $\nu=1$ exponent for all transitions~\cite{Supplemental}.
		
		These findings establish TET as fundamental aspects of non-equilibrium Floquet quantum dynamics. Notably, previous work on Floquet symmetry-protected topological system has demonstrated band crossings in entanglement spectrum \cite{potirniche2017}, however the question of robust universality has been an open question. This is what we have addressed in this work, where our TET arise whenever the driven Hamiltonian and initial state respect global $\mathbb{Z}_2$, with critical times determined by sector-weight crossings $w_+(t_c^{(k)}) = w_-(t_c^{(k)})$, yielding a nonanalytic reorganization (quantum phase transition) of the EH with Ising-class critical exponents $\nu=1$. TET define a class of non-equilibrium criticality distinct from Loschmidt echo singularities and magnetization dynamics (see SM \cite{Supplemental}). The exact Jordan-Wigner equivalence between the TFIM and Kitaev chain suggests that these transitions should also manifest in driven topological superconductors, opening pathways to entanglement-based probes of Floquet Majorana physics \cite{Cadez2019, Soori2022Anomalous}. The requirement of symmetry-preserving initial conditions with critical times selected by sector-weight degeneracies motivates future exploration of how symmetry and entanglement resources jointly control non-equilibrium criticality structures in quantum resource theories \cite{Chitambar2019}. Beyond the TFIM (despite its free-fermion Jordan-Wigner mapping), our symmetry/sector-competition criterion (global symmetry, symmetric initial state, and sector-weight equality $w_+(t)=w_-(t)$) suggests analogous transitions in interacting $\mathbb{Z}_2$-symmetric chains and, more broadly, in other driven systems with discrete symmetries (higher-spin chains, symmetry-protected topological phases, and lattice gauge theories), with experimental prospects in cold-atom and trapped-ion platforms~\cite{Eisert2015Feb, Bernien2017Nov, karch2025probingquantummanybodydynamics, Brydges2019Apr}.

		\textit{Acknowledgments.---} K. G. and R. J. acknowledge financial support by Deutsche Forschungsgemeinschaft (DFG, German Research Foundation) Grants No. 436382789, and No. 493420525, via large equipment grants (GOEGrid). This material is based upon work supported by the U.S. Department of Energy, Office of Science, Office of Advanced Scientific Computing Research via the Exploratory Research for Extreme Scale Science (EXPRESS) program under Award Number DE-SC0026216. This research was supported in part by grant NSF PHY-2309135 to the Kavli Institute for Theoretical Physics (KITP). A. P. thanks the Kavli Institute for Theoretical Physics (KITP) for its hospitality during the program on ``Noise-robust Phases of Quantum Matter'', during which part of this work was completed. 
		
		\textit{Data and code availability.---} All data and code used for data generation are available on Zenodo on reasonable request \cite{Gadge2025_zenodo}.
		
		\textit{Disclaimer.---} This report was prepared as an account of work sponsored by an agency of the United States Government. Neither the United States Government nor any agency thereof, nor any of their employees, makes any warranty, express or implied, or assumes any legal liability or 	responsibility for the accuracy, completeness, or usefulness of any information, apparatus, product, or process disclosed, or represents that its use would not infringe privately owned rights.  Reference herein to any specific commercial product, process, or service by trade name, trademark, manufacturer, or otherwise does not necessarily constitute or imply its endorsement, recommendation, or favoring by the United States Government or any agency thereof. The views and opinions of authors expressed herein do not necessarily state or reflect those of the United States Government or any agency thereof.
		
		%%%%%%%%%%%%%%%%%%%%%%%%%%%%%%%%%
		%%%%%%%%%%%%%%%%%%%%%%%%%%%%%%%%%

		\bibliography{refs.bib}

@ARTICLE{khemani2019review,
       author = {{Khemani}, Vedika and {Moessner}, Roderich and {Sondhi}, S.~L.},
        title = "{A Brief History of Time Crystals}",
      journal = {arXiv e-prints},
         year = 2019,
        month = oct,
          eid = {arXiv:1910.10745},
        pages = {arXiv:1910.10745},
          doi = {10.48550/arXiv.1910.10745},
archivePrefix = {arXiv},
       eprint = {1910.10745},
 primaryClass = {cond-mat.str-el},
       adsurl = {https://ui.adsabs.harvard.edu/abs/2019arXiv191010745K}
}

@ARTICLE{harper2020review,
       author = {{Harper}, Fenner and {Roy}, Rahul and {Rudner}, Mark S. and {Sondhi}, S.~L.},
        title = "{Topology and Broken Symmetry in Floquet Systems}",
      journal = {Annual Review of Condensed Matter Physics},
         year = 2020,
        month = mar,
       volume = {11},
        pages = {345-368},
          doi = {10.1146/annurev-conmatphys-031218-013721}
}

@ARTICLE{else2020review,
       author = {{Else}, Dominic V. and {Monroe}, Christopher and {Nayak}, Chetan and {Yao}, Norman Y.},
        title = "{Discrete Time Crystals}",
      journal = {Annual Review of Condensed Matter Physics},
         year = 2020,
        month = mar,
       volume = {11},
        pages = {467-499},
          doi = {10.1146/annurev-conmatphys-031119-050658}
}

@ARTICLE{lin2025local,
       author = {{Lin}, Zhi-Xing and {Prem}, Abhinav and {Ryu}, Shinsei and {Lapierre}, Bastien},
        title = "{Local-to-Global Entanglement Dynamics by Periodically Driving Impurities}",
      journal = {arXiv e-prints},
         year = 2025,
        month = oct,
archivePrefix = {arXiv},
       eprint = {2510.20908},
 primaryClass = {quant-ph}
}

@article{lihaldane,
  title = {Entanglement Spectrum as a Generalization of Entanglement Entropy: Identification of Topological Order in Non-Abelian Fractional Quantum Hall Effect States},
  author = {Li, Hui and Haldane, F. D. M.},
  journal = {Phys. Rev. Lett.},
  volume = {101},
  issue = {1},
  pages = {010504},
  numpages = {4},
  year = {2008},
  month = {Jul},
  publisher = {American Physical Society},
  doi = {10.1103/PhysRevLett.101.010504},
  url = {https://link.aps.org/doi/10.1103/PhysRevLett.101.010504}
}

@article{regnault1,
  title = {Topological Entanglement and Clustering of Jain Hierarchy States},
  author = {Regnault, N. and Bernevig, B. A. and Haldane, F. D. M.},
  journal = {Phys. Rev. Lett.},
  volume = {103},
  issue = {1},
  pages = {016801},
  numpages = {4},
  year = {2009},
  month = {Jun},
  publisher = {American Physical Society},
  doi = {10.1103/PhysRevLett.103.016801},
  url = {https://link.aps.org/doi/10.1103/PhysRevLett.103.016801}
}

@article{papic,
  title = {Topological Entanglement in Abelian and Non-Abelian Excitation Eigenstates},
  author = {Papi\ifmmode \acute{c}\else \'{c}\fi{}, Z. and Bernevig, B. A. and Regnault, N.},
  journal = {Phys. Rev. Lett.},
  volume = {106},
  issue = {5},
  pages = {056801},
  numpages = {4},
  year = {2011},
  month = {Feb},
  publisher = {American Physical Society},
  doi = {10.1103/PhysRevLett.106.056801},
  url = {https://link.aps.org/doi/10.1103/PhysRevLett.106.056801}
}

@article{chandran2011,
  title = {Bulk-edge correspondence in entanglement spectra},
  author = {Chandran, Anushya and Hermanns, M. and Regnault, N. and Bernevig, B. Andrei},
  journal = {Phys. Rev. B},
  volume = {84},
  issue = {20},
  pages = {205136},
  numpages = {21},
  year = {2011},
  month = {Nov},
  publisher = {American Physical Society},
  doi = {10.1103/PhysRevB.84.205136},
  url = {https://link.aps.org/doi/10.1103/PhysRevB.84.205136}
}

@article{dubail2012,
  title = {Edge-state inner products and real-space entanglement spectrum of trial quantum Hall states},
  author = {Dubail, J. and Read, N. and Rezayi, E. H.},
  journal = {Phys. Rev. B},
  volume = {86},
  issue = {24},
  pages = {245310},
  numpages = {32},
  year = {2012},
  month = {Dec},
  publisher = {American Physical Society},
  doi = {10.1103/PhysRevB.86.245310},
  url = {https://link.aps.org/doi/10.1103/PhysRevB.86.245310}
}

@article{cano2015,
  title = {Interactions along an entanglement cut in $2+1\mathrm{D}$ Abelian topological phases},
  author = {Cano, Jennifer and Hughes, Taylor L. and Mulligan, Michael},
  journal = {Phys. Rev. B},
  volume = {92},
  issue = {7},
  pages = {075104},
  numpages = {31},
  year = {2015},
  month = {Aug},
  publisher = {American Physical Society},
  doi = {10.1103/PhysRevB.92.075104},
  url = {https://link.aps.org/doi/10.1103/PhysRevB.92.075104}
}

@article{qi2012,
  title = {General Relationship between the Entanglement Spectrum and the Edge State Spectrum of Topological Quantum States},
  author = {Qi, Xiao-Liang and Katsura, Hosho and Ludwig, Andreas W. W.},
  journal = {Phys. Rev. Lett.},
  volume = {108},
  issue = {19},
  pages = {196402},
  numpages = {5},
  year = {2012},
  month = {May},
  publisher = {American Physical Society},
  doi = {10.1103/PhysRevLett.108.196402},
  url = {https://link.aps.org/doi/10.1103/PhysRevLett.108.196402}
}

@article{swingle2012,
  title = {Geometric proof of the equality between entanglement and edge spectra},
  author = {Swingle, Brian and Senthil, T.},
  journal = {Phys. Rev. B},
  volume = {86},
  issue = {4},
  pages = {045117},
  numpages = {7},
  year = {2012},
  month = {Jul},
  publisher = {American Physical Society},
  doi = {10.1103/PhysRevB.86.045117},
  url = {https://link.aps.org/doi/10.1103/PhysRevB.86.045117}
}

@article{prodan2010,
  title = {Entanglement Spectrum of a Disordered Topological Chern Insulator},
  author = {Prodan, Emil and Hughes, Taylor L. and Bernevig, B. Andrei},
  journal = {Phys. Rev. Lett.},
  volume = {105},
  issue = {11},
  pages = {115501},
  numpages = {4},
  year = {2010},
  month = {Sep},
  publisher = {American Physical Society},
  doi = {10.1103/PhysRevLett.105.115501},
  url = {https://link.aps.org/doi/10.1103/PhysRevLett.105.115501}
}

@article{ho1,
  title = {Edge-entanglement spectrum correspondence in a nonchiral topological phase and Kramers-Wannier duality},
  author = {Ho, Wen Wei and Cincio, Lukasz and Moradi, Heidar and Gaiotto, Davide and Vidal, Guifre},
  journal = {Phys. Rev. B},
  volume = {91},
  issue = {12},
  pages = {125119},
  numpages = {20},
  year = {2015},
  month = {Mar},
  publisher = {American Physical Society},
  doi = {10.1103/PhysRevB.91.125119},
  url = {https://link.aps.org/doi/10.1103/PhysRevB.91.125119}
}

@article{ho2,
  title = {Universal edge information from wave-function deformation},
  author = {Ho, Wen Wei and Cincio, Lukasz and Moradi, Heidar and Vidal, Guifre},
  journal = {Phys. Rev. B},
  volume = {95},
  issue = {23},
  pages = {235110},
  numpages = {8},
  year = {2017},
  month = {Jun},
  publisher = {American Physical Society},
  doi = {10.1103/PhysRevB.95.235110},
  url = {https://link.aps.org/doi/10.1103/PhysRevB.95.235110}
}

@article{pollmann2010,
  title = {Entanglement spectrum of a topological phase in one dimension},
  author = {Pollmann, Frank and Turner, Ari M. and Berg, Erez and Oshikawa, Masaki},
  journal = {Phys. Rev. B},
  volume = {81},
  issue = {6},
  pages = {064439},
  numpages = {10},
  year = {2010},
  month = {Feb},
  publisher = {American Physical Society},
  doi = {10.1103/PhysRevB.81.064439},
  url = {https://link.aps.org/doi/10.1103/PhysRevB.81.064439}
}

@article{turner2010,
  title = {Entanglement and inversion symmetry in topological insulators},
  author = {Turner, Ari M. and Zhang, Yi and Vishwanath, Ashvin},
  journal = {Phys. Rev. B},
  volume = {82},
  issue = {24},
  pages = {241102},
  numpages = {4},
  year = {2010},
  month = {Dec},
  publisher = {American Physical Society},
  doi = {10.1103/PhysRevB.82.241102},
  url = {https://link.aps.org/doi/10.1103/PhysRevB.82.241102}
}

@article{fidkowski2010,
  title = {Entanglement Spectrum of Topological Insulators and Superconductors},
  author = {Fidkowski, Lukasz},
  journal = {Phys. Rev. Lett.},
  volume = {104},
  issue = {13},
  pages = {130502},
  numpages = {4},
  year = {2010},
  month = {Apr},
  publisher = {American Physical Society},
  doi = {10.1103/PhysRevLett.104.130502},
  url = {https://link.aps.org/doi/10.1103/PhysRevLett.104.130502}
}

@article{alba2012,
  title = {Boundary-Locality and Perturbative Structure of Entanglement Spectra in Gapped Systems},
  author = {Alba, Vincenzo and Haque, Masudul and L\"auchli, Andreas M.},
  journal = {Phys. Rev. Lett.},
  volume = {108},
  issue = {22},
  pages = {227201},
  numpages = {5},
  year = {2012},
  month = {May},
  publisher = {American Physical Society},
  doi = {10.1103/PhysRevLett.108.227201},
  url = {https://link.aps.org/doi/10.1103/PhysRevLett.108.227201}
}

@article{berg2017,
  title = {Edge-entanglement correspondence for a gapped topological phase with symmetry},
  author = {Koch-Janusz, Maciej and Dhochak, Kusum and Berg, Erez},
  journal = {Phys. Rev. B},
  volume = {95},
  issue = {20},
  pages = {205110},
  numpages = {13},
  year = {2017},
  month = {May},
  publisher = {American Physical Society},
  doi = {10.1103/PhysRevB.95.205110},
  url = {https://link.aps.org/doi/10.1103/PhysRevB.95.205110}
}

@article{choo2018,
  title = {Measurement of the Entanglement Spectrum of a Symmetry-Protected Topological State Using the IBM Quantum Computer},
  author = {Choo, Kenny and von Keyserlingk, Curt W. and Regnault, Nicolas and Neupert, Titus},
  journal = {Phys. Rev. Lett.},
  volume = {121},
  issue = {8},
  pages = {086808},
  numpages = {5},
  year = {2018},
  month = {Aug},
  publisher = {American Physical Society},
  doi = {10.1103/PhysRevLett.121.086808},
  url = {https://link.aps.org/doi/10.1103/PhysRevLett.121.086808}
}

@article{premespec,
  title = {Entanglement spectra of stabilizer codes: A window into gapped quantum phases of matter},
  author = {Schmitz, Albert T. and Huang, Sheng-Jie and Prem, Abhinav},
  journal = {Phys. Rev. B},
  volume = {99},
  issue = {20},
  pages = {205109},
  numpages = {25},
  year = {2019},
  month = {May},
  publisher = {American Physical Society},
  doi = {10.1103/PhysRevB.99.205109},
  url = {https://link.aps.org/doi/10.1103/PhysRevB.99.205109}
}

@article{stringnetespec,
  title = {Correspondence between bulk entanglement and boundary excitation spectra in two-dimensional gapped topological phases},
  author = {Luo, Zhu-Xi and Pankovich, Brendan G. and Hu, Yuting and Wu, Yong-Shi},
  journal = {Phys. Rev. B},
  volume = {99},
  issue = {20},
  pages = {205137},
  numpages = {12},
  year = {2019},
  month = {May},
  publisher = {American Physical Society},
  doi = {10.1103/PhysRevB.99.205137},
  url = {https://link.aps.org/doi/10.1103/PhysRevB.99.205137}
}

@ARTICLE{wen2025critical,
       author = {{Tang}, Qicheng and {Wen}, Xueda},
        title = "{A critical state under weak measurement is not critical}",
      journal = {Journal of High Energy Physics},
         year = 2025,
        month = sep,
       volume = {2025},
       number = {9},
          eid = {168},
        pages = {168},
          doi = {10.1007/JHEP09(2025)168}
}

@dataset{Gadge2025_zenodo,
  author    = {K. Gadge and A. Prem and R. Jha},
  year      = {2025},
  title     = {Dynamical Spontaneous Symmetry Breaking and Entanglement Criticality in Periodically Driven Spin Chain [Data set]},
  publisher = {Zenodo},
   doi       = {10.5281/zenodo.18407219},
  url       = {https://doi.org/10.5281/zenodo.17338414}
}

@article{potirniche2017,
  title = {Floquet Symmetry-Protected Topological Phases in Cold-Atom Systems},
  author = {Potirniche, I.-D. and Potter, A. C. and Schleier-Smith, M. and Vishwanath, A. and Yao, N. Y.},
  journal = {Phys. Rev. Lett.},
  volume = {119},
  issue = {12},
  pages = {123601},
  numpages = {6},
  year = {2017},
  month = {Sep},
  publisher = {American Physical Society},
  doi = {10.1103/PhysRevLett.119.123601},
  url = {https://link.aps.org/doi/10.1103/PhysRevLett.119.123601}
}

@article{oka2019floquet,
	title={Floquet Engineering of Quantum Materials},
	author={Oka, Takashi and Kitamura, Sota},
	journal={Annual Review of Condensed Matter Physics},
	volume={10},
	pages={387--408},
	year={2019},
	doi={10.1146/annurev-conmatphys-031218-013423}
}

@article{goldman2014periodically,
	title={Periodically driven quantum systems: effective Hamiltonians and engineered gauge fields},
	author={Goldman, N. and Dalibard, J.},
	journal={Physical Review X},
	volume={4},
	number={3},
	pages={031027},
	year={2014},
	doi={10.1103/PhysRevX.4.031027}
}

@article{rudner2020floquet,
	title={Band structure engineering and non-equilibrium dynamics in Floquet topological insulators},
	author={Rudner, Mark S. and Lindner, Netanel H.},
	journal={Nature Reviews Physics},
	volume={2},
	number={4},
	pages={229--244},
	year={2020},
	doi={10.1038/s42254-020-0170-z}
}

@article{cayssol2013floquet,
	title={Floquet topological insulators},
	author={Cayssol, Jérôme and Dóra, Balázs and Simon, Ferenc and Moessner, Roderich},
	journal={physica status solidi (RRL)},
	volume={7},
	number={1-2},
	pages={101--108},
	year={2013},
	doi={10.1002/pssr.201206451}
}

@article{ippoliti2021many,
	author = {Ippoliti, Matteo and Kechedzhi, Kostyantyn and Moessner, Roderich and Sondhi, S. L. and Khemani, Vedika},
	title = {{Many-Body Physics in the NISQ Era: Quantum Programming a Discrete Time Crystal}},
	journal = {PRX Quantum},
	volume = {2},
	number = {3},
	pages = {030346},
	year = {2021},
	month = sep,
	publisher = {American Physical Society},
	doi = {10.1103/PRXQuantum.2.030346}
}

@article{riera2019time,
	author = {Riera-Campeny, Andreu and Moreno-Cardoner, Maria and Sanpera, Anna},
	title = {{Time crystallinity in open quantum systems}},
	journal = {Quantum},
	volume = {4},
	pages = {270},
	year = {2020},
	month = may,
	publisher = {Verein zur F{\ifmmode\ddot{o}\else\"{o}\fi}rderung des Open Access Publizierens in den Quantenwissenschaften},
	eprint = {1908.11339v4},
	doi = {10.22331/q-2020-05-25-270}
}

@article{tiwari2024dynamical,
	title={Dynamical localization and slow dynamics in quasiperiodically driven quantum systems},
	author={Tiwari, Vatsana and Bhakuni, Devendra Singh and Sharma, Auditya},
	journal={Physical Review B},
	volume={109},
	number={16},
	pages={L161104},
	year={2024},
	doi={10.1103/PhysRevB.109.L161104}
}

@article{nag2014dynamical,
	author = {Nag, Tanay and Roy, Sthitadhi and Dutta, Amit and Sen, Diptiman},
	title = {{Dynamical localization in a chain of hard core bosons under periodic driving}},
	journal = {Phys. Rev. B},
	volume = {89},
	number = {16},
	pages = {165425},
	year = {2014},
	month = apr,
	publisher = {American Physical Society},
	doi = {10.1103/PhysRevB.89.165425}
}

@article{PhysRevLett.109.257201,
	title = {Periodic Steady Regime and Interference in a Periodically Driven Quantum System},
	author = {Russomanno, Angelo and Silva, Alessandro and Santoro, Giuseppe E.},
	journal = {Phys. Rev. Lett.},
	volume = {109},
	issue = {25},
	pages = {257201},
	numpages = {5},
	year = {2012},
	month = {Dec},
	publisher = {American Physical Society},
	doi = {10.1103/PhysRevLett.109.257201},
	url = {https://link.aps.org/doi/10.1103/PhysRevLett.109.257201}
}

@article{Abanin2015Dec,
	author = {Abanin, Dmitry A. and De Roeck, Wojciech and Huveneers, Fran{\ifmmode\mbox{\c{c}}\else\c{c}\fi}ois},
	title = {{Exponentially Slow Heating in Periodically Driven Many-Body Systems}},
	journal = {Phys. Rev. Lett.},
	volume = {115},
	number = {25},
	pages = {256803},
	year = {2015},
	month = dec,
	publisher = {American Physical Society},
	doi = {10.1103/PhysRevLett.115.256803}
}

@article{Machado2019Dec,
	author = {Machado, Francisco and Kahanamoku-Meyer, Gregory D. and Else, Dominic V. and Nayak, Chetan and Yao, Norman Y.},
	title = {{Exponentially slow heating in short and long-range interacting Floquet systems}},
	journal = {Phys. Rev. Res.},
	volume = {1},
	number = {3},
	pages = {033202},
	year = {2019},
	month = dec,
	publisher = {American Physical Society},
	doi = {10.1103/PhysRevResearch.1.033202}
}

@article{zhao2022scaling,
	title={Scaling of entanglement entropy at deconfined quantum criticality},
	author={Zhao, Jiarui and Wang, Yan-Cheng and Yan, Zheng and Cheng, Meng and Meng, Zi Yang},
	journal={Physical Review Letters},
	volume={128},
	number={1},
	pages={010601},
	year={2022},
	doi={10.1103/PhysRevLett.128.010601}
}

@article{cho2017quantum,
	title={Quantum Phase Transition and Entanglement in Topological Quantum Wires},
	author={Cho, Jaeyoon and Kim, Kun Woo},
	journal={Scientific Reports},
	volume={7},
	number={1},
	pages={2745},
	year={2017},
	doi={10.1038/s41598-017-02717-w}
}

@article{redon2024realizing,
	author = {Redon, Quentin and Liu, Qi and Bouhiron, Jean-Baptiste and Mittal, Nehal and Fabre, Aur{\ifmmode\acute{e}\else\'{e}\fi}lien and Lopes, Raphael and Nascimbene, Sylvain},
	title = {{Realizing the entanglement Hamiltonian of a topological quantum Hall system}},
	journal = {Nat. Commun.},
	volume = {15},
	number = {10086},
	pages = {10086},
	year = {2024},
	month = nov,
	issn = {2041-1723},
	publisher = {Nature Publishing Group},
	doi = {10.1038/s41467-024-54085-5}
}

@article{schneider2022entanglement,
	title={Entanglement spectrum and quantum phase diagram of the long-range XXZ chain},
	author={Schneider, J. T. and Thomson, S. J. and Sanchez-Palencia, L.},
	journal={Physical Review B},
	volume={106},
	number={1},
	pages={014306},
	year={2022},
	doi={10.1103/PhysRevB.106.014306}
}

@article{dalmonte2022entanglement,
	author = {Dalmonte, Marcello and Eisler, Viktor and Falconi, Marco and Vermersch, Beno{\ifmmode\hat{\imath}\else\^{\i}\fi}t},
	title = {{Entanglement Hamiltonians: From Field Theory to Lattice Models and Experiments}},
	journal = {Ann. Phys.},
	volume = {534},
	number = {11},
	pages = {2200064},
	year = {2022},
	month = aug,
	issn = {1521-3889},
	publisher = {John Wiley {\&} Sons, Ltd},
	doi = {10.1002/andp.202200064}
}

@article{demidio2024universal,
	author = {D{'}Emidio, Jonathan and Or{\ifmmode\acute{u}\else\'{u}\fi}s, Rom{\ifmmode\acute{a}\else\'{a}\fi}n and Laflorencie, Nicolas and de Juan, Fernando},
	title = {{Universal Features of Entanglement Entropy in the Honeycomb Hubbard Model}},
	journal = {Phys. Rev. Lett.},
	volume = {132},
	number = {7},
	pages = {076502},
	year = {2024},
	month = feb,
	publisher = {American Physical Society},
	doi = {10.1103/PhysRevLett.132.076502}
}

@article{baykusheva2023witnessing,
	title={Witnessing Nonequilibrium Entanglement Dynamics in a Strongly Correlated Fermionic Chain},
	author={Baykusheva, Denitsa R. and Kalthoff, Mona H. and Hofmann, Damian and Claassen, Martin and Kennes, Dante M. and Sentef, Michael A. and Mitrano, Matteo},
	journal={Physical Review Letters},
	volume={130},
	number={10},
	pages={106902},
	year={2023},
	doi={10.1103/PhysRevLett.130.106902}
}

@article{lewis2019unifying,
	title={Unifying scrambling, thermalization and entanglement through measurement of fidelity out-of-time-order correlators in the Dicke model},
	author={Lewis-Swan, R. J. and Safavi-Naini, A. and Bollinger, J. J. and Rey, A. M.},
	journal={Nature Communications},
	volume={10},
	number={1},
	pages={1581},
	year={2019},
	doi={10.1038/s41467-019-09436-y}
}

@article{hahn2024eigenstate,
	title={Eigenstate Correlations, the Eigenstate Thermalization Hypothesis, and Quantum Information Dynamics in Chaotic Many-Body Quantum Systems},
	author={Hahn, Dominik and Luitz, David J. and Chalker, J. T.},
	journal={Physical Review X},
	volume={14},
	number={3},
	pages={031029},
	year={2024},
	doi={10.1103/PhysRevX.14.031029}
}

@misc{wen2018floquet,
	title={Floquet conformal field theory}, 
	author={Xueda Wen and Jie-Qiang Wu},
	year={2018},
	eprint={1805.00031},
	archivePrefix={arXiv},
	primaryClass={cond-mat.str-el},
	url={https://arxiv.org/abs/1805.00031}, 
}

@article{fan2021floquet,
	title={Floquet conformal field theories with generally deformed Hamiltonians},
	author={Fan, Ruihua and Gu, Yingfei and Vishwanath, Ashvin and Wen, Xueda},
	journal={SciPost Physics},
	volume={10},
	number={2},
	pages={049},
	year={2021},
	doi={10.21468/SciPostPhys.10.2.049}
}

@article{berdanier2017floquet,
	title={Floquet Dynamics of Boundary-Driven Systems at Criticality},
	author={Berdanier, William and Kolodrubetz, Michael and Vasseur, Romain and Moore, Joel E.},
	journal={Physical Review Letters},
	volume={118},
	number={26},
	pages={260602},
	year={2017},
	doi={10.1103/PhysRevLett.118.260602}
}

@article{stannigel2012driven,
	title={Driven-dissipative preparation of entangled states in cascaded quantum-optical networks},
	author={Stannigel, K. and Rabl, P. and Zoller, P.},
	journal={New Journal of Physics},
	volume={14},
	number={6},
	pages={063014},
	year={2012},
	doi={10.1088/1367-2630/14/6/063014}
}

@article{zippilli2013entanglement,
	title={Entanglement Replication in Driven Dissipative Many-Body Systems},
	author={Zippilli, S. and Paternostro, M. and Adesso, G. and Illuminati, F.},
	journal={Physical Review Letters},
	volume={110},
	number={4},
	pages={040503},
	year={2013},
	doi={10.1103/PhysRevLett.110.040503}
}

@article{chen2024periodically,
	title={Periodically driven open quantum systems: Spectral properties and nonequilibrium steady states},
	author={Chen, Hao and Hu, Yu-Min and Zhang, Wucheng and Kurniawan, Michael Alexander and Shao, Yuelin and Chen, Xueqi and Prem, Abhinav and Dai, Xi},
	journal={Physical Review B},
	volume={109},
	number={18},
	pages={184309},
	year={2024},
	doi={10.1103/PhysRevB.109.184309}
}

@article{kehrein2024page,
	title={Page curve entanglement dynamics in an analytically solvable model},
	author={Kehrein, Stefan},
	journal={Physical Review B},
	volume={109},
	number={22},
	pages={224308},
	year={2024},
	doi={10.1103/PhysRevB.109.224308}
}

@article{jha2025page,
	author = {Jha, Rishabh and Manmana, Salvatore R. and Kehrein, Stefan},
	title = {{Page curve and entanglement dynamics in an interacting fermionic chain}},
	journal = {Phys. Rev. B},
	volume = {111},
	number = {23},
	pages = {235140},
	year = {2025},
	month = jun,
	publisher = {American Physical Society},
	doi = {10.1103/lt5c-pn14}
}

@article{Li2025Jul,
	author = {Li, Lauren H. and Kehrein, Stefan and Gopalakrishnan, Sarang},
	title = {{Sharp Page transitions in generic Hamiltonian dynamics}},
	journal = {Phys. Rev. B},
	volume = {112},
	number = {1},
	pages = {014307},
	year = {2025},
	month = jul,
	publisher = {American Physical Society},
	doi = {10.1103/vnzh-y22h}
}

@article{wei2018linking,
	title={Linking phase transitions and quantum entanglement at arbitrary temperature},
	author={Wei, Bo-Bo},
	journal={Physical Review A},
	volume={97},
	number={4},
	pages={042115},
	year={2018},
	doi={10.1103/PhysRevA.97.042115}
}

@article{bhattacharyya2015signature,
	author = {Bhattacharyya, Sirshendu and Dasgupta, Subinay and Das, Arnab},
	title = {{Signature of a continuous quantum phase transition in non-equilibrium energy absorption: Footprints of criticality on higher excited states}},
	journal = {Sci. Rep.},
	volume = {5},
	number = {16490},
	pages = {1--9},
	year = {2015},
	month = nov,
	issn = {2045-2322},
	publisher = {Nature Publishing Group},
	doi = {10.1038/srep16490}
}

@article{poyhonen2021entanglement,
	title={Entanglement echo and dynamical entanglement transitions},
	author={Pöyhönen, Kim and Ojanen, Teemu},
	journal={Physical Review Research},
	volume={3},
	number={4},
	pages={L042027},
	year={2021},
	doi={10.1103/PhysRevResearch.3.L042027}
}

@article{dechiara2012entanglement,
	title={Entanglement Spectrum, Critical Exponents, and Order Parameters in Quantum Spin Chains},
	author={De Chiara, G. and Lepori, L. and Lewenstein, M. and Sanpera, A.},
	journal={Physical Review Letters},
	volume={109},
	number={23},
	pages={237208},
	year={2012},
	doi={10.1103/PhysRevLett.109.237208}
}

@article{wald2020closure,
	title={Closure of the entanglement gap at quantum criticality: The case of the quantum spherical model},
	author={Wald, Sascha and Arias, Raúl and Alba, Vincenzo},
	journal={Physical Review Research},
	volume={2},
	number={4},
	pages={043404},
	year={2020},
	doi={10.1103/PhysRevResearch.2.043404}
}

@article{bayat2014order,
	author = {Bayat, Abolfazl and Johannesson, Henrik and Bose, Sougato and Sodano, Pasquale},
	title = {{An order parameter for impurity systems at quantum criticality}},
	journal = {Nat. Commun.},
	volume = {5},
	number = {3784},
	pages = {1--6},
	year = {2014},
	month = may,
	issn = {2041-1723},
	publisher = {Nature Publishing Group},
	doi = {10.1038/ncomms4784}
}

@article{jaeger2003entanglement,
	title={Entanglement, mixedness, and spin-flip symmetry in multiple-qubit systems},
	author={Jaeger, Gregg and Sergienko, Alexander V. and Saleh, Bahaa E. A. and Teich, Malvin C.},
	journal={Physical Review A},
	volume={68},
	number={2},
	pages={022318},
	year={2003},
	doi={10.1103/PhysRevA.68.022318}
}

@article{liu2023symmetry,
	title={Symmetry classification of typical quantum entanglement},
	author={Liu, Yuhan and Kudler-Flam, Jonah and Kawabata, Kohei},
	journal={Physical Review B},
	volume={108},
	number={8},
	pages={085109},
	year={2023},
	doi={10.1103/PhysRevB.108.085109}
}

@article{alba2021entanglement,
	title={Entanglement gap, corners, and symmetry breaking},
	author={Alba, Vincenzo},
	journal={SciPost Physics},
	volume={10},
	number={3},
	pages={056},
	year={2021},
	doi={10.21468/SciPostPhys.10.3.056}
}

@article{Koziol2021Jun,
	author = {Koziol, Jan Alexander and Langheld, Anja and Kapfer, Sebastian C. and Schmidt, Kai Phillip},
	title = {{Quantum-critical properties of the long-range transverse-field Ising model from quantum Monte Carlo simulations}},
	journal = {Phys. Rev. B},
	volume = {103},
	number = {24},
	pages = {245135},
	year = {2021},
	month = jun,
	publisher = {American Physical Society},
	doi = {10.1103/PhysRevB.103.245135}
}

@article{Bukov2015Mar,
	author = {Bukov, Marin and D'Alessio, Luca and Polkovnikov, Anatoli},
	title = {{Universal high-frequency behavior of periodically driven systems: from dynamical stabilization to Floquet engineering}},
	journal = {Adv. Phys.},
	year = {2015},
	month = mar,
	publisher = {Taylor {\&} Francis},
	url = {https://www.tandfonline.com/doi/full/10.1080/00018732.2015.1055918}
}

@article{Sen2021Aug,
	author = {Sen, Arnab and Sen, Diptiman and Sengupta, K.},
	title = {{Analytic approaches to periodically driven closed quantum systems: methods and applications}},
	journal = {J. Phys.: Condens. Matter},
	volume = {33},
	number = {44},
	pages = {443003},
	year = {2021},
	month = aug,
	issn = {0953-8984},
	publisher = {IOP Publishing},
	doi = {10.1088/1361-648X/ac1b61}
}

@article{KUWAHARA201696,
	title = {Floquet–Magnus theory and generic transient dynamics in periodically driven many-body quantum systems},
	journal = {Annals of Physics},
	volume = {367},
	pages = {96-124},
	year = {2016},
	issn = {0003-4916},
	doi = {https://doi.org/10.1016/j.aop.2016.01.012},
	url = {https://www.sciencedirect.com/science/article/pii/S0003491616000142},
	author = {Tomotaka Kuwahara and Takashi Mori and Keiji Saito},
}

@article{tdvp-1,
	author = {Haegeman, Jutho and Cirac, J. Ignacio and Osborne, Tobias J. and Pi{\ifmmode\check{z}\else\v{z}\fi}orn, Iztok and Verschelde, Henri and Verstraete, Frank},
	title = {{Time-Dependent Variational Principle for Quantum Lattices}},
	journal = {Phys. Rev. Lett.},
	volume = {107},
	number = {7},
	pages = {070601},
	year = {2011},
	month = aug,
	publisher = {American Physical Society},
	doi = {10.1103/PhysRevLett.107.070601}
}

@article{tdvp-2,
	author = {Haegeman, Jutho and Lubich, Christian and Oseledets, Ivan and Vandereycken, Bart and Verstraete, Frank},
	title = {{Unifying time evolution and optimization with matrix product states}},
	journal = {Phys. Rev. B},
	volume = {94},
	number = {16},
	pages = {165116},
	year = {2016},
	month = oct,
	publisher = {American Physical Society},
	doi = {10.1103/PhysRevB.94.165116}
}

@article{tdvp-3,
	author = {Li, Jheng-Wei and Gleis, Andreas and von Delft, Jan},
	title = {{Time-Dependent Variational Principle with Controlled Bond Expansion for Matrix Product States}},
	journal = {Phys. Rev. Lett.},
	volume = {133},
	number = {2},
	pages = {026401},
	year = {2024},
	month = jul,
	publisher = {American Physical Society},
	doi = {10.1103/PhysRevLett.133.026401}
}

@article{schollwock2011density,
	title={The density-matrix renormalization group in the age of matrix product states},
	author={Schollw{\"o}ck, Ulrich},
	journal={Annals of Physics},
	volume={326},
	number={1},
	pages={96--192},
	year={2011},
	doi={10.1016/j.aop.2010.09.012}
}

@article{Fishman2022Aug,
	author = {Fishman, Matthew and White, Steven and Stoudenmire, Edwin Miles},
	title = {{The ITensor Software Library for Tensor Network Calculations}},
	journal = {SciPost Phys. Codebases},
	pages = {004},
	year = {2022},
	month = aug,
	issn = {2949-804X},
	doi = {10.21468/SciPostPhysCodeb.4}
}

@article{Heyl2013Mar,
	author = {Heyl, M. and Polkovnikov, A. and Kehrein, S.},
	title = {{Dynamical Quantum Phase Transitions in the Transverse-Field Ising Model}},
	journal = {Phys. Rev. Lett.},
	volume = {110},
	number = {13},
	pages = {135704},
	year = {2013},
	month = mar,
	publisher = {American Physical Society},
	doi = {10.1103/PhysRevLett.110.135704}
}

@article{Heyl2015Oct,
	author = {Heyl, Markus},
	title = {{Scaling and Universality at Dynamical Quantum Phase Transitions}},
	journal = {Phys. Rev. Lett.},
	volume = {115},
	number = {14},
	pages = {140602},
	year = {2015},
	month = oct,
	publisher = {American Physical Society},
	doi = {10.1103/PhysRevLett.115.140602}
}

@article{Zunkovic2018Mar,
	author = {{\ifmmode\check{Z}\else\v{Z}\fi}unkovi{\ifmmode\check{c}\else\v{c}\fi}, Bojan and Heyl, Markus and Knap, Michael and Silva, Alessandro},
	title = {{Dynamical Quantum Phase Transitions in Spin Chains with Long-Range Interactions: Merging Different Concepts of Nonequilibrium Criticality}},
	journal = {Phys. Rev. Lett.},
	volume = {120},
	number = {13},
	pages = {130601},
	year = {2018},
	month = mar,
	publisher = {American Physical Society},
	doi = {10.1103/PhysRevLett.120.130601}
}

@article{Heyl2018Apr,
	author = {Heyl, Markus},
	title = {{Dynamical quantum phase transitions: a review}},
	journal = {Rep. Prog. Phys.},
	volume = {81},
	number = {5},
	pages = {054001},
	year = {2018},
	month = apr,
	issn = {0034-4885},
	publisher = {IOP Publishing},
	doi = {10.1088/1361-6633/aaaf9a}
}

@book{Sachdev_2011,
	author    = {Subir Sachdev},
	title     = {Quantum Phase Transitions},
	edition   = {2},
	publisher = {Cambridge University Press},
	year      = {2011},
	address   = {Cambridge},
	doi       = {10.1017/CBO9780511973765},
	url       = {https://www.cambridge.org/core/books/quantum-phase-transitions/33C1C81500346005E54C1DE4223E5562}
}

@article{Chitambar2019,
	author    = {Chitambar, Eric and Gour, Gilad},
	title     = {Quantum Resource Theories},
	journal   = {Reviews of Modern Physics},
	volume    = {91},
	number    = {2},
	pages     = {025001},
	year      = {2019},
	doi       = {10.1103/RevModPhys.91.025001},
}

@misc{karch2025probingquantummanybodydynamics,
	title={Probing quantum many-body dynamics using subsystem Loschmidt echos}, 
	author={Simon Karch and Souvik Bandyopadhyay and Zheng-Hang Sun and Alexander Impertro and SeungJung Huh and Irene Prieto Rodríguez and Julian F. Wienand and Wolfgang Ketterle and Markus Heyl and Anatoli Polkovnikov and Immanuel Bloch and Monika Aidelsburger},
	year={2025},
	eprint={2501.16995},
	archivePrefix={arXiv},
	primaryClass={cond-mat.quant-gas},
	url={https://arxiv.org/abs/2501.16995}, 
}

@article{Eisert2015Feb,
	author = {Eisert, J. and Friesdorf, M. and Gogolin, C.},
	title = {{Quantum many-body systems out of equilibrium}},
	journal = {Nat. Phys.},
	volume = {11},
	pages = {124--130},
	year = {2015},
	month = feb,
	issn = {1745-2481},
	publisher = {Nature Publishing Group},
	doi = {10.1038/nphys3215}
}

@article{Bernien2017Nov,
	author = {Bernien, Hannes and Schwartz, Sylvain and Keesling, Alexander and Levine, Harry and Omran, Ahmed and Pichler, Hannes and Choi, Soonwon and Zibrov, Alexander S. and Endres, Manuel and Greiner, Markus and Vuleti{\ifmmode\acute{c}\else\'{c}\fi}, Vladan and Lukin, Mikhail D.},
	title = {{Probing many-body dynamics on a 51-atom quantum simulator}},
	journal = {Nature},
	volume = {551},
	pages = {579--584},
	year = {2017},
	month = nov,
	issn = {1476-4687},
	publisher = {Nature Publishing Group},
	doi = {10.1038/nature24622}
}

@article{Cadez2019,
	title={Edge and bulk localization of Floquet topological superconductors},
	author={{\v{C}}ade{\v{z}}, Tilen and Mondaini, Rubem and Sacramento, Pedro D.},
	journal={Physical Review B},
	volume={99},
	number={1},
	pages={014301},
	year={2019},
	doi={10.1103/PhysRevB.99.014301}
}

@article{Soori2022Anomalous,
	author       = {Abhiram Soori},
	title        = {Anomalous Josephson effect and rectification in junctions between Floquet topological superconductors},
	journal      = {Physica E: Low-dimensional Systems and Nanostructures},
	volume       = {146},
	pages        = {115545},
	year         = {2023},
	month        = {jan},
	publisher    = {North-Holland},
	doi          = {10.1016/j.physe.2022.115545},
	issn         = {1386-9477},
	url          = {https://doi.org/10.1016/j.physe.2022.115545}
}

@software{Haegeman_KrylovKit_2024,
	author = {Haegeman, Jutho},
	doi = {10.5281/zenodo.10622234},
	month = mar,
	title = {{KrylovKit}},
	url = {https://github.com/Jutho/KrylovKit.jl},
	version = {0.7.0},
	year = {2024}
}

@misc{Supplemental,
	note = {See Supplemental Material at [URL] for additional details, derivations, and numerical checks.}
}

@article{Berdanier2018Aug,
	author = {Berdanier, William and Kolodrubetz, Michael and Parameswaran, S. A. and Vasseur, Romain},
	title = {{Floquet quantum criticality}},
	journal = {Proc. Natl. Acad. Sci. U.S.A.},
	volume = {115},
	number = {38},
	pages = {9491--9496},
	year = {2018},
	month = aug,
	publisher = {Proceedings of the National Academy of Sciences},
	doi = {10.1073/pnas.1805796115}
}

@article{Pichler2016Nov,
	author = {Pichler, Hannes and Zhu, Guanyu and Seif, Alireza and Zoller, Peter and Hafezi, Mohammad},
	title = {{Measurement Protocol for the Entanglement Spectrum of Cold Atoms}},
	journal = {Phys. Rev. X},
	volume = {6},
	number = {4},
	pages = {041033},
	year = {2016},
	month = nov,
	publisher = {American Physical Society},
	doi = {10.1103/PhysRevX.6.041033}
}

@article{Brydges2019Apr,
	author = {Brydges, Tiff and Elben, Andreas and Jurcevic, Petar and Vermersch, Beno{\ifmmode\hat{\imath}\else\^{\i}\fi}t and Maier, Christine and Lanyon, Ben P. and Zoller, Peter and Blatt, Rainer and Roos, Christian F.},
	title = {{Probing R{\ifmmode\acute{e}\else\'{e}\fi}nyi entanglement entropy via randomized measurements}},
	journal = {Science},
	volume = {364},
	number = {6437},
	pages = {260--263},
	year = {2019},
	month = apr,
	issn = {0036-8075},
	publisher = {American Association for the Advancement of Science},
	doi = {10.1126/science.aau4963}
}

@article{Elben2020Nov,
	author = {Elben, Andreas and Kueng, Richard and Huang, Hsin-Yuan (Robert) and van Bijnen, Rick and Kokail, Christian and Dalmonte, Marcello and Calabrese, Pasquale and Kraus, Barbara and Preskill, John and Zoller, Peter and Vermersch, Beno{\ifmmode\hat{\imath}\else\^{\i}\fi}t},
	title = {{Mixed-State Entanglement from Local Randomized Measurements}},
	journal = {Phys. Rev. Lett.},
	volume = {125},
	number = {20},
	pages = {200501},
	year = {2020},
	month = nov,
	publisher = {American Physical Society},
	doi = {10.1103/PhysRevLett.125.200501}
}
		
		%%%%%%%%%%%%%%%%%%%%%%%%%%%%%%%%%
		%%%%%%%%%%%%%%%%%%%%%%%%%%%%%%%%%

	\end{document}

\preprint{APS/123-QED}
	
	\title{Supplemental Material --- Temporal Entanglement Transitions in the Periodically Driven Ising Chain}% Force line breaks with \\
	%\thanks{A footnote to the article title}%
	
	\author{Karun Gadge}
	\email{karun.gadge@uni-goettingen.de}
	\affiliation{%
		Institute for Theoretical Physics, Georg-August-Universit\"{a}t G\"{o}ttingen, 37077 G\"ottingen, Germany
	}%
	\author{Abhinav Prem}
	\email{aprem@bard.edu}
	\affiliation{%
		Physics Program, Bard College, 30 Campus Road, Annandale-on-Hudson, NY 12504, USA
	}%
	\author{Rishabh Jha}
	%\altaffiliation[Also at ]{Physics Department, XYZ University.}%Lines break automatically or can be forced with \\
	\email{rishabh.jha@uni-goettingen.de}
	\affiliation{%
		Institute for Theoretical Physics, Georg-August-Universit\"{a}t G\"{o}ttingen, 37077 G\"ottingen, Germany
	}%

	\maketitle
	
	\onecolumngrid

	\tableofcontents
	
	%%%%%%%%%%%%%%%%%%%%%%%%%%%%%%%%%
	%%%%%%%%%%%%%%%%%%%%%%%%%%%%%%%%%
	%%%%%%%%%%%%%%%%%%%%%%%%%%%%%%%%%
	%%%%%%%%%%%%%%%%%%%%%%%%%%%%%%%%%

	\section{Numerical Methods and Convergence}
	\label{app. section Numerical Methods and Convergence}
	
	We employ the time-dependent variational principle (TDVP) \cite{tdvp-1, tdvp-2, tdvp-3} for matrix product states (MPS) \cite{schollwock2011density} as implemented in the \texttt{ITensors.jl} library \cite{Fishman2022Aug}, a high-performance tensor network library for Julia (version 1.11.6). Our simulations utilize the two-site TDVP algorithm, which dynamically adapts the bond dimension to maintain a truncation error below $10^{-10}$. All data used in this work and the script to generate them are available via Zenodo upon reasonable request \cite{Gadge2025_zenodo}.
	
	%%%%%%%%%%%%%%%%%%%%%%%%%%%%%%%%%
	%%%%%%%%%%%%%%%%%%%%%%%%%%%%%%%%%

	\subsection{Matrix Product State Implementation} 
	\label{app. subsection Matrix Product State Implementation}
	
	Initial state preparation uses the density matrix renormalization group (DMRG) algorithm to find the ground state of the static Hamiltonian:
	\be
	\mathcal{H}_{\text {static }}=-J \sum_{i=1}^{L-1} \sigma_i^z \sigma_{i+1}^z-\frac{h_0}{2} \sum_{i=1}^L \sigma_i^x
	\ee
	where the DMRG parameters are: (1) Number of sweeps $= 30$; (2) Progressive bond dimensions $= [100, 200, 400, 800]$; Singular value cutoff $=10^{-15}$; and (4) Initial state: N\'eel state $| \uparrow \downarrow \uparrow \downarrow \cdots \rangle$. The ground state energy convergence is monitored to ensure numerical accuracy before time evolution begins.
	
	For time evolution under the periodically driven Hamiltonian 
	\begin{equation}
		\mathcal{H}(t) = -J\sum_{i=1}^{L-1} \sigma_i^z \sigma_{i+1}^z - \frac{h_0}{2}\cos(\omega t)\sum_{i=1}^{L} \sigma_i^x,
	\end{equation}
	we use a time step of $dt = 0.01/J$ for $\omega \leq 10J$ and $dt = 0.002/J$ for higher frequencies $\omega = 30J$ and $50J$ while $dt=0.001/J$ for $\omega = 70J$ and $100J$. The TDVP algorithm preserves unitarity and maintains the MPS in canonical form throughout evolution, with maximum bond dimension $\chi_{\text{max}} = 2000$ and truncation cutoff $10^{-10}$. In the next subsection, we also provide evidence of convergence for $\omega = 10.0J$ for $dt=0.01$ as well as $dt=0.001$ that the physics obtained is not a function of time discretization. 
	
	Key numerical aspects include
	\begin{itemize}
		\item Subsystem analysis: The reduced density matrix $\rho_A$ for subsystem $A$ (first $L_A$ sites) is computed via orthogonalizing the MPS to the subsystem boundary and constructing the density matrix via tensor contractions.
		\item Entanglement spectrum: We compute the top two Schmidt values $\lambda_0, \lambda_1$ and corresponding eigenvectors using KrylovKit (version 0.9.5) \cite{Haegeman_KrylovKit_2024} with tolerance $10^{-8}$, ensuring accurate resolution of near-degenerate states.
		\item Symmetry operators: Subsystem parity $P_A = \prod_{i \in A} \sigma_i^x$ and magnetization $M_A = \sum_{i \in A} \sigma_i^z$ are constructed as explicit matrices for the subsystem Hilbert space.
		\item Convergence: At each time step we enforced the truncation cutoff $10^{-10}$ and tracked the maximum bond dimension $\chi_{\max}$ where we went to a maximum value of 2000 as the need arose for different runs; throughout this work, across all simulations for the entire time evolution, the bond dimension and the truncation cutoff remained below these specified values throughout the time evolution.
	\end{itemize}
	
	The code implements comprehensive diagnostics including entanglement entropy, Schmidt values, entanglement echo $E(t) = \langle\lambda_0(0)|\lambda_0(t)\rangle$, and symmetry expectations, providing multiple consistency checks for the observed temporal entanglement transitions (TET).
	
	\begin{figure}[htbp]
		\centering
		\includegraphics[width=\columnwidth]{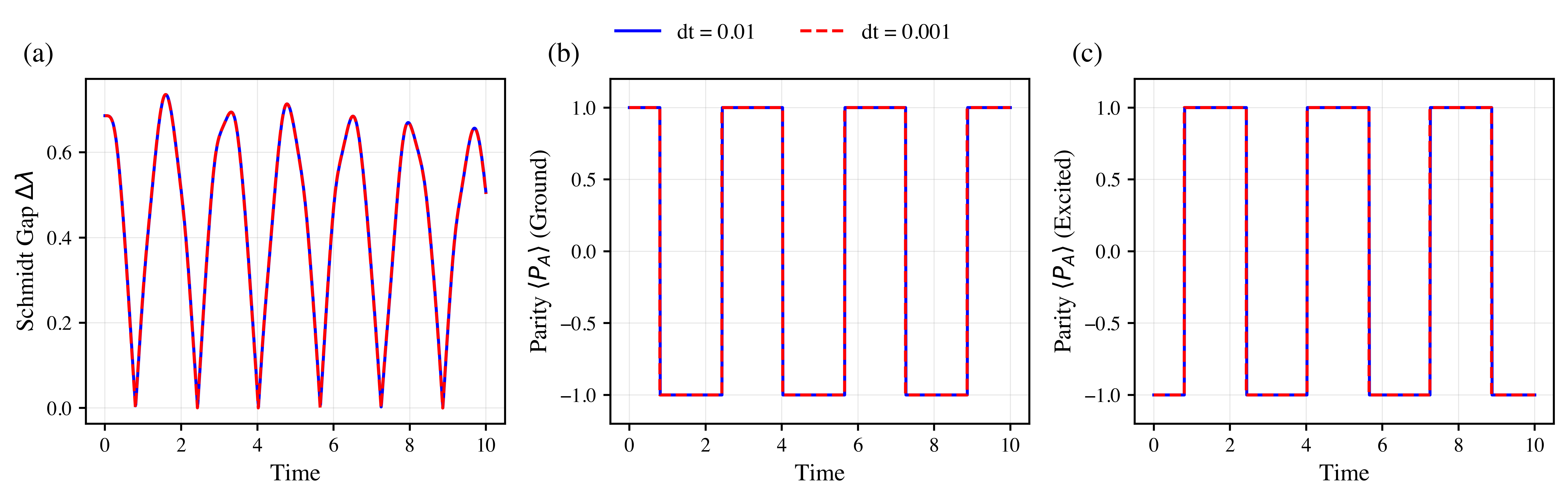}
		\caption{Time-step convergence test for $\omega=10.0$. (a) Schmidt gap dynamics for $dt=0.01$ (solid blue) and $dt=0.001$ (dashed red) show excellent agreement. The overlap of dynamical evolution of subsystem parity expectation value with respect to (b) the largest Schmidt vector and (c) the second largest Schmidt vector also convincingly establish the convergence. Parameters: $L=24$, $L_A=9$, $J=1.0$, $h_0=2.0$.}
		\label{fig:convergence_dt}
	\end{figure}
	
	%%%%%%%%%%%%%%%%%%%%%%%%%%%%%%%%%
	%%%%%%%%%%%%%%%%%%%%%%%%%%%%%%%%%

	\subsection{Time-Step Convergence} 
	\label{app. subsection Time-Step Convergence}

	To ensure that our results are not artifacts of the chosen time step, we performed detailed convergence tests. Figure \ref{fig:convergence_dt} compares simulations with $dt=0.01$ and $dt=0.001$ for $\omega=10.0$. The dynamical evolution, as reflected in the Schmidt eigenvalues and Schmidt vectors, shows an excellent collapse. This confirms that our standard time step of $dt=0.01$ is sufficient to capture the TET for $\omega=10.0$, while in general we take $dt$ to be at least one order of magnitude smaller than $1/\omega$.

	The convergence is particularly important for resolving the rapid oscillations at higher frequencies and the sharp non-analyticities at critical times $t_c^{(k)}$. The TDVP algorithm's structure-preserving properties (unitarity, energy conservation) further enhance numerical stability, ensuring reliable long-time evolution.
	
	%%%%%%%%%%%%%%%%%%%%%%%%%%%%%%%%%
	%%%%%%%%%%%%%%%%%%%%%%%%%%%%%%%%%
	%%%%%%%%%%%%%%%%%%%%%%%%%%%%%%%%%
	%%%%%%%%%%%%%%%%%%%%%%%%%%%%%%%%%

	\section{The Uniqueness of the Entanglement Transition}
	\label{app. section The Uniqueness of the Entanglement Transition}
	
	This section establishes that the observed TET represent a distinct (non-equilibrium) phenomenon not captured by conventional observables, robust across different equilibrium phases provided the three-condition criterion is satisfied: (i) $\mathbb{Z}_2$-symmetric driven Hamiltonian, (ii) $\mathbb{Z}_2$-symmetric initial state, and (iii) the reduced density matrix $\rho_A(t)$ has support in both subsystem-parity sectors, with critical times exactly at sector-weight equalities (this is made more precise in Sec.~\ref{app.:evidence for different initial states and types of driving}).

	To establish that TET represent genuinely novel phenomena, we must demonstrate that conventional quantum many-body probes fail to detect these critical events. Standard observables used to characterize dynamical quantum phase transitions include magnetization dynamics \cite{Sachdev_2011} and Loschmidt echo rate functions \cite{Heyl2013Mar, Heyl2015Oct, Zunkovic2018Mar, Heyl2018Apr}. If TET were simply manifestations of known physics, these conventional measures should exhibit corresponding singularities or anomalies at the critical times $t_c^{(k)}$.

	\begin{figure}[htbp]
		\centering
		\includegraphics[width=\columnwidth]{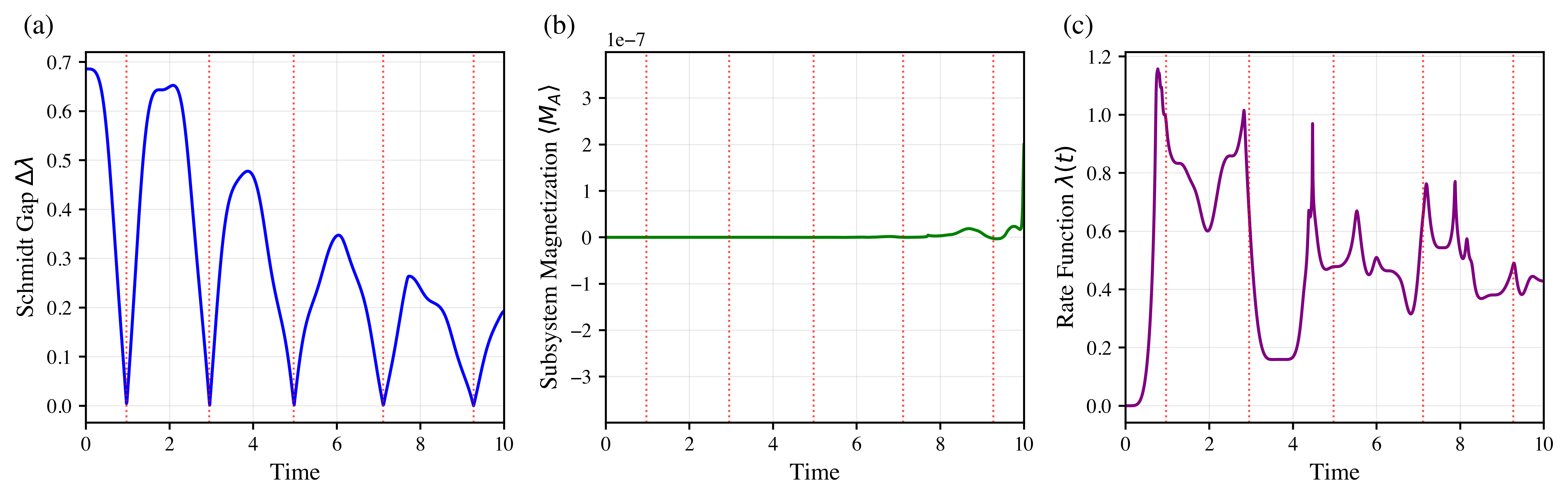}
		\caption{Comparison between entanglement measures and conventional observables for $\omega=5.0$. (a) Schmidt gap $\Delta\lambda = \lambda_0 - \lambda_1$ showing clear non-analyticities at critical times (vertical dashed lines). (b) Subsystem magnetization $\langle M_A \rangle = \frac{1}{L_A}\sum_{i\in A}\langle\sigma_i^z\rangle$ remains smooth through transition points. (c) Loschmidt echo rate function $\lambda(t) = -\ln|\langle\psi(0)|\psi(t)\rangle|^2/L$ shows no singular behavior at critical times. Parameters: $L=24$, $L_A=9$, $J=1.0$, $h_0=2.0$.}
		\label{fig:observables_comparison}
	\end{figure}
	
	%%%%%%%%%%%%%%%%%%%%%%%%%%%%%%%%%
	%%%%%%%%%%%%%%%%%%%%%%%%%%%%%%%%%

	\subsection{Comparison to Conventional Observables} 
	\label{app. subsection Comparison to Conventional Observables}
	
	Figure \ref{fig:observables_comparison} demonstrates that conventional observables fail to capture the TET. While the Schmidt gap exhibits clear non-analyticities at critical times $t_c^{(k)}$, both the subsystem magnetization $\langle M_A \rangle$ and the Loschmidt echo rate function remain smooth throughout the evolution. This establishes that the transitions are uniquely encoded in the entanglement structure rather than local order parameters or global state fidelity.
	
	The smooth behavior of $\langle M_A \rangle$ indicates that no local symmetry breaking occurs in the physical degrees of freedom, while the absence of singularities in the Loschmidt echo distinguishes these transitions from dynamical quantum phase transitions associated with the full system wave-function. This separation confirms that TET represent a new class of non-equilibrium phenomenon specific to entanglement dynamics.

	%%%%%%%%%%%%%%%%%%%%%%%%%%%%%%%%%
	%%%%%%%%%%%%%%%%%%%%%%%%%%%%%%%%%

	\subsection{Robustness Across Equilibrium Phases} 
	\label{app. subsection Robustness Across Equilibrium Phases}
	
	We demonstrate that TET persist across different \emph{equilibrium phases of the initial state}, provided the symmetry and sector-weight equality criteria of the main text are satisfied (also mentioned in depth in Sec.~\ref{app.:evidence for different initial states and types of driving}.
	In particular, all runs in this subsection use (i) a $\mathbb{Z}_2$-symmetric driven Hamiltonian and (ii) an initial state that is a global $\mathbb{Z}_2$ eigenstate, so that $\rho_A(t)$ remains parity block-diagonal, $[\rho_A(t),P_A]=0$, as verified by the symmetry-check diagnostic $C(t)=\|[\rho_A(t),P_A]\|\simeq 0$.
	In the paramagnetic phase ($h_0=2.4$, i.e., $h_0/2=1.2>J$), the ground state naturally respects $\mathbb{Z}_2$ symmetry, satisfying condition (ii). This contrasts with the symmetry-broken MPS/DMRG branch deep in the ferromagnetic phase (e.g., $h_0/2=0.5$) discussed in Sec.~\ref{app.:evidence for different initial states and types of driving}, which fails condition (ii) and correspondingly does not exhibit TET.

\begin{figure}[htbp]
		\centering
		% Replace with your actual filenames
		%	\includegraphics[width=0.95\columnwidth]{h016.png} % (a)-(c)
		\vspace{1mm} % Optional: tweak vertical space
		\includegraphics[width=0.95\columnwidth]{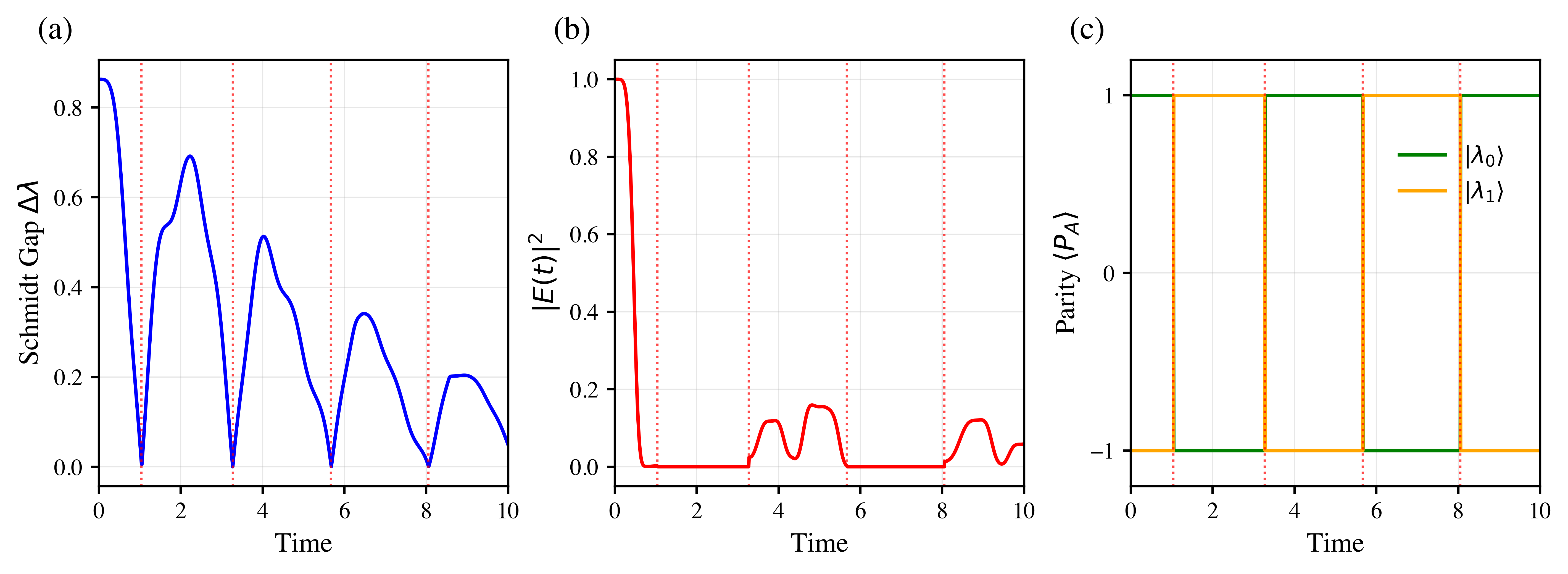} % (d)-(f)
		\caption{Temporal entanglement transitions (TET) in the paramagnetic phase. \textbf{(a)} Schmidt gap $\Delta\lambda$, \textbf{(b)} entanglement echo $|E(t)|^2$, \textbf{(c)} subsystem parity expectations for $h_0=2.4$ (paramagnetic phase, $h_0/2=1.2>J$). The system exhibits clear temporal entanglement transition signatures, demonstrating robustness in an equilibrium phase far from criticality, provided the three-condition criterion is satisfied. Red vertical lines mark critical times as read-off from the parity jumps in (c) (dynamical spontaneous symmetry breaking) and overlaid across all panels. Parameters: $L=24$, $L_A=9$, $J=1.0$, $\omega=5.0$, $dt=0.01$.}
		\label{fig:h0_comparison}
	\end{figure}

	Figure \ref{fig:h0_comparison} demonstrates TET in the paramagnetic phase of the initial state, where the three-condition criterion is satisfied. For $h_0=2.4$ (paramagnetic phase, $h_0/2=1.2>J$), we observe clear signatures: Schmidt gap closures, entanglement echo vanishing, and subsystem parity flips, consistent with the behavior at criticality ($h_0=2.0$) shown in the main text.

\begin{figure}[htbp]
		\centering
		\includegraphics[width=0.85\columnwidth]{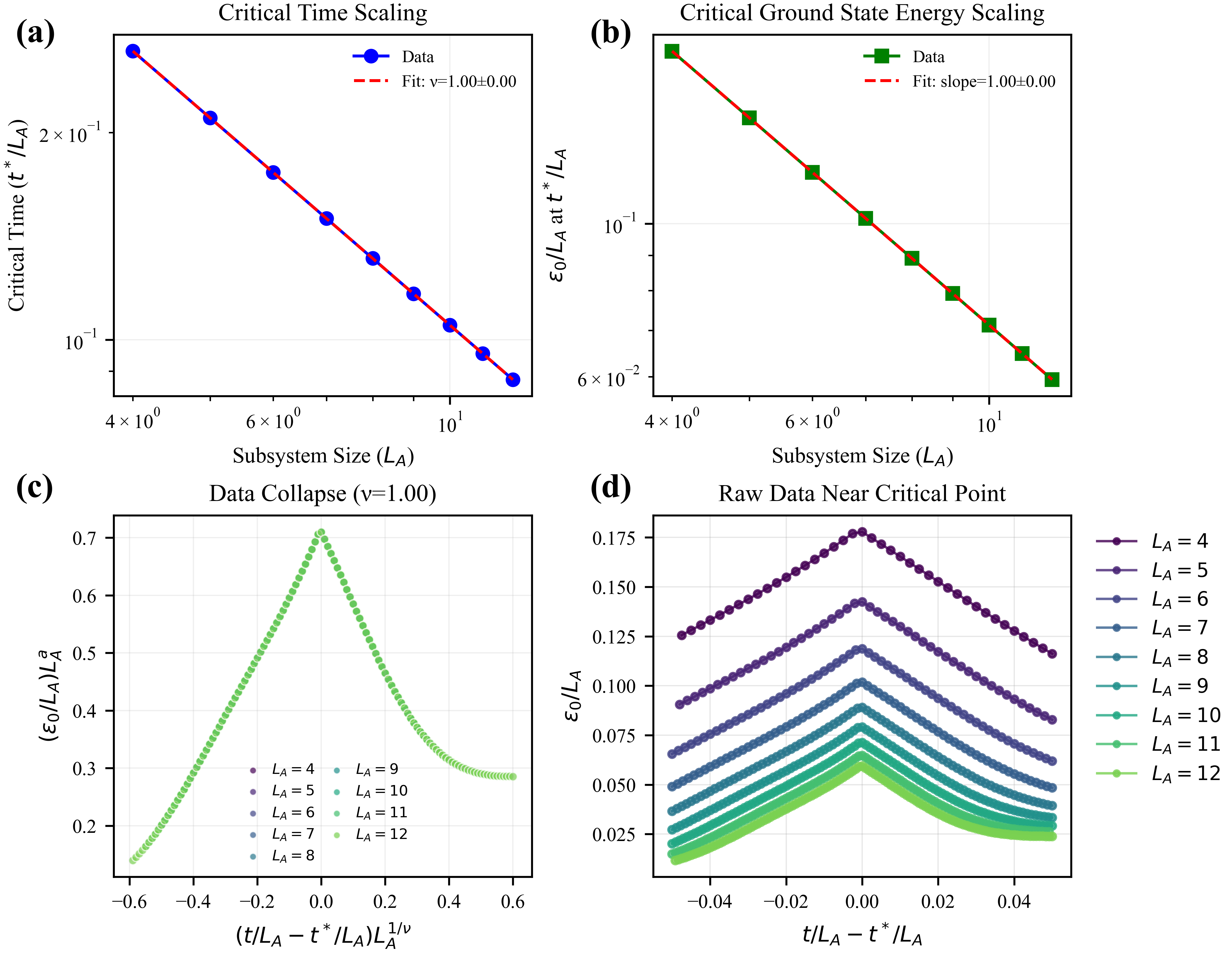}	
		\caption{Finite-size scaling for paramagnetic phase ($h_0= 2.4$) yields $\nu\simeq 1.00$, confirming universality with Fig.~2 of the main text (which uses $h_0=2.0$ at criticality). We again have $a\simeq 1$. The identical exponent in a different equilibrium phase demonstrates that TET are fundamentally non-equilibrium phenomena, decoupled from the equilibrium phase of the initial state. Parameters: $L = 24$, $J = 1.0$, $\omega= 5.0$. The remaining caption and subplot explanations are identical to Fig.~2 of the main text.}
		\label{fig:scaling_h24}
	\end{figure}

	This confirms that TET occur in the paramagnetic phase, far from equilibrium criticality, provided the three-condition criterion is satisfied. Most remarkably, finite-size scaling analyses for the $\mathbb{Z}_2$-symmetric, $h_0=2.4$ (paramagnetic) case in Fig. \ref{fig:scaling_h24} reveal that the correlation length critical exponent remains $\nu \simeq 1.00$, identical to the equilibrium value for the 2D classical Ising/1D transverse-field Ising model universality class. This is particularly surprising because: (i) the initial state is in the paramagnetic phase, away from equilibrium criticality, and (ii) the system is driven far from equilibrium, yet the critical exponent remains unchanged from its equilibrium value. This remarkable invariance suggests that the universal aspects of the TET are decoupled from the specific equilibrium phase of the initial state, thereby having a universality class of its own. Robustness across multiple initial-state phases (ferromagnetic, critical, paramagnetic) is demonstrated in Sec.~\ref{app.:evidence for different initial states and types of driving}. 
	
	We emphasize that while TET occur robustly in paramagnetic and critical initial-state phases satisfying parity symmetry provided the three-condition criterion is satisfied, the \emph{finite-size scaling analyses} presented in this work (Fig.~2 of the main text and Sec.~\ref{app. subsection Finite-Size Scaling for Full Data Set and Universality} below) are performed by initializing near the equilibrium critical point ($h_0/2 = J$). This choice ensures all three conditions are simultaneously and robustly satisfied in the short-time dynamics, thereby optimizing the identification of multiple critical times $t^*_k$ required for universal scaling collapse. Alternative initial-state phases (e.g., deep paramagnetic with high-frequency driving) may fail Condition 3 in short-time windows (Sec.~\ref{sec:necessity_cond3_paramag_pureparamag}) or require careful parameter tuning to achieve sector-weight crossings, making them less suitable for systematic finite-size scaling.

	%%%%%%%%%%%%%%%%%%%%%%%%%%%%%%%%%
	%%%%%%%%%%%%%%%%%%%%%%%%%%%%%%%%%

	\subsection{Role of Initial-State Properties} 
	\label{app. subsection Dependence on Initial Entangled State}
	
	A central message of the main text is that TET occur when three conditions are jointly satisfied: (1) the driven Hamiltonian preserves global $\mathbb{Z}_2$ symmetry, (2) the initial state respects this symmetry, and (3) $\rho_A(t)$ has support in both subsystem-parity sectors, with critical times at sector-weight equalities.
	Conditions (1) and (2) together ensure that $\rho_A(t)$ remains parity block-diagonal, while condition (3) ensures both sectors remain dynamically relevant; in practice, $\mathbb{Z}_2$-symmetric initial states that satisfy these criteria often (though not always) possess substantial bipartite entanglement across the $A|\bar A$ cut, while typical non-entangled states such as random product states or domain wall states violate condition (2) (more on this below).

\begin{figure}[htbp]
		\centering
		\includegraphics[width=0.85\columnwidth]{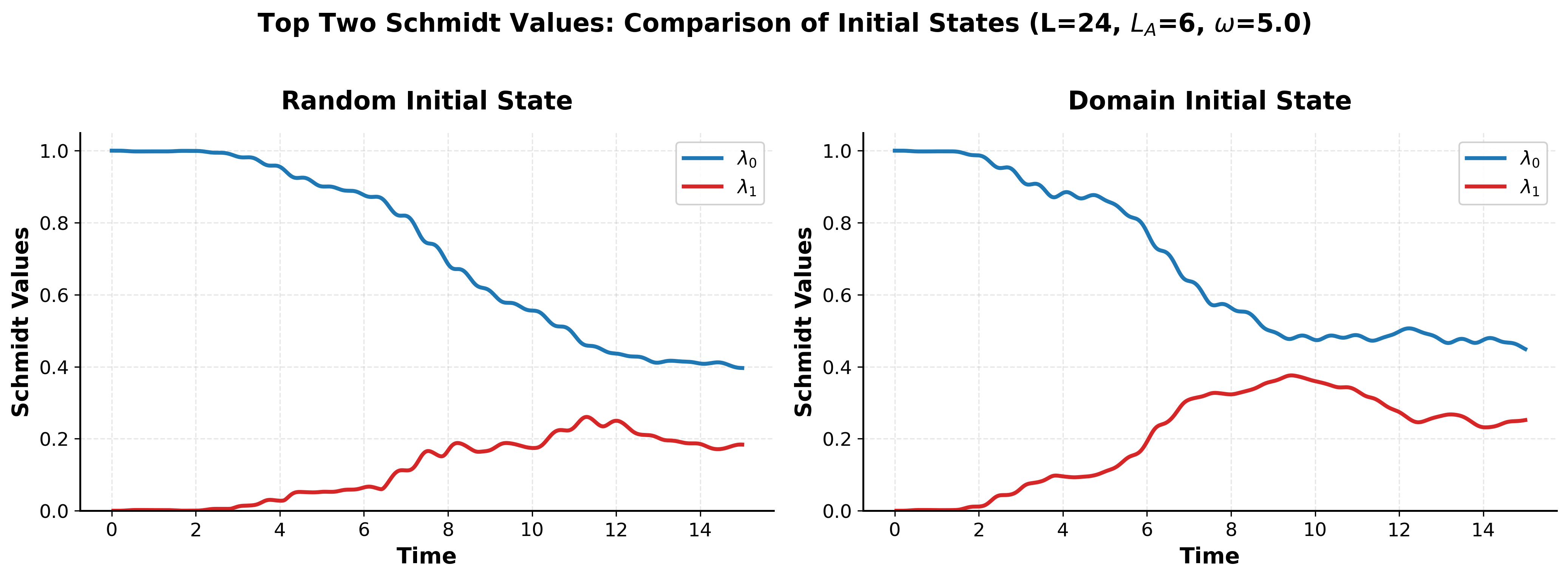}	
		\caption{
			Role of initial-state properties in temporal entanglement transitions (TET).
			The entangled ground state of $H_{\mathrm{static}}$ (prepared as a $\mathbb{Z}_2$ eigenstate) exhibits clear temporal entanglement transition signatures, while the random product state and domain wall state, which violate the three-condition criterion (particularly conditions (2) and (3)), do not.
			Since these product-state preparations are also typically $\mathbb{Z}_2$-breaking, their absence of transitions is consistent with violation of the three-condition criterion; see Sec.~\ref{app.:evidence for different initial states and types of driving} for symmetry-resolved diagnostics using $C(t)$ and $w_\pm(t)$.
			Parameters: $L=24$, $L_A=6$, $J=1.0$, $h_0=2.0$, $\omega=5.0$, $dt=0.01$, $t_{\max}=15.0$.
		}
		\label{fig:initial_states_comparison}
	\end{figure}

	Figure~\ref{fig:initial_states_comparison} compares time evolution from (a) the entangled ground state of the static Hamiltonian (a $\mathbb{Z}_2$-eigenstate in the parameter regime considered) versus (b,c) representative product-state initial conditions (random product and domain wall), which violate the three-condition criterion and therefore do not exhibit TET.
	These product-state preparations also typically break the global $\mathbb{Z}_2$ symmetry, i.e., they violate condition (ii); the symmetry-resolved mechanism and its diagnostics (including $C(t)$ and the sector weights $w_\pm(t)$) are quantified separately in Sec.~\ref{app.:evidence for different initial states and types of driving}.

	The most fundamental question regarding TET concerns their existence conditions: what properties must the initial state and Hamiltonian possess for these transitions to occur? Our investigation reveals that the three-condition criterion—$\mathbb{Z}_2$ symmetry of driven Hamiltonian and initial state, plus sustained sector competition—governs the existence of TET, distinguishing them as phenomena rooted in the interplay between symmetry structure and entanglement dynamics.

	Figure \ref{fig:initial_states_comparison} demonstrates the necessity of the three-condition criterion for TET. The ground state of $\mathcal{H}_{\text{static}}$ (a $\mathbb{Z}_2$ eigenstate that satisfies all three conditions) exhibits clear Schmidt gap closures, while both the random product state and domain wall state (which violate conditions (2) and typically (3)) fail to show these transitions.

	This absence of transitions in product-state initial conditions can be understood through the three-condition framework. Most product states (random product states, domain walls) break the global $\mathbb{Z}_2$ symmetry, violating condition (2); consequently, $[\rho_A(t),P_A]\neq 0$ and the parity block-diagonal structure required for TET is absent. Even when product states happen to respect $\mathbb{Z}_2$, they typically fail condition (3) because one parity sector dominates $\rho_A(t)$, preventing the sector-weight equality to admit any solution which is necessary for the transition mechanism (detailed in the next section). In contrast, the $\mathbb{Z}_2$-symmetric ground state of $\mathcal{H}_{\text{static}}$ satisfies all three conditions and exhibits the transitions.

	%%%%%%%%%%%%%%%%%%%%%%%%%%%%%%%%%
	%%%%%%%%%%%%%%%%%%%%%%%%%%%%%%%%%
	%%%%%%%%%%%%%%%%%%%%%%%%%%%%%%%%%
	%%%%%%%%%%%%%%%%%%%%%%%%%%%%%%%%%
	
	\FloatBarrier
	
	\section{Evidence of Universal Mechanism for Temporal Entanglement Transitions}
	\label{app.:evidence for different initial states and types of driving}
	
	This section provides explicit evidence for the universal ``symmetry + sector-competition'' criterion stated in the main text: TET occur universally across driving frequencies provided 
	\begin{enumerate}
    \item {\bf Symmetry of the driven Hamiltonian:} the driven Hamiltonian preserves the global \(\mathbb{Z}_2\) symmetry (for the TFIM, generated by \(\prod_{i=1}^{L}\sigma_i^x\)).
    \item {\bf Symmetry of the initial state:} the initial state is a global \(\mathbb{Z}_2\)-eigenstate.
\item {\bf Sector-weight equality condition:} The reduced density matrix 
\(\rho_A(t)\) carries appreciable weight in both subsystem-parity sectors 
(quantified by \(w_\pm(t)=\mathrm{Tr}[P_\pm\rho_A(t)]\), with 
\(P_\pm=(\mathds{1}_A\pm P_A)/2\) and \(P_A=\prod_{i\in A}\sigma_i^x\)), 
with critical times \(t_c\) \footnote{Recall that the first critical time is generally denoted by $t^*$ while all critical times are denoted as $t_c$ and if we need to label them, then $t_c^{(k)}$. Therefore $t_c^{(1)} = t^*$.} selected by the sector-weight equality 
\(w_+(t_c) = w_-(t_c)\) (multiple solutions may exist within the evolution window).
    \end{enumerate}
Conditions (1) and (2) together enforce $C(t)  = [\rho_A(t),P_A]\simeq0$, implying that the inter-sector cross-term $\|P_+\rho_A(t)P_-\|\simeq 0$ throughout unitary time evolution, which we verify as a redundancy check. Condition (3) requires that both parity sectors of $\rho_A(t)$ carry comparable weight---quantified via the instantaneous sector weights $w_{\pm}(t)=\mathrm{Tr}[P_{\pm}\rho_A(t)]$---with time-averaged values satisfying $\overline{w}_+\approx\overline{w}_-$, ensuring both sectors remain dynamically 
relevant for \textit{multiple} transitions where transition times are given by the sector-equality $w_+(t) = w_-(t)$.
We demonstrate below that transitions are observed only when the initial state is $\mathbb{Z}_2$-symmetric, while $\mathbb{Z}_2$-breaking ferromagnetic initial conditions do not exhibit transitions, independent of the driving protocol.

\textbf{Operational criterion and universality.}
Condition (3) translates into an operational sufficient condition for TET: whenever the equation $w_+(t^*)=w_-(t^*)$ admits a solution within the evolution window, a temporal entanglement transition occurs at $t=t^*$, provided conditions (1) and (2) are simultaneously satisfied. 
We stress that a sector-weight crossing is not sufficient by itself: without the synchronized, discontinuous parity jumps in the leading Schmidt vectors (which corresponds to Schmidt-gap closure), the dynamics exhibit only a rounded crossover.
This mechanism is universally pronounced in the short-time dynamics across all driving frequencies when the initial state is prepared and the drive is near the equilibrium critical point. Importantly, our finite-size scaling analysis is performed in this regime to ensure all three conditions are robustly satisfied in the short-time window and thereby optimize the resolution of universal critical behavior.
Crucially, however, proximity to criticality is \emph{not} a necessary requirement: as explicitly demonstrated in the examples below (where parameters are chosen to ensure \(w_+(t^*)=w_-(t^*)\) admits solutions within short-time evolution; not all parameter choices yield solutions/crossings in short-time dynamics), TET robustly manifest for initial states deep in the paramagnetic phase subjected to purely paramagnetic driving, as well as for purely ferromagnetic drive protocols, provided the three conditions remain fulfilled and the sector-weight crossing equation admits solutions. While the sector-weight trajectories $w_{\pm}(t)$ may exhibit smooth crossings, the hallmark signatures of TET---namely, the spontaneous $\mathbb{Z}_2$ symmetry breaking evidenced by discontinuous parity flips in the leading Schmidt eigenvectors $|\lambda_0(t)\rangle$ and $|\lambda_1(t)\rangle$, synchronized Schmidt-gap closures, and vanishing entanglement echo---sharply distinguish genuine transitions from mere crossover phenomena.
	
	\textbf{Initial-state symmetry: ED vs MPS/DMRG.}
	A key technical point is that, in the ferromagnetic phase, exact diagonalization (ED) at finite size typically returns a parity-eigenstate ``cat''-like ground state (which respects the global $\mathbb{Z}_2$ symmetry), whereas standard DMRG/MPS ground-state preparation without explicitly enforcing the $\mathbb{Z}_2$ symmetry often converges to a symmetry-broken branch.
	In this section, the label ``symmetry-broken ferromagnetic initial state'' refers precisely to this symmetry-broken MPS/DMRG branch, i.e., an initial state that is \emph{not} a $\mathbb{Z}_2$-eigenstate and is therefore representative of the thermodynamic-limit symmetry-broken phase.
	By contrast, in the paramagnetic phase the ground state is $\mathbb{Z}_2$-symmetric, so condition (2) is satisfied and the symmetry-check diagnostic $C(t)$ defined below remains negligible.
	
	TET occur only when the above three conditions, namely (1) the driven Hamiltonian is $\mathbb{Z}_2$-symmetric, (2) the initial state is a $\mathbb{Z}_2$-eigenstate, and (3) both subsystem-parity sectors carry appreciable weight in $\rho_A(t)$; in contrast, symmetry-broken ferromagnetic MPS initial states fail condition (2), as certified by $C(t)\not\approx 0$, and having a mixed-field Ising model violates condition (1) (both are discussed below).
	
	\subsection{Diagnostics: Sector Weights and Symmetry Check}
	\label{app.:evidence diagnostics}
	
	We quantify condition (3) via sector weights by projecting $\rho_A(t)$ onto the subsystem-parity sectors using
	\begin{equation}
		P_{\pm}=\frac{\mathds{1}_A\pm P_A}{2},
		\qquad
		P_A=\prod_{i\in A}\sigma_i^x,
        \label{eq:definition of P pm}
	\end{equation}
	and defining the instantaneous sector weights
	\begin{equation}
		w_{\pm}(t)=\mathrm{Tr}\!\left[P_{\pm}\rho_A(t)\right],
		\qquad
		w_{+}(t)+w_{-}(t)=1.
		\label{eq:w+ w- defined}
	\end{equation}
	We also monitor the inter-sector block $P_{+}\rho_A(t)P_{-}$ (whose magnitude can be reported via any convenient matrix norm) to verify that both sectors remain dynamically relevant because the inter-sector block remain vanishingly small given $C(t) \approx 0$.
	
	As a consequence of (1) and (2), $\rho_A(t)$ must remain block-diagonal in the $P_A$ sectors, i.e., $[\rho_A(t),P_A]=0$ at all times.
	We therefore define the symmetry-check diagnostic
	\begin{equation}
		C(t)=\left\|[\rho_A(t),P_A]\right\|,
	\end{equation}
	and require $C(t)$ to remain negligibly small throughout evolution in all cases claimed to be symmetry-preserving.
	
	Finally, to connect to the transition phenomenology used throughout the paper, we plot in the same runs the Schmidt gap
	$\Delta\lambda(t)=\lambda_0(t)-\lambda_1(t)$ and the subsystem-parity expectations in the leading Schmidt vectors,
	$\langle P_A\rangle_{\lambda_0(t)}$ and $\langle P_A\rangle_{\lambda_1(t)}$.
	
	\subsection{Drive Protocols: ``Within-Phase'' vs ``Across-Critical''}
	\label{app.:evidence drive protocols}
	
	To test robustness with respect to driving details, we consider a generalized transverse-field protocol
	\begin{equation}
		\mathcal{H}(t)= -J\sum_{i=1}^{L-1}\sigma_i^z\sigma_{i+1}^z
		-\frac{1}{2}\Big(h_{\mathrm{offset}}+h_{\mathrm{ac}}\cos(\omega t)\Big)\sum_{i=1}^{L}\sigma_i^x,
		\label{eq:Ht_general}
	\end{equation}
	which includes the main-text choice as the special case $h_{\mathrm{offset}}=0$ and $h_{\mathrm{offset}}=h_0$.
	Equivalently, one may write $h(t)=h_{\mathrm{offset}}+h_{\mathrm{ac}}\cos(\omega t)$, so that the Hamiltonian contains the transverse field $h(t)/2$.
	
	We separate three representative regimes by the instantaneous field values
	$h_{\min}=h_{\mathrm{offset}}-h_{\mathrm{ac}}$ and $h_{\max}=h_{\mathrm{offset}}+h_{\mathrm{ac}}$:
	
	\begin{itemize}
		\item \textbf{Purely ferromagnetic-phase driving:} $h_{\max}/2 < J$ (the drive stays on the ferromagnetic side at all times).
		\item \textbf{Across-critical driving:} $h_{\min}/2 < J < h_{\max}/2$ (the drive crosses the equilibrium critical point during a cycle).
		\item \textbf{Purely paramagnetic-phase driving (offset drive):} $h_{\min}/2 > J$ (the drive stays on the paramagnetic side at all times).
	\end{itemize}
	
	In all cases below, the driven Hamiltonian is $\mathbb{Z}_2$-symmetric; the only controlled change is whether the \emph{initial state} respects or breaks this symmetry.
	
	\subsection{Symmetry-Broken, Ferromagnetic Initial State}
	\label{app.:evidence ferro init}
	
	We first initialize in the ferromagnetic phase with a symmetry-broken ground state obtained from MPS/DMRG, i.e., an initial state that is not a $\mathbb{Z}_2$-eigenstate (in contrast to the finite-size ED cat-like ground state).
	In this case condition (2) fails, so TET are not expected; correspondingly, $\rho_A(t)$ need not remain strictly parity block-diagonal and $C(t)$ provides a direct diagnostic of this symmetry violation at the reduced-density-matrix level.
	
\begin{figure*}[t]
		\centering
		\includegraphics[width=0.96\textwidth]{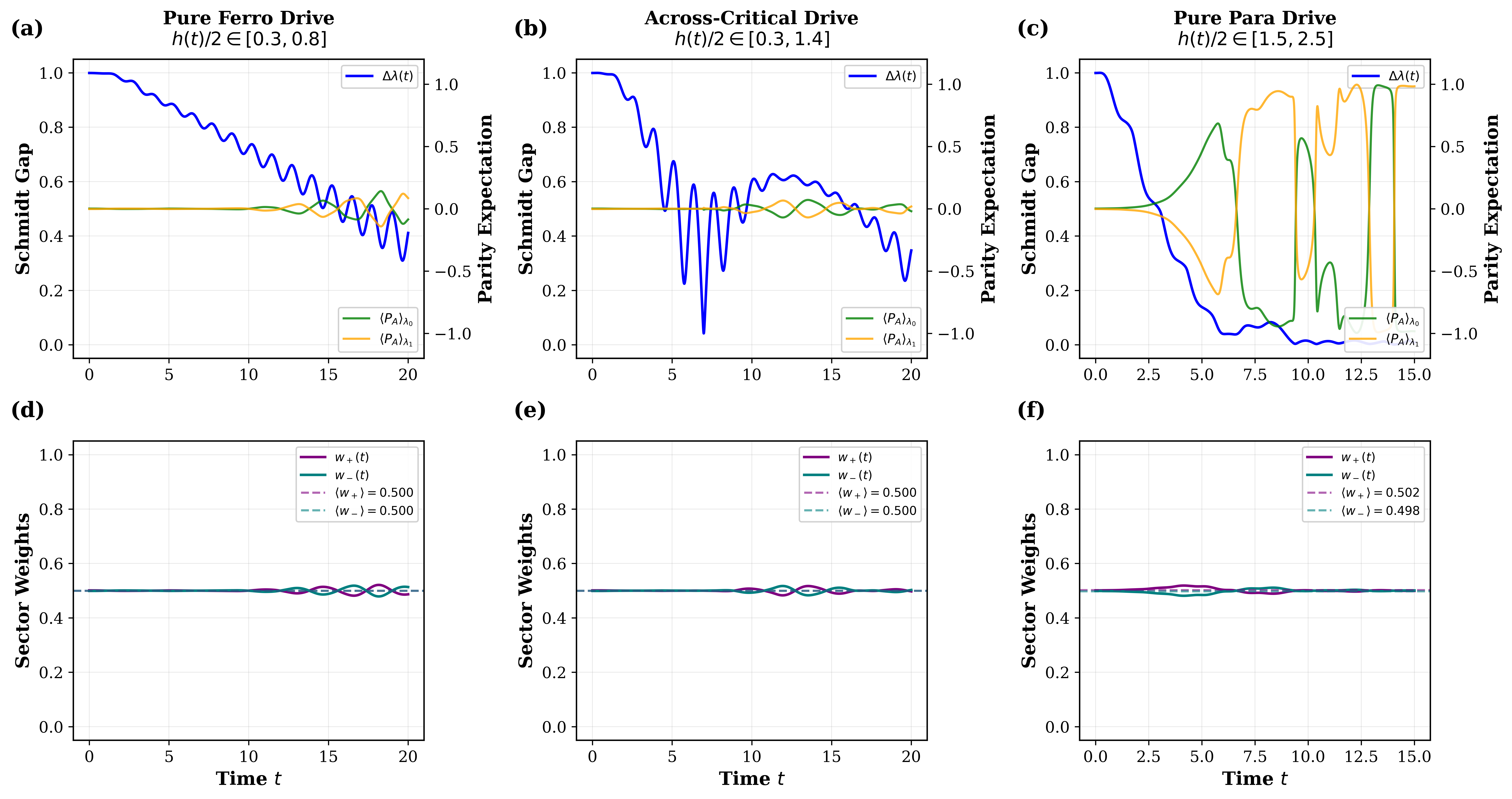}
		\caption{
			\textbf{No temporal entanglement transitions (TET) for a symmetry-broken ferromagnetic initial state at $\omega=5.0$.}
			Three driving protocols tested with initial state from ferromagnetic phase ($h_{\mathrm{init}}/2 = 0.5 < J$): 
			(a--c) Schmidt gap $\Delta\lambda(t)$ (blue, left axis) and parity expectations $\langle P_A\rangle_{\lambda_{0,1}}$ (green/orange, right axis) showing absence of sharp parity exchanges characteristic of TET.
			(d--f) sector weights $w_{\pm}(t)$ with time-averaged values $\overline{ w}_{\pm}$ (dashed lines), demonstrating that even when sector competition occurs and $w_{\pm}(t)$ crosses each other, transitions do not manifest. 
			The symmetry-check diagnostic $C(t) = \|[\rho_A(t), P_A]\|$ exhibits initial value $C(0) \sim 1.4$ and time-averaged $\overline{C} \sim 0.4 - 1.0$ (across three cases), indicating non-negligible symmetry violation due to the initial $\mathbb{Z}_2$-breaking. Note that in (c), although the Schmidt gap becomes parametrically small and the sector weights still cross, the parity evolution remains smooth (rounded) and does not exhibit the discontinuous parity exchange synchronized with a gap closure that characterizes a genuine TET (i.e., a non-analytic quantum phase transition of the entanglement Hamiltonian).
			The key message: symmetry-broken initial states fail condition (2) of the universal mechanism, precluding TET regardless of drive protocol or sector-weight dynamics.
			Parameters: $L=24$, $L_A=9$, $J=1.0$, $\omega=0.5$, $dt=0.01$.
		}
		\label{fig:evidence_ferroInit_omega5}
	\end{figure*}

Figure~\ref{fig:evidence_ferroInit_omega5} explicitly demonstrates the absence of TET for $\omega=5.0$.
In particular, we emphasize that an isolated suppression of the Schmidt gap or a smooth evolution of the Schmidt-vector parities (even when accompanied by a crossing of the sector weights) corresponds to a rounded crossover rather than a genuine entanglement transition. 
This provides direct evidence that condition (2) is a necessary ingredient for TET.
We have also verified the absence of transitions in both the adiabatic and high-frequency regimes; the corresponding data and plots are provided in the Zenodo dataset \cite{Gadge2025_zenodo}.

	\FloatBarrier
	
	\subsection{$\mathbb{Z}_2$-Symmetric, Paramagnetic Initial State}
	\label{app.:evidence para init}
	We initialize in the paramagnetic phase with the unique $\mathbb{Z}_2$-symmetric ground state, satisfying condition (2). Since the driven Hamiltonian preserves this symmetry by construction, satisfying condition (1), $\rho_A(t)$ must remain parity block-diagonal, confirmed by the symmetry-check diagnostic $C(t)\simeq 0$ throughout evolution. 
	
\begin{figure*}[t]
		\centering
		\includegraphics[width=\textwidth]{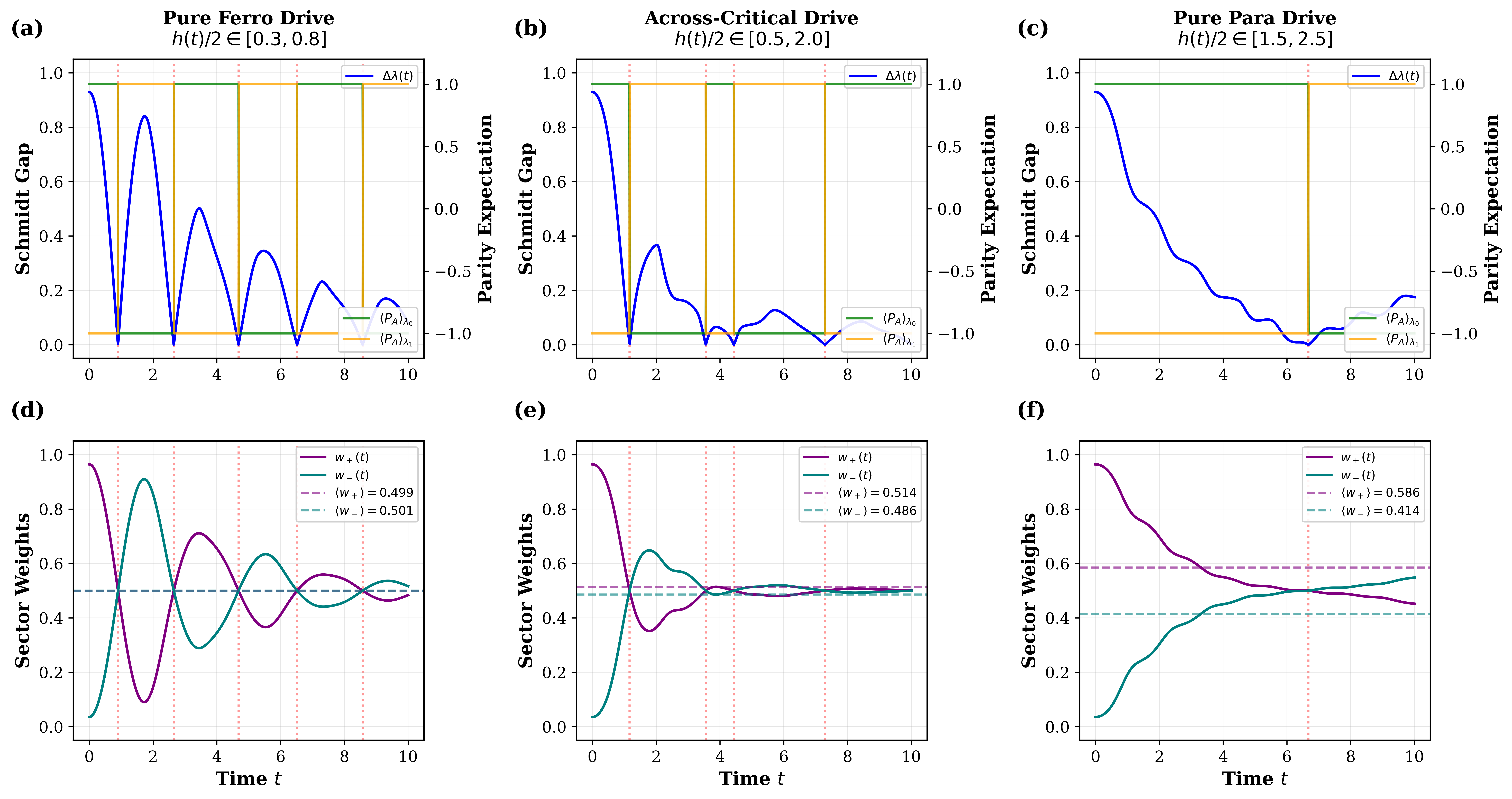}
		\caption{
			\textbf{Temporal entanglement transitions (TET) for a $\mathbb{Z}_2$-symmetric paramagnetic initial state at $\omega=5.0$.}
			Three driving protocols with $\mathbb{Z}_2$-symmetric initial state ($h_{\mathrm{init}}/2 = 1.5 > J$):
			(a--c) Schmidt gap $\Delta\lambda(t)$ (blue, left axis) exhibits sharp closures synchronized with parity exchanges in the leading Schmidt vectors $\langle P_A\rangle_{\lambda_{0,1}}$ (green/orange, right axis), which is the characteristic signatures of TET. Red dotted vertical lines denote parity-flip events, overlaid consistently across all panels to emphasize temporal synchronization;
			(d--f) subsystem-parity sector weights $w_{\pm}(t)$ with time-averaged values $\overline{ w}_{\pm}$ (dashed horizontal lines), confirming sustained balanced competition ($\overline{ w}_+ \approx 0.5$). Red vertical lines coincide with transition events in (a--c), demonstrating simultaneous entanglement reorganization at sector-weight crossings.
			The symmetry-check diagnostic $C(t)=\|[\rho_A(t),P_A]\|$ remains negligible throughout: $C(0) \sim 10^{-14}$ (implying initial inter-sector block $P_+\rho_A(0)P_-$ to remain negligibly small $P_+\rho_A(0)P_- \approx 10^{-14}$) and time-averaged $\overline{ C} \sim 10^{-8}$, verifying that $\rho_A(t)$ stays block-diagonal in parity sectors.
			These results establish that TET occur universally when three conditions are satisfied: (1) $\mathbb{Z}_2$-symmetric driving, (2) $\mathbb{Z}_2$-symmetric initial state, and (3) sustained sector competition. Transitions manifest robustly across all instantaneous-phase regimes (pure ferromagnetic, across-critical, and pure paramagnetic) with precise synchronization between Schmidt-gap closures, parity flips, and sector-weight crossings.
			Parameters: $L=24$, $L_A=9$, $J=1.0$, $\omega=5.0$, $dt=0.01$.
		}
		\label{fig:evidence_paraInit_omega5}
	\end{figure*}

	Figure~\ref{fig:evidence_paraInit_omega5} demonstrates that TET occur when both subsystem-parity sectors carry balanced weight ($w_{\pm} \approx 0.5$), with Schmidt-gap closures and parity exchanges synchronized precisely at sector-weight crossings $w_+(t) = w_-(t)$. This establishes the universal mechanism across all three driving protocols. We show for $\omega=5.0$ but we have also verified the presence of transitions in the adiabatic regime whose data and plots are provided in the dataset \cite{Gadge2025_zenodo}. We discuss the case of high-frequency limit next. Note that the parameters are tuned to ensure \(w_+(t^*)=w_-(t^*)\) admits solutions within short-time evolution; not all parameter choices yield crossings in short-time dynamics.

	\subsection{Marginal Critical Initial State}
	\label{app.:evidence critical init}
	
	The marginal case of initializing exactly at the equilibrium critical point (for our convention, $h_0=2.0$ so that $h_0/2=1.0$) is $\mathbb{Z}_2$-symmetric and is already discussed in the main text; we do not repeat it here.

    We have summarized a comprehensive table capturing the physics and the universal three requirements in Table~\ref{tab:scenarios_six}.

\subsection{Necessity of Condition 3: Paramagnetic Initial State with Pure Paramagnetic Driving}
\label{sec:necessity_cond3_paramag_pureparamag}

In this subsection we emphasize that, even when the driven dynamics preserve global $Z_2$ symmetry (Condition 1) and the initial state is a global $Z_2$ eigenstate (Condition 2), temporal entanglement transitions (TET) additionally require \emph{sector competition} in the reduced density matrix $\rho_A(t)$ (Condition 3). Concretely, if the subsystem-parity sector weights never become equal within the simulation window, then the sector-equality equation has no solution and no TET can occur in that window.

\paragraph{Subsystem-parity sectors and sector weights.}
Let $A$ be a contiguous subsystem of length $L_A$ (e.g.\ the first $L_A$ sites), and define the subsystem parity operator and its associated projectors
\begin{equation}
	P_A := \prod_{i\in A}\sigma_i^x,
	\qquad
	P_\pm := \frac{\mathbb{1}_A \pm P_A}{2}.
\end{equation}
For the full density matrix $\rho(t)$ evolving unitarily under $H(t)$, the reduced density matrix is $\rho_A(t):=\mathrm{Tr}_{\bar A}\rho(t)$, and the instantaneous sector weights are
\begin{equation}
	w_\pm(t) := \mathrm{Tr}\bigl(P_\pm\,\rho_A(t)\bigr),
	\qquad
	w_+(t)+w_-(t)=1.
\end{equation}
Their imbalance is
\begin{equation}
	w_+(t)-w_-(t) = \mathrm{Tr}\bigl(P_A\,\rho_A(t)\bigr) =: \langle P_A\rangle_t.
	\label{eq:imbalance_PA}
\end{equation}
When Conditions 1 and 2 are satisfied, $\rho_A(t)$ remains block-diagonal in the parity basis, $[\rho_A(t),P_A]=0$, ensuring that the two sectors evolve independently and the sector-equality condition
\begin{equation}
	w_+(t)=w_-(t)
	\quad\Longleftrightarrow\quad
	\langle P_A\rangle_t = 0
	\label{eq:sector_equality}
\end{equation}
selects the candidate critical times for TET.

\paragraph{Logical necessity of Condition 3.}
Within the symmetry-resolved mechanism studied in this work, the candidate critical times are selected by the sector-equality condition \eqref{eq:sector_equality}. Hence, if the trajectories $w_\pm(t)$ do not admit a crossing within the observation window $t\in[0,t_{\max}]$, i.e.\ $w_+(t)\neq w_-(t)$ for all $t\in[0,t_{\max}]$, then there exists no candidate time at which the entanglement reorganization can occur, and correspondingly no TET is expected in that window.

\paragraph{High-frequency suppression via the $\omega\to\infty$ effective Hamiltonian.}
Consider the driven TFIM protocol
\begin{equation}
	H(t) = -J\sum_{i=1}^{L-1}\sigma_i^z\sigma_{i+1}^z
	-\frac{h(t)}{2}\sum_{i=1}^{L}\sigma_i^x,
	\qquad
	h(t)=h_{\rm offset}+h_{\rm ac}\cos(\omega t).
\end{equation}
Let $P:=\prod_{i=1}^{L}\sigma_i^x$ be the global $Z_2$ generator. Since $P\sigma_i^xP=\sigma_i^x$ and $P\sigma_i^zP=-\sigma_i^z$, one has $[P,H(t)]=0$ for all $t$ (Condition 1).

In the strict infinite-frequency limit $\omega\to\infty$, the leading Floquet-Magnus effective Hamiltonian is the time average
\begin{equation}
	H_\infty \equiv \frac{1}{T}\int_0^T H(t)\,dt
	= -J\sum_{i=1}^{L-1}\sigma_i^z\sigma_{i+1}^z
	-\frac{h_{\rm offset}}{2}\sum_{i=1}^{L}\sigma_i^x,
	\label{eq:H_infty}
\end{equation}
which is manifestly global-$Z_2$ symmetric, $[P,H_\infty]=0$.

\paragraph{Boundary localization of the subsystem-parity dynamics.}
Under unitary evolution generated by $H_\infty$, the sector imbalance $\langle P_A\rangle_t := \mathrm{Tr}(P_A\rho_A(t))$ satisfies
\begin{equation}
	\frac{d}{dt}\langle P_A\rangle_t
	= i\,\mathrm{Tr}\Bigl([H_\infty,P_A]\rho(t)\Bigr).
	\label{eq:PA_eom}
\end{equation}
The transverse-field term in \eqref{eq:H_infty} commutes with $P_A$ exactly, $[\sum_i\sigma_i^x, P_A]=0$, so only the Ising interaction contributes. Moreover, for a contiguous subsystem $A$, all bonds $(i,i{+}1)$ strictly inside or strictly outside $A$ satisfy $[\sigma_i^z\sigma_{i+1}^z, P_A]=0$; only bonds crossing the boundary $\partial A$ contribute:
\begin{equation}
	[H_\infty,P_A]
	= -J\sum_{(i,i+1)\in \partial A}[\sigma_i^z\sigma_{i+1}^z,P_A],
	\label{eq:boundary_only}
\end{equation}
where $|\partial A|$ counts the number of cut bonds (for a bulk interval in an open chain, $|\partial A|=2$; if $A$ touches a physical edge, $|\partial A|=1$). Crucially, $|\partial A|$ does \emph{not} scale with the system size $L$ or the subsystem size $L_A$—it depends only on the topology of the bipartition.

Using H\"older's inequality for Schatten norms, $|\Tr(XY)| \le \|X\|_\infty\,\|Y\|_1$, and the fact that $\|\rho(t)\|_1=1$ for any density matrix, we obtain
$\left|\frac{d}{dt}\langle P_A\rangle_t\right| = \bigl|\Tr(\rho(t)\, i[H_\infty,P_A])\bigr| \le \|[H_\infty,P_A]\|_\infty \equiv \|[H_\infty,P_A]\|$. So we have
\begin{equation}
	\biggl|\frac{d}{dt}\langle P_A\rangle_t\biggr|
	\le \|[H_\infty,P_A]\|
	\le c\,J\,|\partial A|,
	\label{eq:rate_bound_c}
\end{equation}
for some constant $c$ independent of $L$ and $L_A$. Integrating yields
\begin{equation}
	\bigl|\langle P_A\rangle_t-\langle P_A\rangle_0\bigr|
	\le c\,J\,|\partial A|\,t.
	\label{eq:int_bound_c}
\end{equation}

\paragraph{Paramagnetic initial condition and suppression of sector competition.}
For the paramagnetic initial state (the ground state of the TFIM at large transverse field $h_{\rm init}/2 > J$), the reduced density matrix $\rho_A(0)$ is fully polarized in the $+$ parity sector:
\begin{equation}
	w_+(0)=1,
	\qquad
	w_-(0)=0,
	\qquad
	\text{i.e.}\quad \langle P_A\rangle_0 = 1.
	\label{eq:paramag_init}
\end{equation}
Combining \eqref{eq:int_bound_c} with \eqref{eq:paramag_init}, reaching the sector-equality condition $\langle P_A\rangle_t=0$ requires a change of order unity and hence
\begin{equation}
	|\langle P_A\rangle_t - 1| \ge 1
	\quad\Rightarrow\quad
	t \ge \frac{1}{c\,J\,|\partial A|}.
	\label{eq:time_lower_bound}
\end{equation}

\paragraph{Numerical observation at high frequency.}
Consistent with the above reasoning, we numerically observe that when Conditions 1 and 2 are satisfied but the driving frequency is sufficiently high, the sector weights $w_\pm(t)$ fail to achieve competition in the short-time dynamics, and correspondingly no TET occur within the observation window. We emphasize that the critical time bound derived above applies to the effective infinite-frequency generator $H_\infty$ and therefore provides a lower bound on the timescale over which sector weights can change appreciably in that limiting description. In a periodically driven system at large but finite $\omega$, the actual onset time for sector competition (and thus for TET) can be parametrically larger than this lower bound, and may lie beyond the accessible observation window even when the bound itself is satisfied. This is what we observe in the high-frequency limit, for instance, for the parameter set $h_{\rm init}=3.0$, $h_{\rm offset}=4.0$, $h_{\rm ac}=1.0$, $L=24$, $L_A=9$, $\omega=10$, $t_{\max}=20.0$, where conditions 1 and 2 are satisfied and yet the absence of sector-weight crossings (and hence absence of TET) is explicitly verified. The full numerical data are available on Zenodo \cite{Gadge2025_zenodo}.

\paragraph{Relation to finite-size scaling protocol.}
Throughout this work, our finite-size scaling analyses (Fig.~2 of the main text, and Sec.~\ref{app. subsection Finite-Size Scaling for Full Data Set and Universality} below) are performed by initializing near the equilibrium critical point ($h_{\rm init}/2 = J$) and driving with parameters that ensure sustained sector competition in the short-time dynamics. This choice optimally satisfies all three conditions simultaneously: Condition 1 (global $Z_2$ symmetry of the drive) and Condition 2 (initial $Z_2$-eigenstate) are exact, while Condition 3 (sector-weight crossings $w_+(t^*)=w_-(t^*)$ within the observation window) is ensured by proximity to criticality where entanglement buildup is most efficient. The high-frequency paramagnetic-drive protocol discussed in this subsection deliberately violates Condition 3 in the short-time regime, demonstrating its necessity; correspondingly, such parameters are not used for finite-size scaling, as they would not yield multiple well-defined critical times within accessible evolution windows.

\noindent\rule{\linewidth}{0.5pt}

Now we move forward to further demonstrate the robustness and necessity of the three-condition criterion through two complementary controlled-violation experiments, each isolating the failure of a single condition while preserving the others. In Sec.~\ref{app.:evidence controlled Z2 violation}, we construct a one-parameter family of initial states via controlled admixture of opposite-parity GHZ components, systematically violating condition (2)—initial-state $\mathbb{Z}_2$ symmetry, while maintaining the driven Hamiltonian's symmetry (condition (1) satisfied) and sustained sector competition with crossings $w_+(t^*)=w_-(t^*)$ (condition (3) satisfied). Despite identical driving protocols and system parameters that produce robust TET in the fully symmetric case, the $\mathbb{Z}_2$-breaking admixture destroys the parity block-diagonal structure of $\rho_A(t)$ (quantified by $C(t)=\|[\rho_A(t),P_A]\|\neq 0$), precluding sharp entanglement transitions and establishing condition (2) as individually necessary. Conversely, in Sec.~\ref{app. subsection Mixed-Field Ising Model}, we violate condition (1) by introducing a longitudinal field term in the driven Hamiltonian that explicitly breaks $\mathbb{Z}_2$ symmetry, while initializing in the $\mathbb{Z}_2$-symmetric paramagnetic ground state of the TFIM (condition (2) satisfied) and verifying sustained sector-weight competition and crossings (condition (3) satisfied). Remarkably, despite using identical parameters and driving protocols that yield unambiguous TET in the $\mathbb{Z}_2$-symmetric TFIM, the longitudinal-field-induced symmetry violation completely suppresses temporal entanglement transitions across all three instantaneous-phase regimes (pure ferromagnetic, across-critical, and pure paramagnetic driving), thereby isolating condition (1) as fundamentally necessary. Together, these controlled experiments provide a definitive demonstration that all three conditions—Hamiltonian symmetry, initial-state symmetry, and sector competition—are individually necessary and jointly sufficient for the emergence of temporal entanglement transitions, establishing the universal mechanism underlying this non-equilibrium quantum critical phenomenon.

\subsection{Necessity of Condition 2: Controlled Violation of $\mathbb{Z}_2$ Symmetry of Initial GHZ State}
\label{app.:evidence controlled Z2 violation}

A central requirement for temporal entanglement transitions (TET) is condition (2): the initial state must be an eigenstate of the global $\mathbb{Z}_2$ symmetry generated by the spin-flip operator $P=\prod_{i=1}^L \sigma_i^x$, rather than an eigenstate of the static TFIM Hamiltonian itself.
Indeed, throughout the paper we observe TET for a broad class of TFIM \emph{eigenstate} initial conditions, provided they are (i) $\mathbb{Z}_2$-eigenstates (i.e. paramagnetic or critical) and (ii) generate sustained competition between the two subsystem-parity sectors of $\rho_A(t)$.
To isolate the role of condition (2) as cleanly as possible, we therefore employ a controlled initial-state family built around the maximally symmetry-sensitive reference state: the $\mathbb{Z}_2$-symmetric Greenberger–Horne–Zeilinger (GHZ) (cat) state, which is not an eigenstate of TFIM.

Specifically, we take the exact $\mathbb{Z}_2$-eigenstate
\begin{equation}
    \ket{\psi_{\mathbb{Z}_2}}
    \equiv \ket{\mathrm{GHZ}^+}
    = \frac{1}{\sqrt{2}}\Big(\ket{\uparrow\uparrow\cdots\uparrow}+\ket{\downarrow\downarrow\cdots\downarrow}\Big),
\end{equation}
and define $\ket{\phi_{\mathrm{broken}}}$ as an orthogonal component in the opposite $\mathbb{Z}_2$ sector,
\begin{equation}
    \ket{\phi_{\mathrm{broken}}}
    \equiv \ket{\mathrm{GHZ}^-}
    = \frac{1}{\sqrt{2}}\Big(\ket{\uparrow\uparrow\cdots\downarrow}-\ket{\downarrow\downarrow\cdots\downarrow}\Big),
\end{equation}
so that $\ket{\psi_{\mathbb{Z}_2}}$ and $\ket{\phi_{\mathrm{broken}}}$ carry opposite eigenvalues under the global spin-flip operator $P=\prod_{i=1}^L \sigma_i^x$, i.e.$P\ket{\mathrm{GHZ}^\pm}=\pm \ket{\mathrm{GHZ}^\pm}$.

We then construct the one-parameter family
\begin{equation}
	\ket{\psi_\epsilon}
	= \sqrt{1-\epsilon^2}\,\ket{\psi_{\mathbb{Z}_2}} + \epsilon\,\ket{\phi_{\mathrm{broken}}},
\end{equation}
where $\epsilon\in[0,1]$ controls the degree of initial symmetry violation.  Equivalently, one may write $\epsilon=\sin\theta$, where $\theta=0$ corresponds to an exactly $\mathbb{Z}_2$-symmetric GHZ initial state.
Note that the \emph{full} state $\ket{\psi_\epsilon}$ is an eigenstate of $P$ (and hence exactly $\mathbb{Z}_2$-symmetric) \textit{only} at the selected points $\epsilon\in\{0,1\}$ (equivalently, $\theta\in\{0,\pi/2\}$); for any $\epsilon\in(0,1)$ it is a coherent superposition of the $P=\pm 1$ sectors and therefore not an eigenstate of $P$, i.e.\ it explicitly violates $\mathbb{Z}_2$ symmetry.  In this work we focus on the weak-violation regime $\epsilon\ll 1$ (equivalently, $\theta\approx 0$), and we therefore refer to $\ket{\phi_{\mathrm{broken}}}\equiv\ket{\mathrm{GHZ}^-}$ as the symmetry-violating admixture relative to the $\mathbb{Z}_2$-even reference state $\ket{\psi_{\mathbb{Z}_2}}\equiv\ket{\mathrm{GHZ}^+}$.

We choose three representative cases: (i)~strong violation with $\theta=0.1$ yielding $C(0)\sim 2.8\times 10^{-1}$, (ii)~weak violation with $\theta=0.01$ yielding $C(0)\sim 2.8\times 10^{-2}$, and (iii)~exact $\mathbb{Z}_2$ symmetry with $\theta=0$ giving $C(0)=0$.  All cases use the across-critical driving protocol ($h_{\mathrm{offset}}=2.5$, $h_{\mathrm{ac}}=1.5$, $\omega=5.0$).

\begin{figure*}[htbp]
	\centering
	\includegraphics[width=\textwidth]{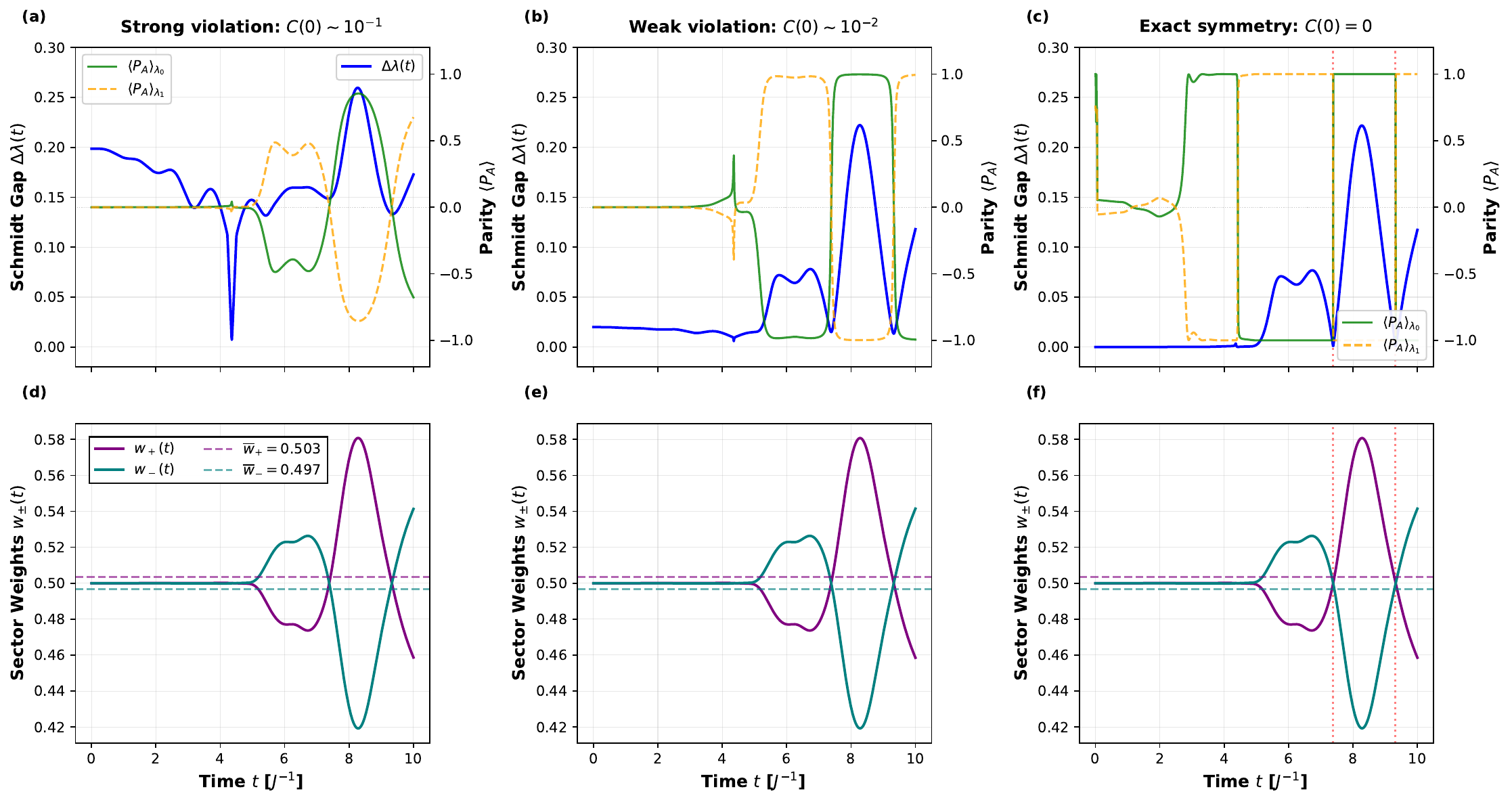}
	\caption{
		\textbf{Controlled violation of initial-state $\mathbb{Z}_2$ symmetry progressively destroys temporal entanglement transition criticality.}
		Systematic tuning of initial-state symmetry via $\ket{\psi_\epsilon}=\sqrt{1-\epsilon^2}\ket{\psi_{\mathbb{Z}_2}}+\epsilon\ket{\phi_{\mathrm{broken}}}$ (where $\epsilon = \sin \theta$ is a useful parametrization) demonstrates that condition (2), namely the initial state is a global $\mathbb{Z}_2$ eigenstate, governs both the existence and sharpness of TET.
		Here the $\mathbb{Z}_2$-symmetric reference state $\ket{\psi_{\mathbb{Z}_2}}$ is chosen to be the GHZ (cat) state, which is not an eigenstate of the static TFIM Hamiltonian but is an \emph{exact} eigenstate of the global parity symmetry.
		All panels use across-critical driving ($h_{\mathrm{offset}}=2.5$, $h_{\mathrm{ac}}=1.5$, $\omega=5.0$).
		(a,d)~\textbf{Strong violation} ($\theta=0.1$, $C(0)\sim 2.8\times 10^{-1}$): Schmidt gap $\Delta\lambda(t)$ narrows smoothly without sharp closures; parity expectations $\langle P_A\rangle_{\lambda_{0,1}}$ evolve continuously; sector weights $w_{\pm}(t)$ cross gradually. Time-averaged $\overline{C}\sim 0.22$ confirms sustained symmetry violation and only crossover behavior.
		(b,e)~\textbf{Weak violation} ($\theta=0.01$, $C(0)\sim 2.8\times 10^{-2}$): Transitions sharpen but remain imperfect. Gap closures become more abrupt, parity flips steepen, and sector-weight crossings grow more pronounced. Residual symmetry breaking ($\overline{C}\sim 0.022$) softens the criticality.
		(c,f)~\textbf{Exact $\mathbb{Z}_2$ symmetry} ($\theta=0$, $C(0)=0$): TET manifest as sharp non-analytic reorganizations in the entanglement spectrum. The GHZ initial state yields a perfectly degenerate leading entanglement spectrum at $t=0$ ($\lambda_0(0)=\lambda_1(0)=0.5$), reflecting equal subsystem-parity sector weights; under driving the degeneracy is lifted while maintaining $[\rho_A(t),P_A]=0$. At critical times synchronized exactly with sector-weight crossings, Schmidt gaps vanish to numerical precision ($\sim 10^{-14}$) and parity expectations flip discontinuously (red dotted vertical lines). $C(t)\simeq 0$ throughout ($\overline{C}\sim 6\times 10^{-4}$) confirms block-diagonal structure.
		Dashed horizontal lines in (d--f) indicate time-averaged sector weights $\overline{w}_{\pm}\approx 0.50$ (the numerical values are the same across all three subplots in row 2), confirming sustained competition.
		Parameters: $L=24$, $L_A=9$, $J=1.0$, $h_{\mathrm{offset}}=2.5$, $h_{\mathrm{ac}}=1.5$, $\omega=5.0$, $dt=0.01$, $t_{\max}=10.0$.
	}
	\label{fig:controlled_Z2_violation}
\end{figure*}

Figure~\ref{fig:controlled_Z2_violation} demonstrates that as the initial $\mathbb{Z}_2$ symmetry is progressively restored ($\epsilon\to 0$), the entanglement dynamics sharpen from smooth crossovers to increasingly abrupt temporal criticality.
For strong violation [Fig.~\ref{fig:controlled_Z2_violation}(a,d)], the Schmidt gap exhibits gradual narrowing rather than sharp closures, parity expectations evolve continuously without discontinuous flips, and sector weights cross smoothly.
The time-averaged symmetry-check diagnostic $\overline{C}\sim 0.22$ confirms appreciable violation of the parity block structure of $\rho_A(t)$.
At weak violation [Fig.~\ref{fig:controlled_Z2_violation}(b,e)], transitions begin to sharpen: gap closures become more abrupt, parity flips steepen, and sector-weight crossings grow more pronounced, though residual symmetry breaking ($\overline{C}\sim 0.022$) softens the criticality.

Most strikingly, for the exactly symmetric GHZ initial state (satisfying $C(0)=0$ exactly) [Fig.~\ref{fig:controlled_Z2_violation}(c,f)], the entanglement Hamiltonian exhibits the cleanest and sharply resolved TET.
At $t=0$, the reduced density matrix $\rho_A(0)$ inherits the cat-state structure and yields a perfectly degenerate leading entanglement spectrum with $\lambda_0(0)=\lambda_1(0)=0.5$, reflecting equal subsystem-parity sector weights.
Under driving, this degeneracy is dynamically lifted as the entanglement Hamiltonian evolves, while $\rho_A(t)$ remains rigorously block-diagonal in the $P_A$ basis due to exact symmetry preservation.
At critical times synchronized precisely with sector-weight crossings $w_+(t)=w_-(t)$, the entanglement reorganization triggers sharp TET: Schmidt gaps vanish to numerical precision ($\Delta\lambda\sim 10^{-14}$), and parity expectations flip discontinuously.

The synchronization between parity flips (red dotted vertical lines) and sector-weight crossings is exact only when $C(0)=0$, establishing that condition (2) governs the \emph{sharpness} of TET and that it's not a crossover.
Overall, this controlled symmetry-violation protocol using GHZ as a symmetry-pure reference state shows that temporal entanglement transitions acquire true non-analytic critical character only when the initial state exactly respects the global $\mathbb{Z}_2$ symmetry preserved by the driven Hamiltonian, while approximate symmetry yields crossover behavior.

\subsection{Necessity of Condition 1: Mixed-Field Ising Model: Driven Hamiltonian Violates $\mathbb{Z}_2$ Symmetry}
\label{app. subsection Mixed-Field Ising Model}

To rigorously demonstrate the necessity of condition (1)—namely, that the driven Hamiltonian must preserve the global $\mathbb{Z}_2$ symmetry—we examine the mixed-field Ising model where this requirement is explicitly violated. The driven Hamiltonian is given by
\begin{equation}
	\mathcal{H}(t) = -J\sum_{i=1}^{L-1} \sigma_i^z\sigma_{i+1}^z - \frac{g}{2}\sum_{i=1}^{L}\sigma_i^z - \frac{h(t)}{2}\sum_{i=1}^{L}\sigma_i^x,
	\label{eq:mixed_field_hamiltonian}
\end{equation}
where $h(t) = h_{\mathrm{offset}} + h_{\mathrm{amp}}\cos(\omega t)$ is the time-dependent transverse field and $g$ denotes the longitudinal field coupling. The presence of the longitudinal field term $-\frac{g}{2}\sum_i\sigma_i^z$ explicitly breaks the $\mathbb{Z}_2$ symmetry generated by the global spin-flip operator $\prod_{i=1}^L\sigma_i^x$, as $[\mathcal{H}(t), \prod_i\sigma_i^x] \neq 0$ for any $g \neq 0$. This stands in direct contrast to the transverse-field Ising model (TFIM) studied throughout this work, where $[\mathcal{H}_{\mathrm{TFIM}}(t), \prod_i\sigma_i^x] = 0$ is exact for all $t$.

We initialize the system in the ground state of the TFIM with transverse field $h_{\mathrm{init}}/2 = 1.5 > J$ (where $J=1$ throughout), placing the initial state in the paramagnetic phase where it is manifestly a $\mathbb{Z}_2$ eigenstate (condition (2) satisfied). The system is then evolved under the mixed-field Hamiltonian \eqref{eq:mixed_field_hamiltonian} with longitudinal coupling $g=1.0$ at driving frequency $\omega=5.0$—the same frequency that produces robust temporal entanglement transitions in the $\mathbb{Z}_2$-symmetric TFIM. We examine three distinct driving protocols that comprehensively span the phase diagram: (i) pure ferromagnetic driving with $h(t) \in [0.3, 0.8]$ (both field values below the TFIM critical point $h_c=1.0$), (ii) across-critical-point driving with $h(t) \in [0.5, 2.0]$ (field crosses $h_c$ during evolution), and (iii) pure paramagnetic driving with $h(t) \in [1.5, 2.5]$ (both values above $h_c$). These protocols are deliberately chosen to match the parameter regimes where the $\mathbb{Z}_2$-symmetric TFIM exhibits clear TET, enabling direct comparison. Note that we also have sector competition where $w_+$ and $w_-$ crosses each other, therefore condition (3) is satisfied as well.

\begin{figure}[htbp]
	\centering
	\includegraphics[width=\textwidth]{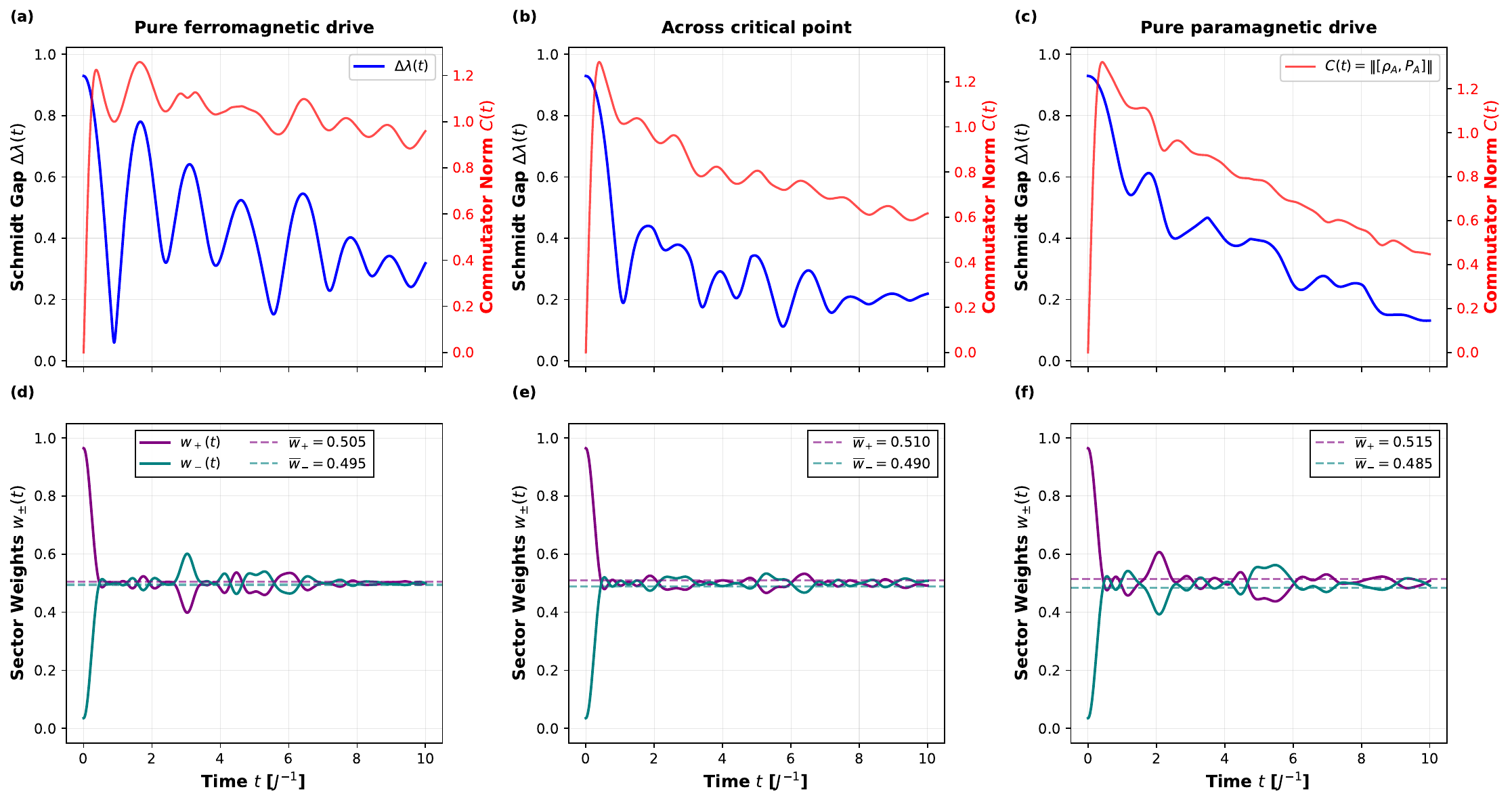}
	\caption{Absence of temporal entanglement transitions in the mixed-field Ising model where condition (1)—$\mathbb{Z}_2$ symmetry of the driven Hamiltonian—is violated. The system evolves under $\mathcal{H}(t) = -J\sum_i \sigma_i^z\sigma_{i+1}^z - \frac{g}{2}\sum_i\sigma_i^z - \frac{h(t)}{2}\sum_i\sigma_i^x$ with longitudinal field coupling $g=1.0$ breaking the global $\mathbb{Z}_2$ symmetry generated by $\prod_i\sigma_i^x$, and time-dependent transverse field $h(t) = h_{\mathrm{offset}} + h_{\mathrm{amp}}\cos(\omega t)$. All three cases initialize from the paramagnetic ground state of the transverse-field Ising model (TFIM) at $h_{\mathrm{init}}/2 = 1.5 > J=1.0$ (satisfying $\mathbb{Z}_2$ symmetry for the initial state) and drive at frequency $\omega=5.0$. \textbf{(a,d)} Pure ferromagnetic driving with $h(t) \in [0.3, 0.8]$ (both below the TFIM critical point $h_c=1.0$). \textbf{(b,e)} Driving across the critical point with $h(t) \in [0.5, 2.0]$. \textbf{(c,f)} Pure paramagnetic driving with $h(t) \in [1.5, 2.5]$ (both above $h_c$). \textbf{Top row:} Schmidt gap $\Delta\lambda(t)$ (blue, left axis) remains smooth throughout evolution without sharp closures, while the symmetry-breaking diagnostic $C(t) = \|[\rho_A(t), P_A]\|$ (red, right axis) exhibits large values $C(t) \sim \mathcal{O}(1)$, confirming that the longitudinal field destroys the parity block-diagonal structure of $\rho_A(t)$. \textbf{Bottom row:} Parity sector weights $w_\pm(t)$ still competes and crosses, satisfying condition (3). Yet, there is an absence of sharp TET across all three driving protocols, despite using identical parameters ($L=24$, $L_A=9$, $\omega=5.0$) that produce robust TET in the $\mathbb{Z}_2$-symmetric TFIM. This demonstrates that condition (1) is a fundamental requirement: TET require the driven Hamiltonian to preserve the global $\mathbb{Z}_2$ symmetry. Time-averaged commutator norms $\overline{C} \sim \mathcal{O}(1)$ confirm persistent symmetry violation throughout evolution, precluding the emergence of entanglement criticality.}
	\label{fig:mixed_field_no_TET}
\end{figure}

Figure~\ref{fig:mixed_field_no_TET} demonstrates the complete absence of temporal entanglement transitions across all three driving scenarios. The symmetry-breaking diagnostic $C(t) = \|[\rho_A(t), P_A]\|$ exhibits large values $C(t) \sim \mathcal{O}(1)$ throughout the evolution, with time-averaged norms $\overline{C} \sim \mathcal{O}(1)$, confirming that the longitudinal field destroys the parity block-diagonal structure of the subsystem reduced density matrix: $\rho_A(t) \neq \frac{1}{2}(P_+ \rho_A(t) P_+ + P_- \rho_A(t) P_-)$. The cross-terms are dominating significantly (see below Eq.~\eqref{eq:w+ w- defined}) where $w_+ + w_- \neq 1$. 
Correspondingly, the Schmidt gap $\Delta\lambda(t) = \lambda_0(t) - \lambda_1(t)$ remains smooth without the sharp closures characteristic of TET, and the parity sector weights $w_\pm(t)$ exhibit no synchronized crossings or temporal criticality. This stark contrast with the driven TFIM $\mathbb{Z}_2$-symmetric case, where identical parameters produce unambiguous TET, establishes that condition (1) as a fundamental requirement: the driven Hamiltonian must preserve the global $\mathbb{Z}_2$ symmetry for temporal entanglement transitions to emerge. The persistent symmetry violation quantified by $\overline{C} \sim \mathcal{O}(1)$ (accordingly, cross-terms across sectors in $\rho_A$ being significant) precludes the formation of well-defined parity sectors in $\rho_A(t)$, despite the sectors $w_+$ and $w_-$ competing and crossing (as can be seen in Fig.~\ref{fig:mixed_field_no_TET}).

	\begin{table}[htbp]
		\centering
		\caption{Seven driving/initial-state scenarios tested in Sec.~\ref{app.:evidence for different initial states and types of driving}. The initial state is the ground state (GS) of the system with initial field $h_{\rm init}/2$, which is either paramagnetic ($Z_2$-symmetric) or ferromagnetic (symmetry-broken as considered here; the critical case $h_{\rm init}/2=1.0$ is considered throughout the main text). The transverse field entering the Hamiltonian is $h(t)/2$ [Eq.~\eqref{eq:Ht_general}], so the instantaneous equilibrium phase classification is determined by whether $h(t)/2$ is below/above $J$ (with $J=1$ throughout). For all parameter sets, the Hamiltonian remains $Z_2$-symmetric [generated by $\prod_{i=1}^{L}\sigma_i^x$]; the distinction is whether the initial state is a $Z_2$-eigenstate. The final column reports whether TET (TET) are observed according to the synchronized entanglement diagnostics used in this work (Schmidt-gap closure and parity exchange in the leading Schmidt vectors), in the regime where the symmetry-check diagnostic remains negligible, $\mathcal{C}(t)\equiv\|[\rho_A(t),P_A]\|\approx 0$, and both subsystem-parity sectors carry appreciable weight. \emph{Note:} The finite-size scaling analyses performed in the main text (Fig.~2 in the main text) as well as below in Sec.~\ref{app. subsection Finite-Size Scaling for Full Data Set and Universality} are performed using critical initial states ($h_{\rm init}/2 = J$; Scenario 7 or equivalently the main-text setup) to ensure all three conditions are robustly satisfied in short-time dynamics, optimizing the resolution of universal critical behavior. Scenarios 1--3 (paramagnetic initial states with various drives) also exhibit TET but require parameter tuning to achieve sector-weight crossings in accessible short-time windows; Scenario 1 at high frequency (Sec.~\ref{sec:necessity_cond3_paramag_pureparamag}) demonstrates the necessity of Condition 3 by showing its violation whose data is available on Zenodo \cite{Gadge2025_zenodo}.}
		\label{tab:scenarios_six}
		
		\setlength{\tabcolsep}{5pt}
		\begin{ruledtabular}
			\begin{tabular}{lccccc}
				Scenario & $h_{\rm init}/2$ for GS & $(h_{\rm offset},h_{\rm amp})$ & $h(t)/2$ range & Class & TET observed \\ \hline
				1 & 1.5 & (4.0,1.0) & [1.5,2.5] & para init, para drive & Yes \\
				2 & 1.5 & (2.5,1.5) & [0.5,2.0] & para init, cross-critical drive & Yes \\
				3 & 1.5 & (1.1,0.5) & [0.3,0.8] & para init, ferro drive & Yes \\
				4 & 0.5 & (1.1,0.5) & [0.3,0.8] & ferro init, ferro drive & No \\
				5 & 0.5 & (1.7,1.1) & [0.3,1.4] & ferro init, cross-critical drive & No \\
				6 & 0.5 & (4.0,1.0) & [1.5,2.5] & ferro init, para drive & No \\
				7 & 1.0 & (4.0,1.0) & [1.5,2.5] & critical init, any drive & Yes \\
			\end{tabular}
		\end{ruledtabular}
	\end{table}

	%%%%%%%%%%%%%%%%%%%%%%%%%%%%%%%%%
	%%%%%%%%%%%%%%%%%%%%%%%%%%%%%%%%%
	%%%%%%%%%%%%%%%%%%%%%%%%%%%%%%%%%
	%%%%%%%%%%%%%%%%%%%%%%%%%%%%%%%%%
	
	\begin{figure}[htbp]
		\centering
		
		% First row: Low frequency ω = 1.0 (panels a,b,c)
		\begin{subfigure}[b]{0.9\textwidth}
			\centering
			\includegraphics[width=\textwidth]{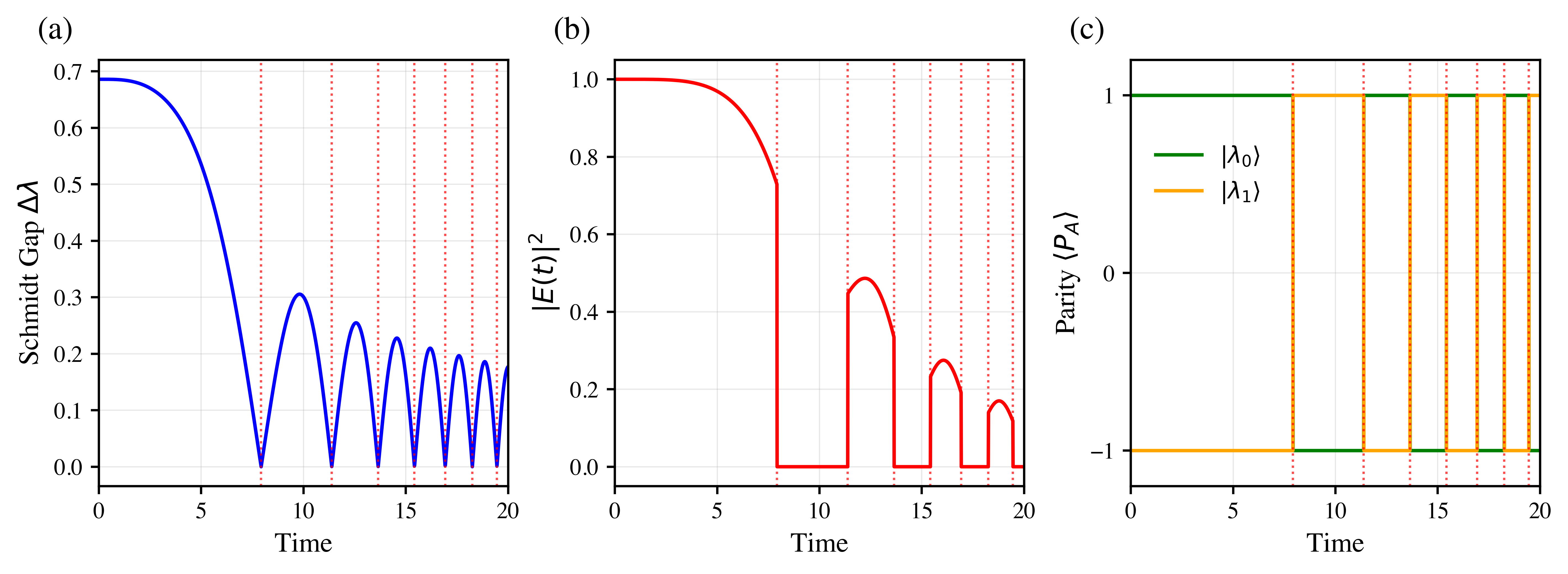}
			\caption*{(A) $\quad \omega = 0.1$}
			\label{fig:freq_low}
		\end{subfigure}
		\hfill
		% Second row: Intermediate frequency ω = 5.0 (panels d,e,f)  
		\begin{subfigure}[b]{0.9\textwidth}
			\centering
			\includegraphics[width=\textwidth]{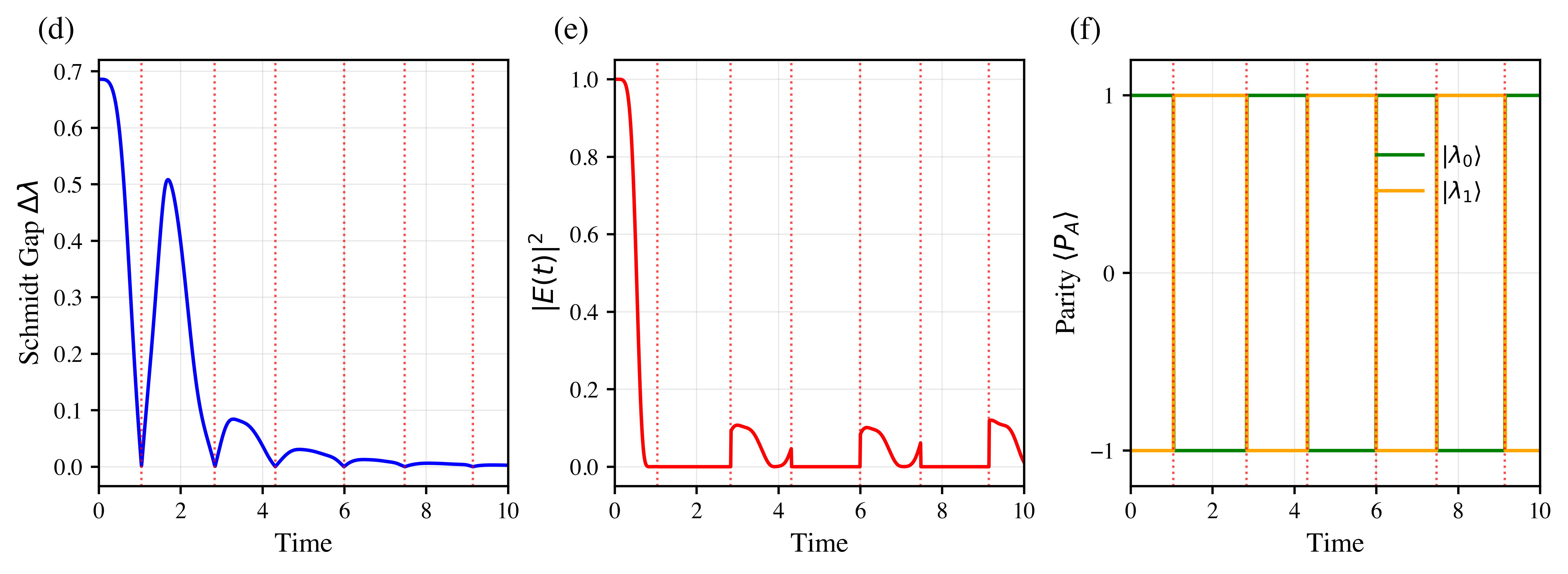}
			\caption*{(B) $\quad \omega = 4.0$}
			\label{fig:freq_mid}
		\end{subfigure}
		\hfill
		% Third row: High frequency ω = 50.0 (panels g,h,i)
		\begin{subfigure}[b]{0.9\textwidth}
			\centering
			\includegraphics[width=\textwidth]{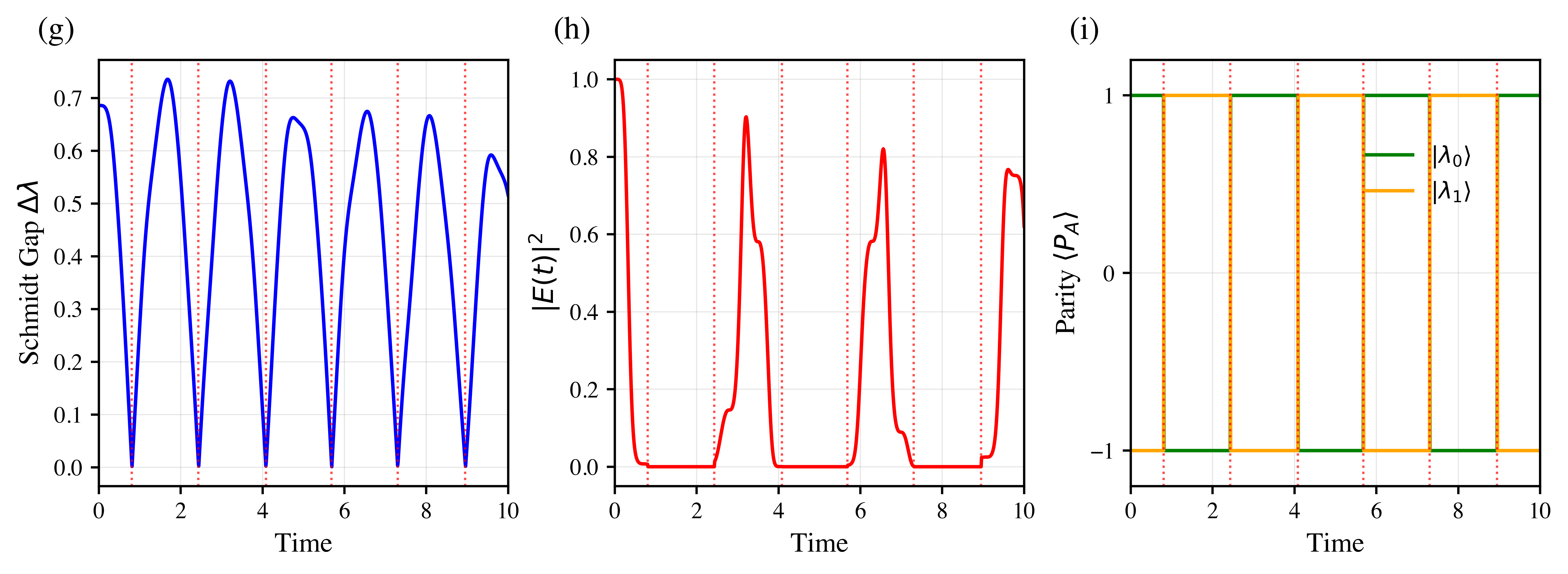}
			\caption*{(C) $\quad \omega = 9.0$}
			\label{fig:freq_high}
		\end{subfigure}
		
		\caption{Frequency dependence of temporal entanglement transitions (TET). \textbf{(a-c)} Low frequency regime ($\omega = 0.1$): Left panel shows Schmidt gap $\Delta\lambda$, middle shows entanglement echo $|E(t)|^2$, right shows subsystem parity expectations. Transitions occur irregularly with varying critical times, reflecting poor synchronization with slow driving. \textbf{(d-f)} Intermediate frequency ($\omega = 4.0$): Transitions begin to regularize as the system starts to lock into the driving frequency. \textbf{(g-i)} High frequency regime ($\omega = 9.0$): Perfectly periodic transitions emerge, demonstrating Floquet inheritance where the entanglement Hamiltonian synchronizes with the driving period. Red vertical lines mark critical times determined from parity discontinuities. Parameters: $L=24$, $L_A=9$, $J=1.0$, $h_0=2.0$.}
		\label{fig:frequency_dependence}
	\end{figure}
	
	\begin{figure}[htbp]
		\centering
		\includegraphics[width=0.9\textwidth]{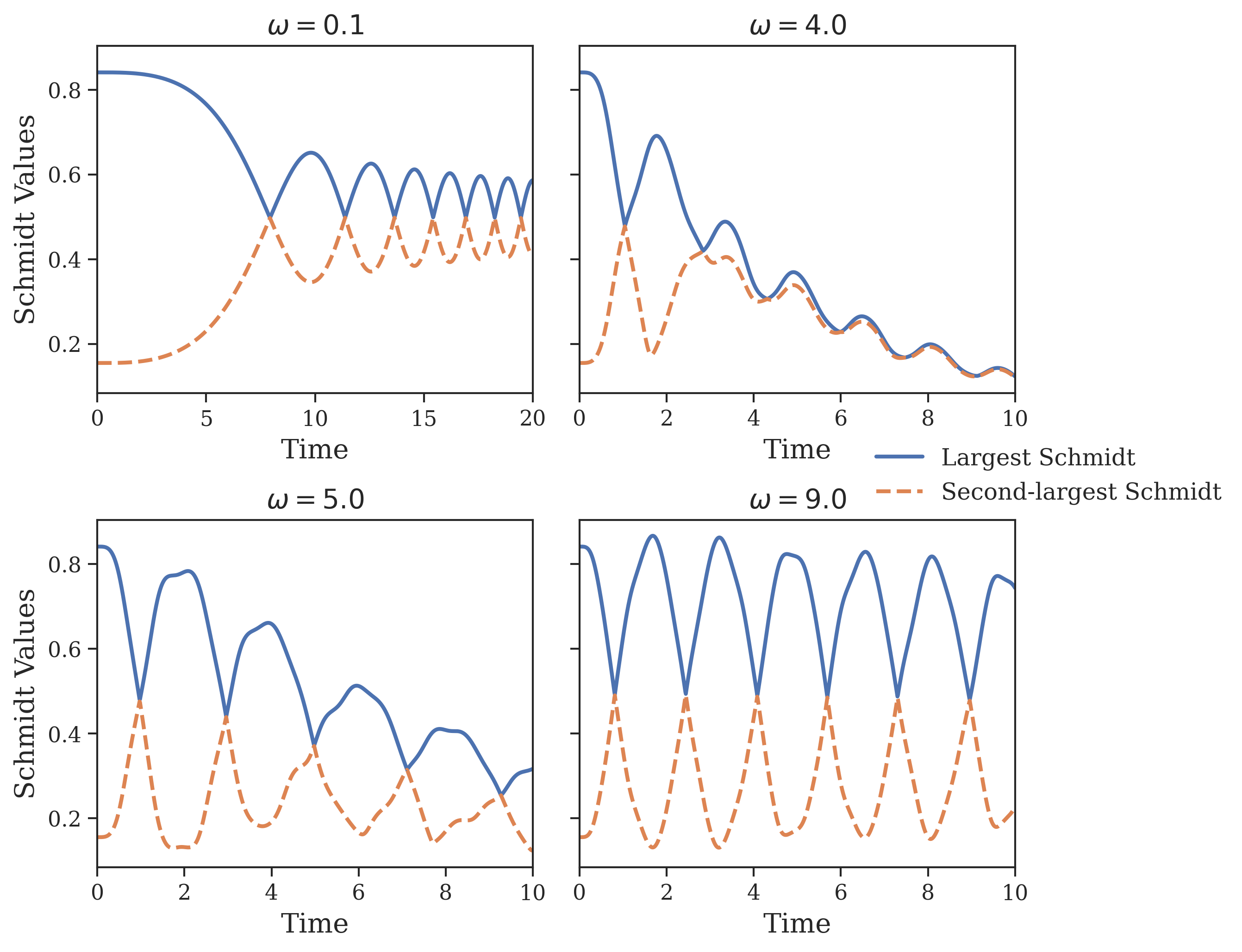}
		\caption{Time evolution of the two largest Schmidt values $\lambda_1(t)$ (solid) and $\lambda_2(t)$ (dashed) for different driving frequencies whose Schmidt gap closing is already plotted in Fig.~1 of the main text ($\omega=5.0$) and in Fig.~\ref{fig:frequency_dependence} ($\omega=0.1, 4.0, 9.0$) here. (a) $\omega = 0.1$: Irregular gap closing events reflect poor synchronization with slow driving. (b) $\omega = 4.0$ and (c) $\omega = 5.0$: Intermediate frequencies show increasingly regular gap closing. (d) $\omega = 9.0$: Perfectly periodic gap closing emerges, demonstrating Floquet inheritance by the entanglement Hamiltonian. The near-degeneracy of Schmidt values at critical times $t_c^{(k)}$ corresponds to entanglement transitions in Fig.~1 of the main text and Fig.~\ref{fig:frequency_dependence} here. Parameters: $L=24$, $L_A=9$, $J=1.0$, $h_0=2.0$, $dt=0.01$.}
		\label{fig:schmidt_values}
	\end{figure}

	\begin{figure}[htbp]
		\centering
		\includegraphics[width=0.9\textwidth]{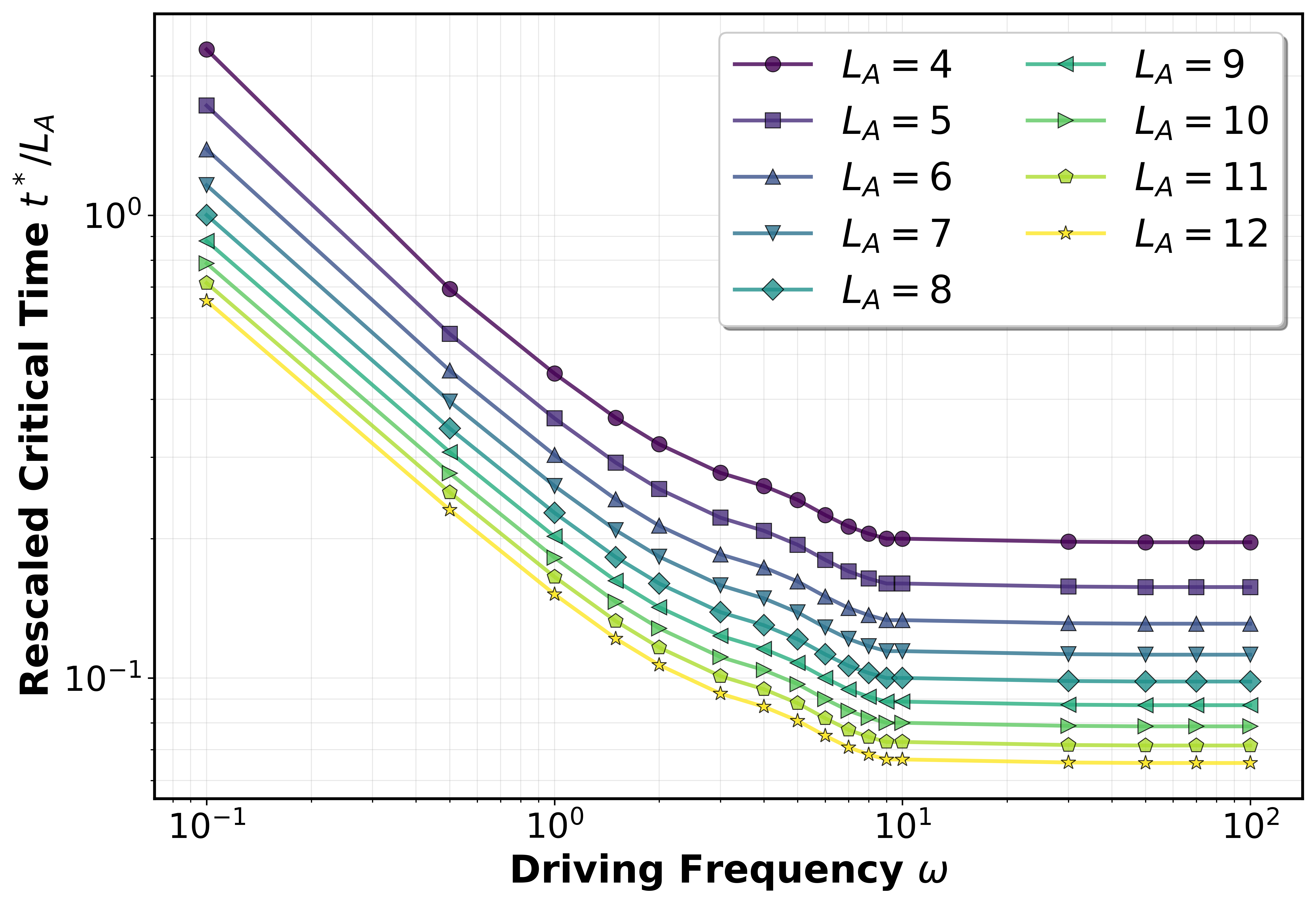}
		\caption{Rescaled first critical time $t^*/L_A$ is plotted against the driving frequency. Parameters: $L=24$, $L_A = 4-12$, $J=1.0$, $h_0=2.0$. Time steps are chosen as $dt=0.01$ for $\omega=[0.1, 10.0]$, $dt=0.002$ for $\omega=\{30, 50\}$ and $dt=0.001$ for $\omega=\{70,100\}$. This is a total of 153 data points that are analyzed in this work. A perfect collapse of these data points occur with the finite-size scaling, as explicitly shown in Fig.~3 of the main text.}
		\label{fig:rescaled_critical_time_vs_omega}
	\end{figure}

	%%%%%%%%%%%%%%%%%%%%%%%%%%%%%%%%%
	%%%%%%%%%%%%%%%%%%%%%%%%%%%%%%%%%
	%%%%%%%%%%%%%%%%%%%%%%%%%%%%%%%%%
	%%%%%%%%%%%%%%%%%%%%%%%%%%%%%%%%%
	
	\FloatBarrier
	
	\section{Frequency Dependence and Entanglement Hamiltonian's Floquet Inheritance}
	\label{app. section Frequency Dependence and Entanglement Hamiltonian's Floquet Inheritance}
	
	This section explores how TET evolve with driving frequency, revealing a fundamental connection to Floquet theory. We demonstrate that the entanglement Hamiltonian inherits Floquet-like periodicity from the driven system, enabling universal scaling across multiple transitions at high frequencies.

	\subsection{Entanglement Dynamics at Different Frequencies} 
	\label{app. subsection Entanglement Dynamics at Different Frequencies}
	
	We systematically investigate the frequency dependence of TET across three regimes: low frequency ($\omega \ll \min(J, h_0)$), intermediate frequency ($\omega \sim J, h_0$), and high frequency ($\omega \gg \max{J, h_0}$). Figure \ref{fig:frequency_dependence} shows representative dynamics for $\omega = 0.1$, $4.0$, and $30$ (we have already shown the plot for another intermediate frequency $\omega=5.0$ in Fig.~1 of the main text).
	
	At low frequencies (Fig.~\ref{fig:frequency_dependence}(a-c)), transitions occur irregularly with varying critical times $t_c^{(k)}$. The entanglement echo exhibits deep but non-periodic dips, and parity flips occur at seemingly random intervals. This reflects the system's inability to synchronize with the slow driving, leading to aperiodic entanglement reorganizations.
	
	At intermediate frequencies (Fig.~\ref{fig:frequency_dependence}(d-f)), transitions begin to regularize. The critical times $t_c^{(k)}$ approach periodicity, and the entanglement echo shows more regular vanishing points. This represents a crossover regime where the system starts to lock into the driving frequency.
	
	Most strikingly, at high frequencies (Fig.~\ref{fig:frequency_dependence}(g-i)), transitions become perfectly periodic with period $T_c \approx \text{constant}$ independent of $\omega$. The entanglement echo vanishes at precisely regular intervals, and parity flips occur with clock-like regularity. This high-frequency behavior demonstrates that the entanglement Hamiltonian $\mathcal{H}{\text{ent}}(t)$ inherits Floquet periodicity from the driven system, even though $\mathcal{H}{\text{ent}}(t)$ itself is not periodic.
	
	We explicitly show the time evolution of the two largest Schmidt values $\lambda_1(t)$ and $\lambda_2(t)$ in Fig.~\ref{fig:schmidt_values}. These correspond to the Schmidt gap closing for frequencies already plotted in Fig.~1 of the main text ($\omega =5.0$) as well as in Fig. \ref{fig:frequency_dependence} ($\omega=0.1, 4.0, 9.0$) here. The closing of the Schmidt gap $\Delta\lambda = \lambda_1 - \lambda_2$ at critical times $t_c^{(k)}$ manifests as near-degeneracy between these values. At low frequencies, the gap closing events occur irregularly, consistent with the aperiodic transitions observed in Fig.~\ref{fig:frequency_dependence}(a-c). At intermediate frequencies (Fig.~\ref{fig:schmidt_values}(d-f) and Fig.~1 of the main text), the gap closing becomes more regular, while at high frequency (Fig.~\ref{fig:schmidt_values}(g-i)), perfect periodicity emerges with $\lambda_1(t)$ and $\lambda_2(t)$ crossing at precisely regular intervals.
	
	Finally, we provide the plot of the re-scaled first critical time $t^*/L_A$ against the driving frequency in Fig.~\ref{fig:rescaled_critical_time_vs_omega}. As can be seen in the plot, there are a total of 153 data points (as analyzed in this work), that perfectly collapse as shown in Fig.~3 of the main text.
	
	The emergence of Floquet-like periodicity in the entanglement spectrum suggests that TET become intrinsic features of the driven steady state rather than transient effects. This inheritance mechanism explains why universal critical behavior persists across driving frequencies: at high frequencies, the entanglement Hamiltonian effectively samples from a time-independent ensemble described by the Floquet-Magnus expansion. We will explore this later in Sec. \ref{app. section Magnus-Floquet Effective Theory Comparison Plots}.

	%%%%%%%%%%%%%%%%%%%%%%%%%%%%%%%%%
	%%%%%%%%%%%%%%%%%%%%%%%%%%%%%%%%%
	
	\subsection{A Note on Periodicity of Transitions}
	\label{app. subsection A Note on Periodicity of Transitions}
	
	In order to quantify the alternating nature of TET, it is useful to track the expectation values of subsystem parity $P_A$ with respect to the two largest Schmidt vectors. At $t=0$, the leading Schmidt vector begins in the $\langle P_A\rangle=+1$ sector while the second-largest occupies $\langle P_A\rangle=-1$. The first nontrivial flip, denoted $t^*$, occurs when the leading vector spontaneously transitions to $\langle P_A\rangle=-1$ (and the subleading vector to $\langle P_A\rangle=+1$). This initial event establishes the baseline for two distinct periodicities:
	\begin{enumerate}
		\item \textit{Negative-parity residence time}: the interval during which the leading Schmidt vector remains in $\langle P_A\rangle=-1$ (and the second-largest in $\langle P_A\rangle=+1$), referred to as (odd) Period 1, 3, 5, \ldots. We call them \textit{odd periodicities}.
		\item \textit{Positive-parity residence time}: the subsequent interval before the leading vector returns to $\langle P_A\rangle=+1$ (and the second-largest to $\langle P_A\rangle=-1$), referred to as (even) Period 2, 4, 6, \ldots. We call them \textit{even periodicities}. 
	\end{enumerate}
	
	These two alternating periods characterize the flip-flop dynamics inherent to entanglement evolution under periodic driving. As the driving frequency $\omega$ increases, several key regimes emerge:
	\begin{itemize}
		\item Intermediate frequencies ($\omega \gtrsim 7.0$): Both odd and even periodicities stabilize individually, namely each becomes uniform within reasonable accuracy in duration across successive flips, although they remain unequal to one another. This partial locking indicates that the entanglement Hamiltonian has inherited aspects of the Floquet period without fully synchronizing to it.
		
		\item High frequencies ($\omega \gtrsim 10.0$): Not only do both alternating periods remain uniform, but odd and even periodicities converge to the same value, signifying complete Floquet-period inheritance by the entanglement Hamiltonian. In this regime, the effective Floquet-Magnus expansion (see Section \ref{app. section Magnus-Floquet Effective Theory Comparison Plots} below) governs the dynamics, endowing the entanglement Hamiltonian with an intrinsic timescale independent of the time-period of the drive.
	\end{itemize}
	
	Concomitantly, the total number of transitions within a fixed elapsed time also locks in at high frequencies, reflecting a mutual matching of alternating and overall periodicities. By contrast, at lower $\omega$, neither the residence times nor the transition counts align, as evidenced in Table \ref{tab:periodicity}. Note that $t^*$ marks the first flip event and is not itself counted as Period 1, since a true period requires both a beginning and an end within the simulation window.
	
	% Requires: \usepackage{graphicx} in the preamble
	\begin{table}[t]
		\centering
		\caption{Periodicity data as a function of driving frequency $\omega$. Note that $t^*$ marks the first flip event and is not itself counted as Period 1, since a true period requires both a beginning and an end within the simulation window. The first critical time $t^*$ and alternating periods (the leading Schmidt vector begins in the $\langle P_A\rangle=+1$ sector at $t=0$ $\longrightarrow$ Period 1: leading Schmidt vector in $\langle P_A\rangle=-1$ $\longrightarrow$ Period 2: leading Schmidt vector back in $\langle P_A\rangle=+1$; so on) are listed. At intermediate $\omega$, each residence time becomes uniform but unequal; above $\omega \gtrsim 10.0$, both periods equalize, indicating complete Floquet-period inheritance by the entanglement Hamiltonian. Parameters: $L = 24$, $L_A = 9$, $J = 1.0$, $h_0 = 2.0$. Maximum time and time steps are chosen as $T_{\text{max}}=20.0, dt=0.01$ for $\omega=\{0.1, 0.5\}$, $T_{\text{max}}=10.0, dt=0.01$ for $\omega=[1.0, 10.0]$, $T_{\text{max}}=10.0, dt=0.002$ for $\omega=\{30, 50\}$ and $T_{\text{max}}=10.0, dt=0.001$ for $\omega=\{70,100\}$.}
		\label{tab:periodicity}
		\setlength{\tabcolsep}{3pt}
		\resizebox{\columnwidth}{!}{%
			\begin{ruledtabular}
				\begin{tabular}{lrrrrrrrrrrrrrrrrr}
					Period & $\omega=$0.1 & 0.5 & 1.0 & 1.5 & 2.0 & 3.0 & 4.0 & 5.0 & 6.0 & 7.0 & 8.0 & 9.0 & 10.0 & 30.0 & 50.0 & 70.0 & 100.0 \\
					\hline
					First $t^*$  & 7.92 & 2.77 & 1.82 & 1.46 & 1.28 & 1.11 & 1.04 & 0.97 & 0.90 & 0.85 & 0.82 & 0.80 & 0.80 & 0.79 & 0.79 & 0.79 & 0.79 \\
					Period 1  & 3.46 & 1.36 & 0.98 & 0.87 & 0.82 & 1.40 & 1.79 & 1.98 & 1.72 & 1.70 & 1.64 & 1.63 & 1.63 & 1.57 & 1.57 & 1.57 & 1.57 \\
					Period 2  & 2.26 & 0.96 & 0.81 & 0.99 & 2.20 & 0.93 & 1.48 & 2.02 & 1.79 & 1.67 & 1.64 & 1.65 & 1.59 & 1.57 & 1.57 & 1.57 & 1.57 \\
					Period 3  & 1.78 & 0.82 & 0.95 & 2.18 & 0.83 & 1.83 & 1.68 & 2.14 & 1.85 & 1.65 & 1.64 & 1.60 & 1.63 & 1.57 & 1.57 & 1.57 & 1.57 \\
					Period 4  & 1.50 & 0.80 & 3.53 & 1.92 & 2.44 & 2.01 & 1.48 & 2.16 & 1.67 & 1.67 & 1.64 & 1.62 & 1.60 & 1.58 & 1.57 & 1.57 & 1.57 \\
					Period 5  & 1.33 & 0.82 & 0.98 & 1.88 & 0.81 & 1.19 & 1.67 &  & 1.86 & 1.69 & 1.64 & 1.65 & 1.62 & 1.57 & 1.57 & 1.57 & 1.57 \\
					Period 6  & 1.20 & 0.87 & 0.75 &  &  & 1.18 &  &  &  &  &  &  &  &  &  &  &  \\
					Period 7  &  & 1.48 &  &  &  &  &  &  &  &  &  &  &  &  &  &  &  \\
					Period 8  &  & 5.45 &  &  &  &  &  &  &  &  &  &  &  &  &  &  &  \\
					Period 9 &  & 1.32 &  &  &  &  &  &  &  &  &  &  &  &  &  &  &  \\
					Period 10 &  & 0.88 &  &  &  &  &  &  &  &  &  &  &  &  &  &  &  \\
				\end{tabular}
			\end{ruledtabular}
		}%
	\end{table}

	\subsection{Finite-Size Scaling for Full Data Set and Universality} 
	\label{app. subsection Finite-Size Scaling for Full Data Set and Universality}
	
	The Floquet inheritance demonstrated in Sec.~\ref{app. subsection A Note on Periodicity of Transitions} enables rigorous finite-size scaling analysis across multiple transitions. Throughout this subsection and in Fig.~2 of the main text, we perform scaling analyses by initializing the system at the equilibrium critical point ($h_0/2 = J = 1.0$), which ensures all three necessary conditions are robustly satisfied in the short-time dynamics: (1) the driven Hamiltonian preserves global $Z_2$ symmetry, (2) the critical ground state is a $Z_2$-eigenstate, and (3) the reduced density matrix $\rho_A(t)$ exhibits sustained sector competition with multiple sector-weight crossings $w_+(t^*)=w_-(t^*)$ within accessible short-time evolution windows. This initialization protocol optimizes the resolution of universal critical behavior by maximizing the number of well-defined temporal entanglement transitions available for scaling collapse.

\begin{figure}[htbp]
		\centering
		\includegraphics[width=\columnwidth]{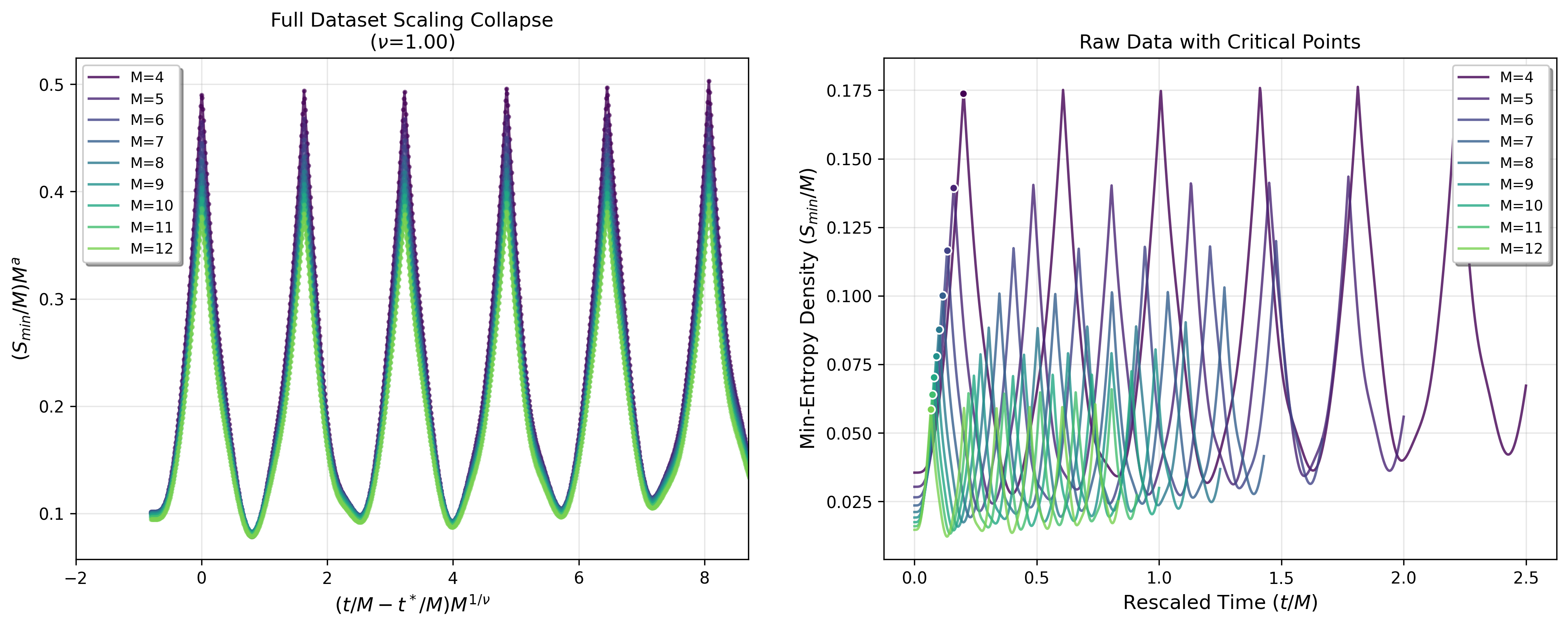}
		\caption{Finite-size scaling collapse of temporal entanglement transitions (TET) at high driving frequency ($\omega = 10.0$) for full data set. \textbf{Top:} Scaling collapse using Eq.~(2) of the main text with $\nu = 1.00$ across subsystem sizes $L_A = 4$--$12$. The universal data collapse demonstrates that the entanglement Hamiltonian inherits Floquet-like periodicity from the driven system, enabling scaling across multiple critical times $t_c^{(k)}$. \textbf{Bottom:} Raw data showing finite-size-dependent critical times before scaling. Parameters: $L=24$, $J=1.0$, $h_0=2.0$, $dt=0.01$.}
		\label{fig:scaling_full_data_om10}
	\end{figure}

	The Floquet inheritance enables scaling analysis across multiple transitions. Fig.~\ref{fig:scaling_full_data_om10} demonstrates finite-size scaling for the full dataset at $\omega = 10.0$, where we analyze all critical times $t_c^{(k)}$ for $k = 1, 2, 3, \ldots$ simultaneously.
	
	Remarkably, the same scaling ansatz $\Big($Eq.~(2) from the main text, namely $	\frac{\epsilon_0}{L_A} = \frac{1}{L_A^{a}} \mathcal{F}\left[ \left( \frac{t}{L_A} - \frac{t_c}{L_A} \right) L_A^{1/\nu } \right]\Big)$ with identical critical exponent $\nu \simeq 1.00$ describes all transitions across subsystem sizes $L_A = 4-12$. The universal data collapse (Fig.~\ref{fig:scaling_full_data_om10}a) confirms that TET belong to a single universality class regardless of their temporal order $k$ in $t_c^{(k)}$.
	
	This multi-transition scaling reveals several key insights:
	\begin{enumerate}
		\item Universality across transitions: The identical exponent for all $t_c^{(k)}$ indicates a common underlying fixed point governing entanglement reorganizations.
		
		\item Floquet steady state: The successful scaling across multiple periods demonstrates that transitions represent steady-state features rather than transient phenomena where the entanglement Hamiltonian develops an intrinsic timescale independent of the drive (see Fig.~3 of the main text). 
		
		\item Subsystem independence: The scaling holds for all $L_A$, suggesting the critical behavior is intrinsic to the entanglement Hamiltonian structure rather than specific to particular subsystem sizes.
	\end{enumerate}
	
\begin{figure}[htbp]
		\centering
		\includegraphics[width=\columnwidth]{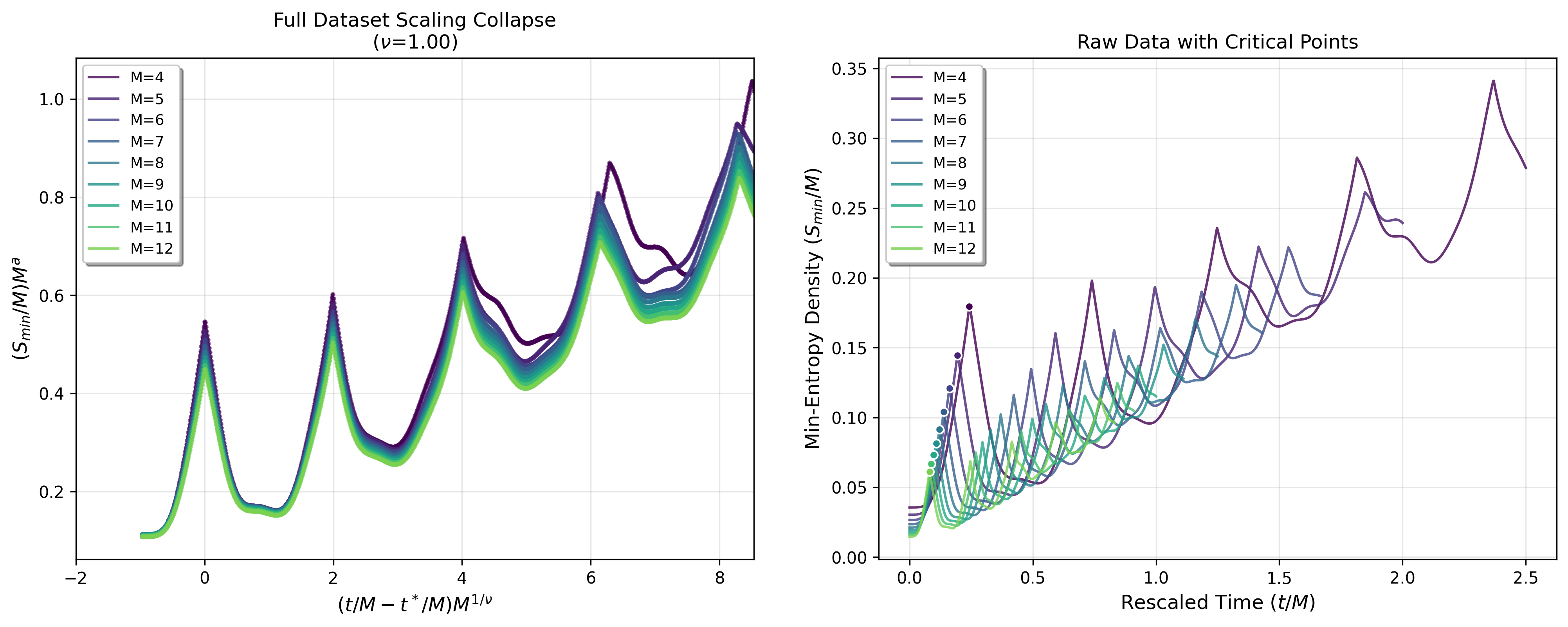}
		\caption{Finite-size scaling collapse at intermediate frequency ($\omega = 5.0$) showing partial Floquet inheritance. \textbf{Top:} Scaling collapse using Eq.~(2) of the main text with $\nu = 1.00$ across subsystem sizes $L_A = 4$--$12$. The collapse quality is intermediate between high-frequency perfection and low-frequency deterioration, reflecting the crossover regime where Floquet inheritance begins to emerge. \textbf{Bottom:} Raw data showing less regular critical times compared to $\omega=10.0$. Parameters: $L=24$, $J=1.0$, $h_0=2.0$, $dt=0.01$.}
		\label{fig:scaling_full_data_om5}
	\end{figure}

\begin{figure}[htbp]
		\centering
		\includegraphics[width=\columnwidth]{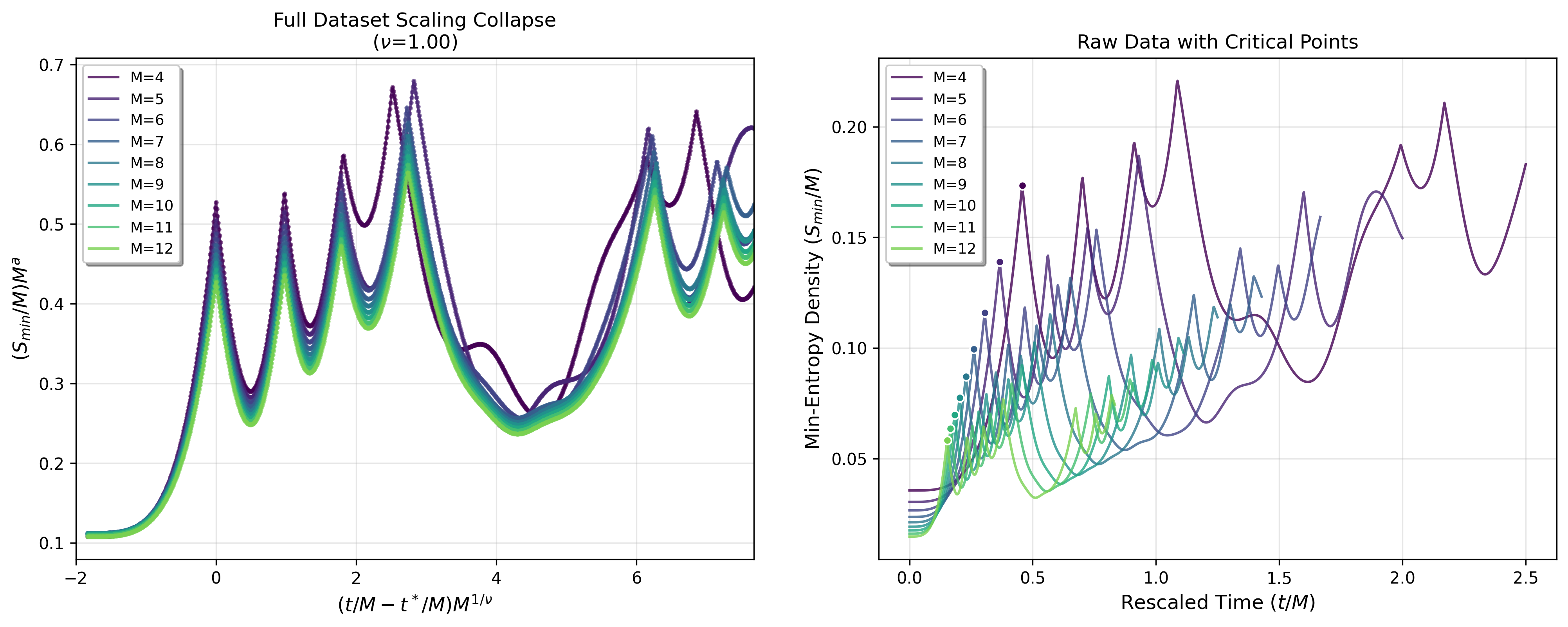}
		\caption{Finite-size scaling collapse at low frequency ($\omega = 1.0$) demonstrating breakdown of Floquet inheritance. \textbf{Top:} Attempted scaling collapse using Eq.~(2) of the main text with $\nu = 1.00$ shows significant deterioration, reflecting the absence of Floquet synchronization at low driving frequencies. \textbf{Bottom:} Raw data showing irregular, non-periodic critical times that prevent universal scaling across multiple transitions. Parameters: $L=24$, $J=1.0$, $h_0=2.0$, $dt=0.01$.}
		\label{fig:scaling_full_data_om1}
	\end{figure}

	The scaling collapse quality systematically deteriorates with decreasing frequency (Figs.~\ref{fig:scaling_full_data_om5} and \ref{fig:scaling_full_data_om1}), mirroring the progression from complete Floquet inheritance ($\omega=10.0$) through partial inheritance ($\omega=5.0$) to its absence ($\omega=1.0$) across all critical times $t_c^{(k)}$. This frequency-dependent scaling quality provides additional evidence for the Floquet inheritance mechanism: universal critical behavior emerges most clearly when the entanglement Hamiltonian can synchronize with the driving.
	
	These results establish TET as fundamental aspects of Floquet quantum matter, where entanglement spectra inherit dynamical symmetries and exhibit universal critical behavior distinct from conventional local observables.
	
	%%%%%%%%%%%%%%%%%%%%%%%%%%%%%%%%%
	%%%%%%%%%%%%%%%%%%%%%%%%%%%%%%%%%
	%%%%%%%%%%%%%%%%%%%%%%%%%%%%%%%%%
	%%%%%%%%%%%%%%%%%%%%%%%%%%%%%%%%%

	\section{Magnus-Floquet Effective Theory Comparison Plots}
	\label{app. section Magnus-Floquet Effective Theory Comparison Plots}
	
	The Floquet-Magnus expansion \cite{Bukov2015Mar} provides a powerful framework for understanding high-frequency driven systems through an effective time-independent Hamiltonian $\mathcal{H}_{\text{eff}}$ (provided in Eq.~(3) of the main text). We reproduce here the expression:
	\be
	\Hh_{\mathrm{eff}}=-J\left(1+\frac{h_0^2}{2 \omega^2}\right) \sum_{i=1}^{L-1} \sigma_i^z \sigma_{i+1}^z+\frac{h_0^2 J}{2 \omega^2} \sum_{i=1}^{L-1} \sigma_i^y \sigma_{i+1}^y -\frac{2 h_0 J^2}{\omega^2}\left(\sigma_1^x+\sigma_L^x\right)-\frac{4 h_0 J^2}{\omega^2} \sum_{i=2}^{L-1} \sigma_i^x  -\frac{4 h_0 J^2}{\omega^2} \sum_{i=2}^{L-1} \sigma_{i-1}^z \sigma_i^x \sigma_{i+1}^z + \Oo(\omega^{-3}).
	\label{eq:floquet appendix}
	\ee
	
	The systematic frequency dependence revealed in Figs.~\ref{fig:effective_ham_comparison_om100}--\ref{fig:effective_ham_comparison_om10} exposes a fundamental relationship between TET and the Floquet driving period $T = 2\pi/\omega$. At low-to-intermediate frequencies, the first critical time satisfies $t^* < T$, indicating that entanglement reorganization occurs \textit{within} a single Floquet cycle and is governed by the instantaneous time-dependent Hamiltonian $\mathcal{H}(t)$. Conversely, at high frequencies where the entanglement Hamiltonian also inherits Floquet-properties (as discussed above), we observe $t^* > T$ with critical times saturating to frequency-independent values (see Fig.~3 in the main text), demonstrating that transitions occur on timescales \textit{beyond} the driving period and are controlled by the effective Hamiltonian $\mathcal{H}_{\text{eff}}$. This crossover from sub-period to super-period critical dynamics reveals that the entanglement Hamiltonian $\mathcal{H}_{\text{ent}}(t)$ develops its own intrinsic timescales that decouple from the external driving frequency. The saturation phenomenon confirms that TET at high frequencies represent genuine steady-state properties of Floquet quantum matter rather than transient effects, with the effective theory accurately capturing both the transition timescales and universal critical behavior as demonstrated by the exceptional many-body fidelity ($|\langle\psi_{\text{exact}}(t)|\psi_{\text{eff}}(t)\rangle|^2 > 0.99$) across multiple Floquet cycles.
	
	To systematically validate the Floquet-Magnus effective theory across different frequency regimes, we examine the dynamics for five representative driving frequencies: $\omega = 100, 70, 50, 30, 10$ (Figs.~\ref{fig:effective_ham_comparison_om100}--\ref{fig:effective_ham_comparison_om10}). These frequencies span from the asymptotic high-frequency limit where the effective theory is expected to be highly accurate, down to lower frequencies where deviations become significant. The time step $dt$ is chosen appropriately for each frequency to ensure numerical stability while capturing the relevant dynamics: $dt = 0.001$ for $\omega = 100, 70$; $dt = 0.002$ for $\omega = 50, 30$; and $dt = 0.01$ for $\omega = 10.0$.
	
	At asymptotic frequencies ($\omega = 100, 70$ in Figs.~\ref{fig:effective_ham_comparison_om100} and \ref{fig:effective_ham_comparison_om70}), the Floquet-Magnus expansion achieves exceptional accuracy. The many-body state fidelity $|\langle\psi_{\text{exact}}(t)|\psi_{\text{eff}}(t)\rangle|^2$ remains above 99.9\% for $\omega = 100$ and above 99.5\% for $\omega = 70$ throughout the evolution, confirming that $\mathcal{H}_{\text{eff}}$ captures the essential physics with negligible deviations. Both entanglement entropy dynamics and Schmidt gap $\Delta\lambda (t) = \lambda_0 (t) - \lambda_1 (t)$ ($\lambda_0$ and $\lambda_1$ are the largest and second largest Schmidt values, respectively) show near-perfect overlap between exact time evolution and effective Hamiltonian evolution, with temporal entanglement transition critical times $t_c^{(k)}$ matching precisely.
	
	At intermediate-to-asymptotic frequencies ($\omega = 50, 30$ in Figs.~\ref{fig:effective_ham_comparison_om50} and \ref{fig:effective_ham_comparison_om30}), the agreement remains excellent at $\omega = 50$, with fidelity above 96\% throughout evolution. At $\omega = 30$ (Fig.~\ref{fig:effective_ham_comparison_om30}), while the short-time dynamics show good agreement, subtle deviations emerge in the long-time behavior where fidelity gradually decays but remains above 80\% by the end of evolution. Nevertheless, the effective Hamiltonian continues to accurately capture both the entanglement entropy growth patterns and the temporal entanglement transition critical times, confirming these are genuine steady-state features rather than transient artifacts of the driving protocol.
	
	Finally at $\omega = 10.0$ in Fig.~\ref{fig:effective_ham_comparison_om10}, the limitations of the truncated Floquet-Magnus expansion become evident. While the effective description qualitatively reproduces the exact evolution and captures transition times with reasonable accuracy in the short-time dynamics while capturing the first critical time $t^*$ quantitatively, substantial deviations emerge at later times. The state fidelity exhibits systematic decay from unity, approaching zero by the end of the evolution window. This deteriorating agreement indicates that higher-order corrections in the Floquet-Magnus expansion ($\mathcal{O}(\omega^{-3})$ and beyond) become comparable to the leading-order terms, significantly limiting the long-term predictive power while still providing valuable insight into the initial transition dynamics.
	
	Crucially, across all frequencies, the fidelity begins at unity as theoretically required, since both exact and effective evolutions start from identical initial states. The systematic frequency-dependent decay pattern validates both our numerical implementation and the theoretical expectation that Floquet-Magnus accuracy deteriorates as higher-order $1/\omega$ corrections become comparable to the leading-order terms in Eq.~\eqref{eq:floquet appendix}.
	
	These results collectively demonstrate that TET represent genuine steady-state features of the driven system, accurately captured by the Floquet-Magnus effective Hamiltonian across a broad frequency range. The persistence of universal critical behavior in both exact (for all frequencies) and effective dynamics (at higher frequencies) establishes these transitions as fundamental aspects of Floquet quantum matter, independent of specific driving protocol details. Importantly, the maintained accuracy at high frequencies confirms that these phenomena are not transient effects but rather intrinsic properties of the asymptotic steady-state dynamics.

	\begin{figure}[htbp!]
		\centering
		\includegraphics[width=\columnwidth]{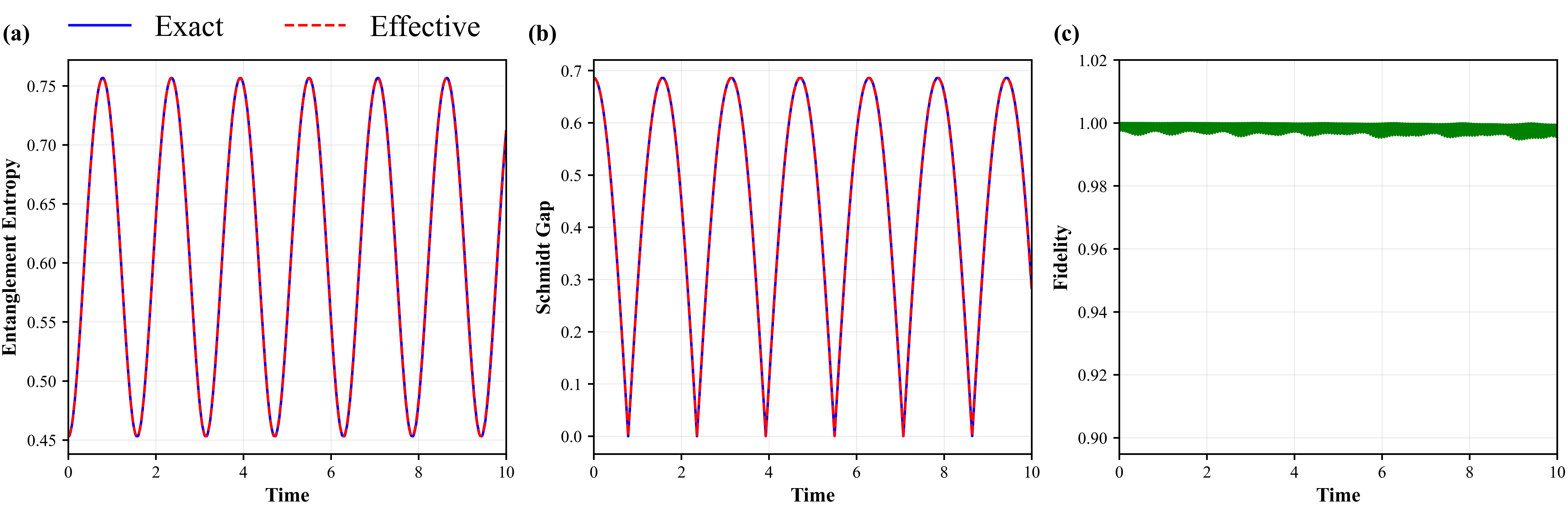}
		\caption{Systematic validation of Floquet-Magnus effective Hamiltonian across driving frequencies.
			\textbf{(a)} Entanglement entropy dynamics comparing exact evolution under $\mathcal{H}(t)$ (solid blue) and effective evolution under $\mathcal{H}_{\text{eff}}$ (dashed red).
			\textbf{(b)} Schmidt gap $\Delta\lambda = \lambda_0 - \lambda_1$ dynamics showing temporal entanglement transition critical times.
			\textbf{(c)} Many-body state fidelity $|\langle\psi_{\text{exact}}(t)|\psi_{\text{eff}}(t)\rangle|^2$ demonstrating frequency-dependent accuracy of the effective theory.
			From top to bottom: $\omega = 100$ ($dt = 0.001$) (this figure), $\omega = 70$ ($dt = 0.001$) (Fig.~\ref{fig:effective_ham_comparison_om70}), $\omega = 50$ ($dt = 0.002$) (Fig.~\ref{fig:effective_ham_comparison_om50}), $\omega = 30$ ($dt = 0.002$) (Fig.~\ref{fig:effective_ham_comparison_om30}), $\omega = 10.0$ ($dt = 0.01$) (Fig.~\ref{fig:effective_ham_comparison_om10}).
			Parameters: $L=24$, $L_A=9$, $J=1.0$, $h_0=2.0$.}
		\label{fig:effective_ham_comparison_om100}
	\end{figure}
	
	\begin{figure}[htbp!]
		\centering
		\includegraphics[width=\columnwidth]{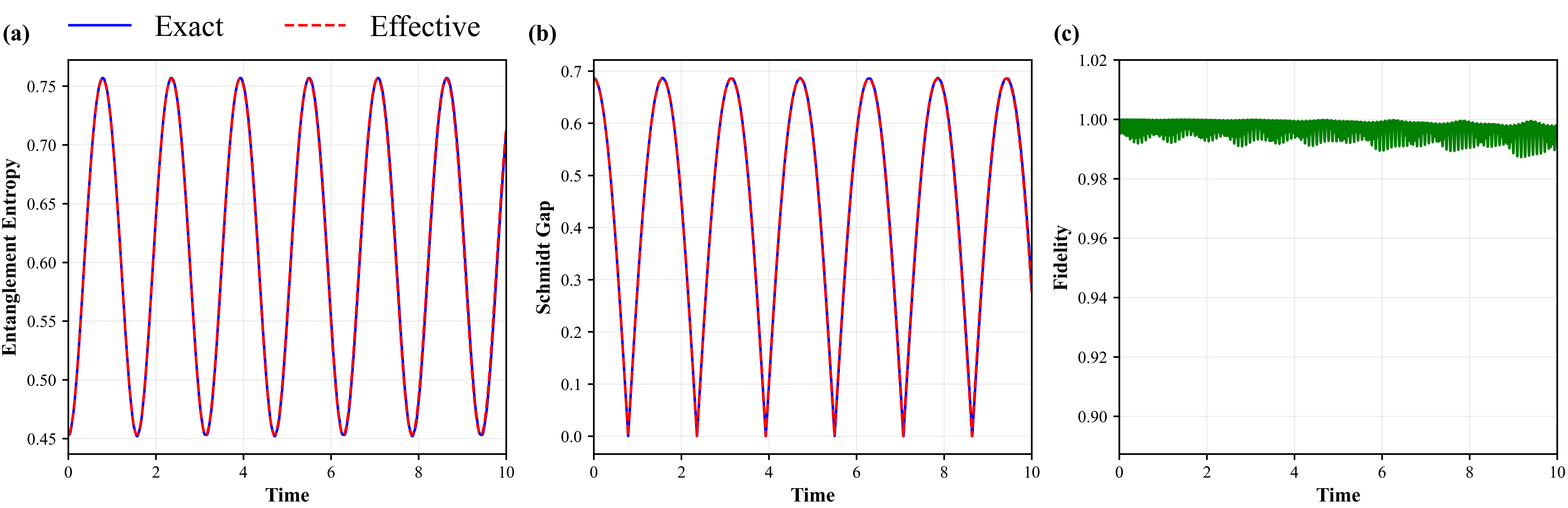}
		\caption{Continued from Fig.~\ref{fig:effective_ham_comparison_om100}: $\omega = 70$.}
		\label{fig:effective_ham_comparison_om70}
	\end{figure}
	
	\begin{figure}[htbp!]
		\centering
		\includegraphics[width=\columnwidth]{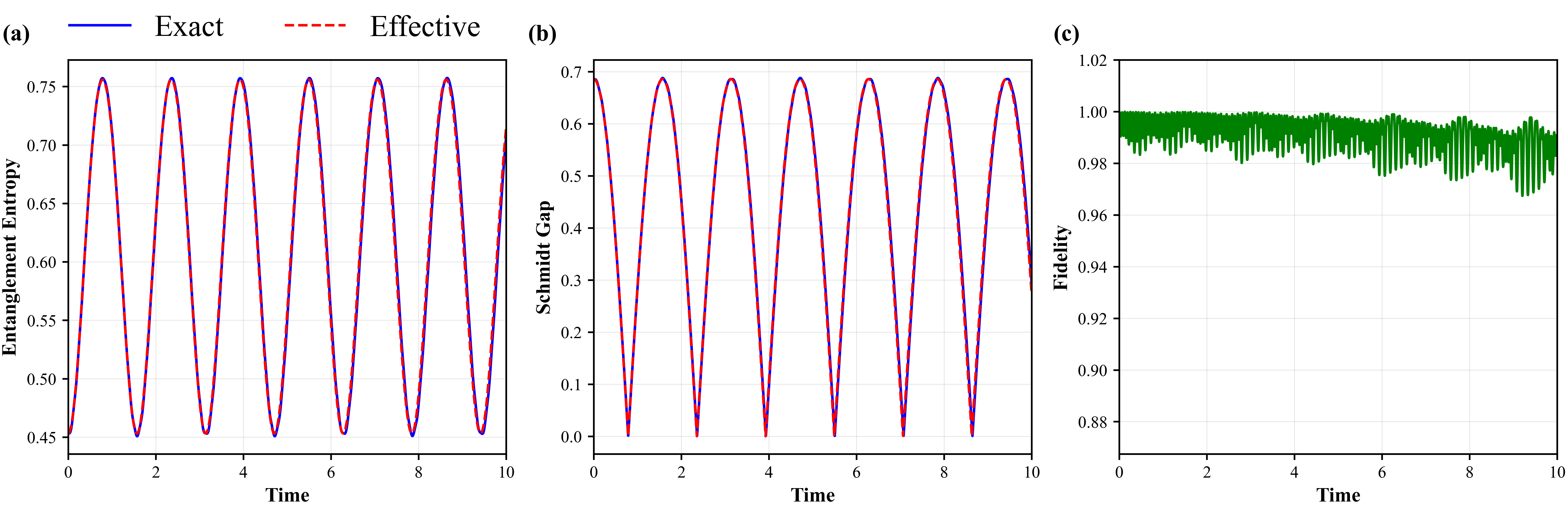}
		\caption{Continued from Fig.~\ref{fig:effective_ham_comparison_om100}: $\omega = 50$.}
		\label{fig:effective_ham_comparison_om50}
	\end{figure}
	
	\begin{figure}[htbp!]
		\centering
		\includegraphics[width=\columnwidth]{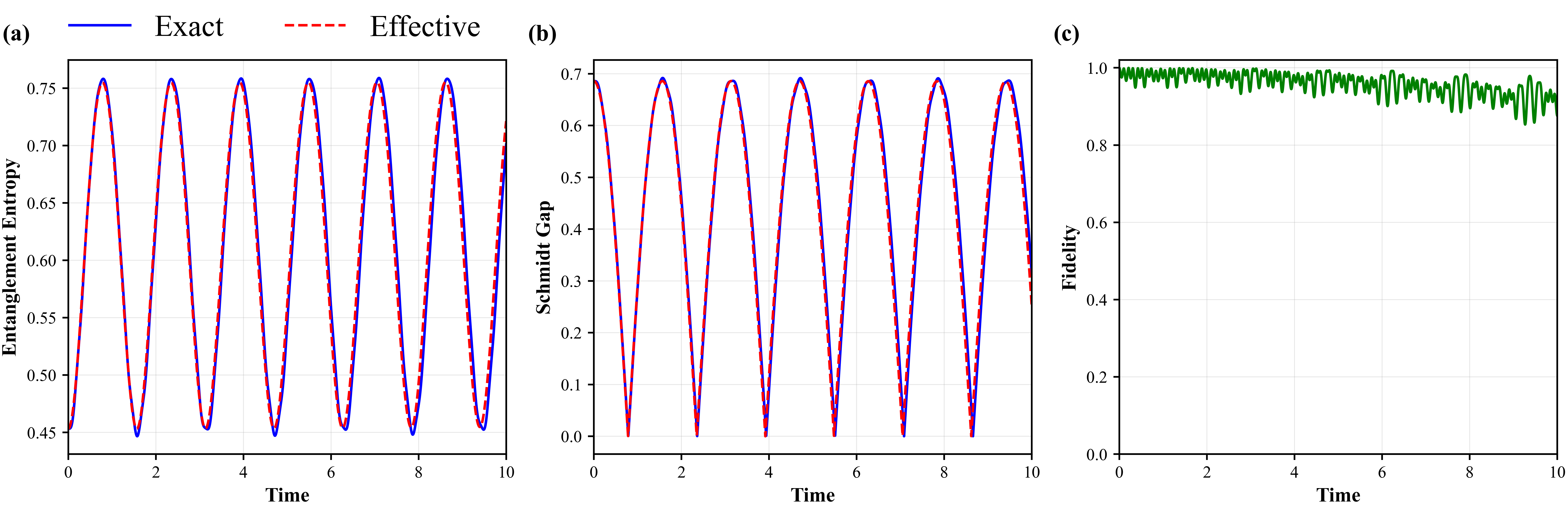}
		\caption{Continued from Fig.~\ref{fig:effective_ham_comparison_om100}: $\omega = 30$.}
		\label{fig:effective_ham_comparison_om30}
	\end{figure}
	
	\begin{figure}[htbp!]
		\centering
		\includegraphics[width=\columnwidth]{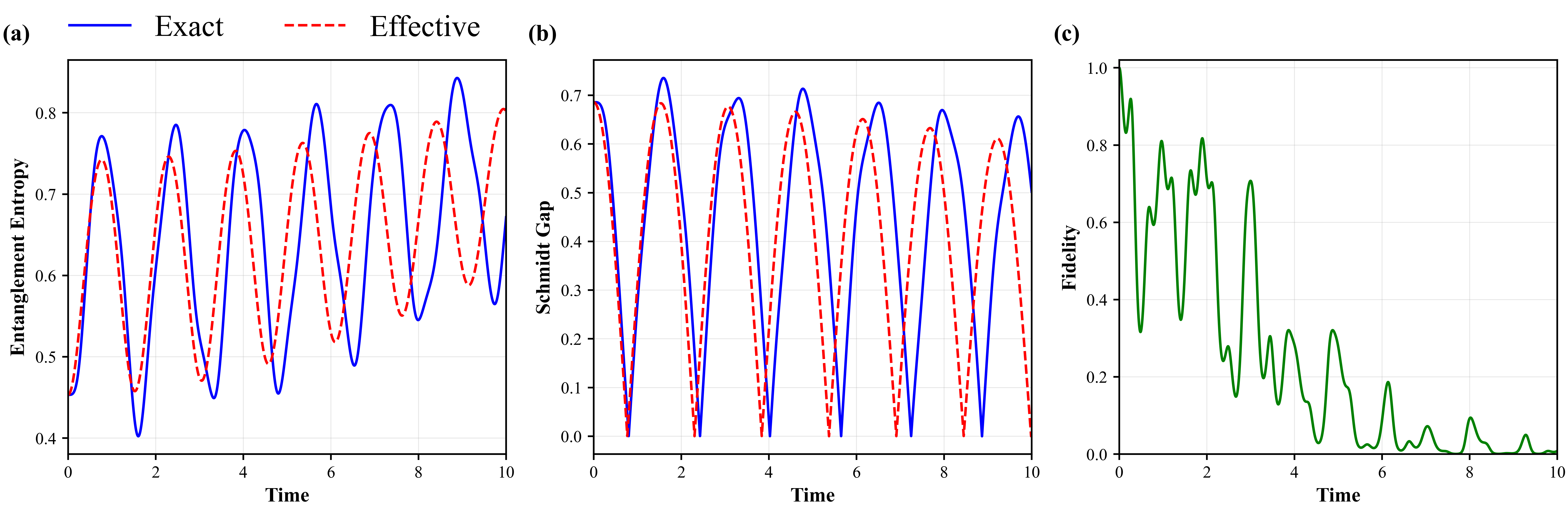}
		\caption{Continued from Fig.~\ref{fig:effective_ham_comparison_om100}: $\omega = 10.0$.}
		\label{fig:effective_ham_comparison_om10}
	\end{figure}

	\bibliography{refs.bib}